\documentclass[11pt]{article}
\usepackage{jheppub}
\usepackage{hyperref}
\usepackage{amsmath,amssymb,amsthm,amscd}
\usepackage[all,cmtip]{xy}
\usepackage{xcolor}
\usepackage{blkarray}
\usepackage{float}
\usepackage{multirow}

\newtheorem{definition}{Definition}[section]

\usepackage{tikz,textcomp}
\usetikzlibrary{decorations.pathreplacing,calc,fadings,fit,shapes,arrows,positioning,chains,matrix}

\DeclareFontFamily{U}{wncy}{}
\DeclareFontShape{U}{wncy}{m}{n}{<->wncyr10}{}
\DeclareSymbolFont{mcy}{U}{wncy}{m}{n}
\DeclareMathSymbol{\Sh}{\mathord}{mcy}{"58} 

\numberwithin{equation}{section}

\title{In search of almost generic Calabi-Yau 3-folds}
\abstract{
  We call a projective Calabi-Yau (CY) 3-fold almost generic if it has only isolated nodes as singularities and the homology classes of all of the exceptional curves in an analytic small resolution are non-trivial but torsion. Such a Calabi-Yau supports a topologically non-trivial flat B-field and the corresponding A-model topological string partition function encodes a torsion refinement of the Gopakumar-Vafa invariants of the smooth deformation. Our goal in this paper is to find new examples of almost generic CY 3-folds, using both conifold transitions as well as the integral structure of the periods of the mirrors. In this way we explicitly construct two quintic CY 3-folds with $\mathbb{Z}_2$-torsion, two octics with $\mathbb{Z}_3$-torsion and deduce the existence of a complete intersection $X_{(6,6)}\subset\mathbb{P}^5_{1,1,2,2,3,3}$ with $\mathbb{Z}_5$-torsion. Via mirror symmetry, the examples give new geometric interpretations to several AESZ Calabi-Yau operators. The mirror periods of the almost generic $X_{(6,6)}$ with non-trivial B-field topology are annihilated by an irrational Picard-Fuchs operator. We describe how the usual integral structure of the periods has to be modified and in all of the cases we calculate the monodromies around the singular points to verify integrality. Additional points of maximally unipotent monodromy in the moduli spaces lead us to find several more examples of smooth or almost generic CY 3-folds and to conjecture new twisted derived equivalences. We integrate the holomorphic anomaly equations and extract the torsion refined Gopakumar-Vafa invariants up to varying genus. For our construction of the almost generic octic CY 3-folds, we also give a short introduction to the subject of hypermatrices and hyperdeterminants.

}

\author[a]{Thorsten Schimannek}
\affiliation[a]{
Institute for Theoretical Physics \& Department of Mathematics, Utrecht University,
3584 CC Utrecht, The Netherlands
}
\emailAdd{thorsten.schimannek@gmail.com}
\begin{document}

\maketitle
\flushbottom

\section{Introduction}
The calculation of genus zero Gromov-Witten invariants of the quintic Calabi-Yau threefold using mirror symmetry in~\cite{CANDELAS199121} gave rise to a rich and beautiful interplay between string theory and mathematics, the consequences of which still appear far from exhausted.
The genus zero calculation has been quickly generalized to other Calabi-Yau threefolds~\cite{Hosono:1993qy,Hosono:1994ax,Candelas:1993dm,Candelas:1994hw}.
Subsequently, using the holomorphic anomaly equations derived in~\cite{Bershadsky:1993ta,Bershadsky:1993cx} has in many cases allowed to obtain results to relatively high genus~\cite{Katz:1999xq,Huang:2006hq,Alexandrov:2023zjb,Alexandrov:2023ltz}.

Certain singular Calabi-Yau varieties can support a flat but topologically non-trivial B-field that regularizes the string worldsheet theory and obstructs deformations that remove the singularities~\cite{Vafa:1994rv,Aspinwall:1995rb}.
The A-model topological string partition function depends on the topology of the B-field and it has more recently been observed~\cite{Schimannek:2021pau,Katz:2022lyl,Katz:2023zan} that the partition functions for different choices of the topology encode a torsion refinement of the Gopakumar-Vafa invariants~\cite{Gopakumar:1998ii,Gopakumar:1998jq} of the smooth deformation of the underlying geometry.

For the examples considered in~\cite{Aspinwall:1995rb,Schimannek:2021pau,Katz:2022lyl,Katz:2023zan}, the phenomenon arises when a projective Calabi-Yau threefold $X$ has isolated nodal singularities and, given any complex analytic small resolution $\widehat{X}$ of $X$, every exceptional curve represents a non-trivial torsion class in $H_2(\widehat{X},\mathbb{Z})$.
We will refer to such geometries as~\textit{almost generic Calabi-Yau threefolds} (AGCY--3).
We refer to
\begin{align}
    B(X)=\text{Tors}\,H^3(\widehat{X},\mathbb{Z})\simeq \text{Tors}\,H_2(\widehat{X},\mathbb{Z})\,,    
    \label{eqn:almostGenericTorsion}
\end{align}
as the torsion of the geometry, where the second isomorphism in~\eqref{eqn:almostGenericTorsion} follows from the universal coefficient theorem.
The torsion is independent of the choice of small resolution $\widehat{X}$ up to isomorphism.
The precise definition will be formulated in Section~\ref{sec:almostGeneric} where we also review some of the relevant geometric background.

The name ``almost generic'' is motivated by the fact that an AGCY--3 $X$ always has a smooth deformation $\widetilde{X}$ and never admits a small resolution that is projective or K\"ahler.
This will also be elaborated on in Section~\ref{sec:almostGeneric}.
Almost generic Calabi-Yau threefolds therefore correspond to a special sublocus in the moduli space of complex structures on the Calabi-Yau $\widetilde{X}$ where the geometry is not generic enough to be smooth but at the same time not singular enough to admit a projective small resolution.~\footnote{We exclude nodes that are resolved by exceptional curves that are homologically trivial because they do not lead to interesting physical effects and can always be removed by a complex structure deformation. From the perspective of M-theory such nodes lead to localised uncharged matter, as has also been discussed in~\cite{Arras:2016evy,Grassi:2018rva}.}

It has been argued in~\cite{Schimannek:2021pau,Katz:2022lyl,Katz:2023zan}, building on~\cite{Aspinwall:1995rb}, that the possible topologies of flat B-fields on an AGCY--3 $X$ are in one-to-one correspondence with elements $\alpha\in B(X)$.
In the following, we will denote by $(X,\alpha)$ an almost generic Calabi-Yau threefold $X$ together with a choice of B-field topology $\alpha \in B(X)$.
We call the tuple $(X,\alpha)$ a \textit{Calabi-Yau background}.
The B-field stabilizes the geometry in the sense that the nodes that support a non-trivial B-field holonomy are protected against deformation and the string worldsheet theory is expected to be regular.

Examples of almost generic Calabi-Yau threefolds have been obtained in~\cite{Schimannek:2021pau} by considering relative moduli spaces of line bundles on genus one fibered Calabi-Yau threefolds, using results from~\cite{Caldararu2002}.
Due to the fibration structure, the complex dimension $b_2(X)$ of the corresponding stringy K\"ahler moduli is at least two.
On the other hand, examples with one-dimensional moduli spaces, i.e. $b_2(X)=1$, have been obtained in~\cite{Aspinwall:1995rb,Katz:2022lyl,Katz:2023zan} by considering Calabi-Yau double covers of $\mathbb{P}^3$ that are ramified over symmetric determinantal octic surfaces.
The torus fibered examples studied in~\cite{Schimannek:2021pau}, and subsequently in~\cite{Dierigl:2022zll}, exhibit torsion $B(X)= \mathbb{Z}_N$ with $N=2,\ldots,5$.~\footnote{We will always write $\mathbb{Z}_N=\mathbb{Z}/N\mathbb{Z}$.}
The symmetric determinantal double covers that have been constructed in~\cite{Aspinwall:1995rb,Katz:2022lyl,Katz:2023zan} yield examples with torsion $B(X)=\mathbb{Z}_2$.

This naturally leads to the question if there exist AGCY--3 $X$ with $b_2(X)=1$ that are not double covers and/or have torsion $B(X)=\mathbb{Z}_N$ with $N>2$.

In this paper we will show that the response in both cases is positive.
To this end, we will construct almost generic nodal quintic Calabi-Yau threefolds $X_1,X_2$ with torsion $\mathbb{Z}_2$, almost generic nodal octics $X_3,X_4$ with torsion $\mathbb{Z}_3$ and we will argue that there exists an almost generic Calabi-Yau threefold $X_5$, that is a degeneration of a smooth complete intersection $X_{(6,6)}\subset \mathbb{P}^5_{1,1,2,2,3,3}$, with torsion $\mathbb{Z}_5$.
The geometries are summarized in Table~\ref{tab:geometries}.
In each of these examples we will use mirror symmetry and integrate the holomorphic anomaly equations in order to calculate the corresponding torsion refined Gopakumar-Vafa invariants.

Our construction of the geometries $X_a$ for $a=1,\ldots,4$ makes use of conifold transitions between Calabi-Yau threefolds, in the spirit of~\cite{Candelas1990,Candelas:1989js,Batyrev:1998kx}.
Such transitions have already been used to study the properties of almost generic Calabi-Yau double covers of $\mathbb{P}^3$~\cite{Katz:2022lyl}.
By identifying the dual transitions under mirror symmetry, we can extract the Picard-Fuchs operators that annihilate the periods of the mirrors of the Calabi-Yau backgrounds $(X_a,\alpha)$.
This allows us to calculate the topological string free energies at genus $g=0,1$ and by integrating the holomorphic anomaly equations we obtain results up to varying genus $g\ge 2$.
On the other hand, we deduce the existence of the almost generic Calabi-Yau threefold $X_5$ by first constructing an irrational differential operator, using a certain Hadamard product, and then deducing the topological invariants of a geometric interpretation by studying the monodromies of the corresponding periods.

Providing new geometric interpretations for such differential operators is in fact one of the motivations for our work.
The genus zero topological string free energy associated to the Calabi-Yau background $(X,\alpha)$ is related to the variation of Hodge structure of the mirror Calabi-Yau $W_\alpha$.
The periods of the holomorphic 3-form on $W_\alpha$ are annihilated by a differential operator, the so-called Picard-Fuchs operator.~\footnote{For cases with $b_2(X)>1$ one has to consider Picard-Fuchs ideals that are generated by multiple operators but we are focusing on examples with $b_2(X)=1$.}
Differential operators that have certain characteristic properties of such Picard-Fuchs operators are called Calabi-Yau operators~\cite{Almkvist:2004kj,vanstraten2017calabiyauoperators}.
A large number of Calabi-Yau operators have been constructed independently of an underlying geometry, see for example~\cite{Almkvist:2004kj,almkvist2010tablescalabiyauequations,vanstraten2017calabiyauoperators,Almkvist:2023yja}, and in many cases a geometric realization is still not known.
The operators that have been constructed in~\cite{almkvist2010tablescalabiyauequations} are referred to as AESZ operators, while more operators can be found in the corresponding online database~\cite{cyopdatabase}.
In some of the cases, it is clear that the operator does not admit an integral period basis that is compatible with the usual expressions for the prepotential in terms of the topological invariants of a smooth Calabi-Yau $X$ with $b_2(X)=1$.~\footnote{This is because the coefficient of the term $\zeta(3)/(2\pi{\rm i})^3$, that can often be fixed by demanding that the imaginary part of the monodromy around the nearest conifold point vanishes, is positive and can therefore not be the Euler characteristic of a smooth Calabi-Yau with Picard rank one. We observe this for the mirror periods of our examples $(Y_2,[1]_2)$, $(X_3,[\pm 1]_3)$ and $(Y_4,[1]_2)$, with the respective Picard-Fuchs operators being AESZ~$225^*$, AESZ~199 and $4.70^*$, see Appendix~\ref{sec:dataY2},~\ref{sec:dataX3} and~\ref{sec:dataY4}.}
Using a modified expression for the integral structure, that will be discussed in Section~\ref{sec:BmodelGenus0}, for cases when $X$ is almost generic and carries a flat but topologically non-trivial B-field, we will resolve this puzzle and provide a new geometric interpretation for several Calabi-Yau operators.

In cases where the torsion is $\mathbb{Z}_5$, the modified structure of the periods also allows to understand the underlying integrality and to give a geometric interpretation to certain Picard-Fuchs operators with irrational coefficients.
Inhomogeneous irrational Picard-Fuchs operators, related to open topological strings, have appeared in the literature already in~\cite{Walcher:2012zk,Laporte:2012hv,Jefferson:2013vfa}.
More generally, the integrality properties of irrational Picard-Fuchs operators have subsequently been studied for example in~\cite{Schwarz:2013zua,Schwarz:2017gnp,haney2024infinityinnerproductsopen}.~\footnote{As was pointed out to us by Johannes Walcher, after we have presented our results at the workshop ``The Arithmetic of Calabi-Yau Manifolds'' in Mainz 2025, March 17--28, examples of irrational Picard-Fuchs operators with coefficients in $\mathbb{Q}[\sqrt{5}]$ have also been presented by him in previous talks and will be discussed in~\cite{WalcherWIP}. To our knowledge, both the interpretation of the integral structure and the concrete examples are different from the one discussed in Section~\ref{sec:example3}.}
However, a clear enumerative interpretation in the homogeneous case, that also extends to higher genus, has long been missing.
One such interpretation was provided in~\cite{Schimannek:2021pau}, where an almost generic Calabi-Yau threefold that is torus fibered and exhibits $\mathbb{Z}_5$-torsion has been constructed, building on the study of genus one fibrations with 5-sections in~\cite{Knapp:2021vkm}.
In that example, the periods of the mirror are annihilated by a system of Picard-Fuchs operators with coefficients in $\mathbb{Q}[\sqrt{5}]$.
Nevertheless, the corresponding torsion refined Gopakumar-Vafa invariants turned out to be integral.

We describe the, to our knowledge, first example of an irrational Picard-Fuchs operator in one variable that admits such an interpretation.
We argue that it corresponds to the mirror of an almost generic degeneration of the Calabi-Yau threefold $X_{(6,6)}\subset \mathbb{P}^5_{1,1,2,2,3,3}$ with $\mathbb{Z}_5$-torsion and a non-trivial B-field topology.
This allows us to find a basis of periods that has integral monodromies around all singular points in the moduli space.
We also show the integrality of the $\mathbb{Z}_5$-refined Gopakumar-Vafa invariants for genus $g=0,1,2$ and check the modified expression for the constant map contributions at genus $g=2$.

In all of our examples, the complex structure moduli space of the mirror exhibits a second limit point of maximally unipotent monodromy (MUM-point).
In each case this admits an interpretation as being mirror to another Calabi-Yau background.
In some cases the dual Calabi-Yau is smooth while in others it is almost generic and carries itself a non-trivial B-field topology.~\footnote{The properties of the geometries $Y_a$ for $a=1,\ldots,4$ can again be inferred by following sequences of flops and conifold transitions. We have decided to omit this somewhat tedious discussion as the properties can also be deduced from the integral periods.}
We denote the geometries that are associated to a second limit point in the moduli space obtained from $X_a$ for $a=1,\ldots,4$ respectively by $Y_a$, $a=1,\ldots,4$.
Their properties are also summarized in Table~\ref{tab:geometries}.
D-brane transport between the path is known to imply equivalences of the associated categories of topological B-branes.
This leads us to conjecture the following twisted derived equivalences
\begin{align}
\begin{split}
    D^b(\widehat{X}_1,[1]_2)&\simeq D^b(Y_1)\,,\quad D^b(\widehat{X}_2,[1]_2)\simeq D^b(\widehat{Y}_2,[1]_2)\,,\\
    D^b(\widehat{X}_3,[\pm 1]_3)&\simeq D^b(Y_3)\,,\quad D^b(\widehat{X}_4,[\pm 1]_3)\simeq D^b(\widehat{Y}_4,[1]_2)\,,
\end{split}
\end{align}
where by $[k]_N$ we denote the equivalence class of $k$ in $\mathbb{Z}_N$.
A brief introduction to twisted sheaves can be found, for example, in~\cite[Section 4.4]{Katz:2022lyl} but will not be necessary for the remainder of this paper.

A further motivation of our work is that one might ask if almost generic degenerations are an exception that only occurs for special Calabi-Yau threefold.
This is of course closely related to the question whether torus fibrations and double covers are somehow privileged in this regard.
With our construction of almost generic Calabi-Yau quintics we provide evidence for the idea that almost generic degenerations are, in fact, generic and that for any given Calabi-Yau threefold with a sufficiently large number of complex structure parameters one can expect that there exists a multitude of almost generic degeneration loci in the moduli space.

Finally, the fact that the nodes in an almost generic Calabi-Yau threefold can be protected against deformations by a non-trivial B-field topology means that the Calabi-Yau background $(X,\alpha)$ deforms in a moduli space of its own.
The corresponding string vacua form a corner of the string landscape that is largely unexplored.
In particular, the fact that the B-field effectively reduces the number of complex structure moduli without increasing the number of K\"ahler moduli has potential phenomenological interest.
However, exploring these questions will be beyond the scope of this work.

\begin{table}[ht!]\renewcommand{\arraystretch}{1.2}
    \centering
    \begin{tabular}{|c|c|c|c|c|c|c|c|c|}\hline
    CY $X$ & $\kappa$ & $\# S$ & $\chi(\widehat{X})$ & $B(X)$ & AESZ & $\widetilde{X}$ & $\chi(\widetilde{X})$ & $\tilde{b}$ \\\hline
    $X_1$ & 5 & 54 & -92 & $\mathbb{Z}_2$ & 203~\eqref{eqn:AESZ203} & $X_{(5)}\subset\mathbb{P}^4$ & $-200$ & 50 \\
    $Y_1$ & 38 & 0 & -92 &  & $202$~\eqref{eqn:AESZ202} & $Y_1$ & $-92$ & 80 \\\hline
    $X_2$ & 5 & 48 & -104 & $\mathbb{Z}_2$ & 222~\eqref{eqn:AESZ222} & $X_{(5)}\subset\mathbb{P}^4$ & $-200$ & 50 \\
    $Y_2$ & 2 & 96 & -104 & $\mathbb{Z}_2$ & $225^*~\eqref{eqn:AESZ225}$ & $X_{(8)}\subset\mathbb{P}^4_{1,1,1,1,4}$ & $-296$ & 44 \\\hline
    $X_3$ & 2 & 104 & -88 & $\mathbb{Z}_3$ & 199~\eqref{eqn:AESZ199} & $X_{(8)}\subset\mathbb{P}^4_{1,1,1,1,4}$ & $-296$ & 44 \\
    $Y_3$ & 34 &  0 & -88 &  & $194$~\eqref{eqn:AESZ194} & $Y_3$ & $-88$ & 76 \\\hline
    $X_4$ & 2 & 100 & -96 & $\mathbb{Z}_3$ & 350~\eqref{eqn:AESZ350} & $X_{(8)}\subset\mathbb{P}^4_{1,1,1,1,4}$ & $-296$ & 44 \\
    $Y_4$ & 2 & 100 & -96 & $\mathbb{Z}_2$ & $ 4.70^*$~\eqref{eqn:op470} & $X_{(8)}\subset\mathbb{P}^4_{1,1,1,1,4}$  & $-296$ & 44 \\\hline
    $X_5$ & 1 & 30+20 & -20 & $\mathbb{Z}_5$ & {\eqref{eqn:irrationalOpOrder4}} & $X_{(6,6)}\subset\mathbb{P}^5_{1,1,2,2,3,3}$ & $-120$ & 22 \\\hline
    \end{tabular}
    \label{tab:geometries}
    \caption{The different almost generic or smooth Calabi-Yau threefolds that appear in our examples, together with some of the relevant topological invariants. Here $\kappa$ is the triple intersection number $\kappa=\widetilde{J}\cdot\widetilde{J}\cdot \widetilde{J}$ of the smooth deformation $\widetilde{X}$ while $\tilde{b}=\widetilde{J}\cdot c_2(T\widetilde{X})$. For the Picard-Fuchs operators we list the AESZ number, except for the cases $X_4$, for which the operator can be found in the database~\cite{cyopdatabase}, and $X_5$ for which the operator is irrational and we construct it in Section~\ref{sec:example3}. The $*$ indicates that $225^*$ and $4.70^*$ are obtained via pullback along a two-to-one covering map from the operators $225$ and $4.70$.}
\end{table}

We recently learned about work in progress that, among other things, also discusses an example of an almost generic Calabi-Yau threefold $X$ with $b_2(X)=1$ and $\mathbb{Z}_3$-torsion~\cite{Knapp:2025}.
To our knowledge, the example is different from the ones discussed in this paper and provides a geometric interpretation for yet another AESZ operator.
We thank the authors for informing us about their work and for coordinating our submissions.

\paragraph{Outline}
The outline of this work is as follows.
In Section~\ref{sec:almostGenericDef}, we will first review some relevant geometric background and then introduce the notion of an almost generic Calabi-Yau threefold.
In Section~\ref{sec:conifoldTransitions} we then outline our general strategy to construct examples of almost generic type by using conifold transitions, starting with Calabi-Yau threefolds that are complete intersections in products of projective spaces.
In Section~\ref{sec:topologicalStrings} we review the properties of the topological string A-model on a Calabi-Yau background $(X,\alpha)$ and of the B-model on the mirror Calabi-Yau $W_\alpha$.
In particular, in Section~\ref{sec:BmodelGenus0} we discuss the modified integral structure of the periods, generalizing previous results from~\cite{Katz:2022lyl,Katz:2023zan}.
Our first class of examples, leading to the almost generic quintics $X_1,X_2$ as well as the dual geometries $Y_1,Y_2$, is discussed in Section~\ref{sec:example1}.
The second class of examples, containing the almost generic octics $X_3,X_4$ and their dual geometries $Y_3,Y_4$, is constructed in Section~\ref{sec:example2}.
In Section~\ref{sec:hyperdeterminants} we give a brief introduction to the subject of hyperdeterminants.
The example $X_{5}$ is discussed in Section~\ref{sec:example3}.
The construction of the propagators, that are used to integrate the holomorphic anomaly equations, as well as the BCOV ring are reviewed in Appendix~\ref{sec:BCOVring}.
While the B-model data and the $\mathbb{Z}_5$-refined invariants for $X_5$ are discussed directly in Section~\ref{sec:example3}, the corresponding expressions for the examples $X_a,Y_a$ for $a=1,\ldots,4$ are collected in Appendix~\ref{sec:dataXs}.
The corresponding torsion refined Gopakumar-Vafa invariants are collected in Appendix~\ref{sec:gvinvariants}.

\subsection*{Acknowledgements}{\small
We thank Amir-Kian Kashani-Poor, Sheldon Katz, Johanna Knapp, Joseph McGovern, Boris Pioline, Emanuel Scheidegger and Johannes Walcher for useful discussions.
Furthermore, we want to thank Albrecht Klemm and Eric Sharpe for collaboration on previous related work as well as Amir-Kian Kashani-Poor and David Jaramillo Duque for collaboration on related work in progress.
We are particularly grateful to Sheldon Katz for insightful comments on the geometries and the torsion refined invariants.
Similarly, we are grateful to Emanuel Scheidegger for explaining to us a new method to fix the normalization of the flat coordinate around a conifold point (thus allowing the genus $g=2$ results in the example $X_5$), among many other insightful discussions.
}

\section{Geometric preliminaries}
\label{sec:almostGeneric}

In this section we will first review some basic properties of projective threefolds with isolated nodes and introduce the notion of an almost generic Calabi-Yau threefold.
We then discuss how conifold transitions can be used in order to generate examples.

\subsection{Almost generic Calabi-Yau threefolds}
\label{sec:almostGenericDef}

Locally, a threefold nodal singularity, also known as threefold ordinary double point or conifold singularity, takes the form
\begin{align}
    V=\{\,uv-zw=0\,\}\subset\mathbb{C}^4\,.
    \label{eqn:conifold}
\end{align}
In order to remove the singularity, one can either deform the defining equations to be $uv-zw=\epsilon$ for some $\epsilon\ne 0$ -- we will refer to this as a smoothing -- or consider the resolutions
\begin{align}
    \widehat{V}=\text{Bl}_{\{u=z=0\}}V\,,\quad \widehat{V}'=\text{Bl}_{\{u=w=0\}}V\,,
\end{align}
that are in fact isomorphic to each other and related by the Atiyah flop.

Since $\widehat{V}$ is isomorphic to $V$ over the complement of a subset of codimension greater than one -- in this case over the complement of the origin $u=v=z=w=0$ -- it is called a \textit{small resolution}.
This implies that the canonical class is unaffected by the blowup and $K_{\widehat{V}}\sim 0$.
Moreover, since $\widehat{V}$ is obtained from $V$ by blowing up along a divisor, we know that $\widehat{V}$ is quasi-projective.~\footnote{Recall that a projective variety is necessarily compact and a quasi-projective variety can be embedded as a locally closed subset into a projective variety.}

Consider now a projective Calabi-Yau threefold $X$ which contains at most isolated nodes as singularities.
Let us denote the set of nodes by $S\subset X$ and their number by $\# S$.
Around each of the nodes $p\in S$ one can choose an open neighborhood and perform an analytic change of coordinates to obtain the geometry~\eqref{eqn:conifold}.
This implies that there exist $2^{\#S}$ small resolutions $\rho:\widehat{X}\rightarrow X$ where $\widehat{X}$ is a complex manifold.
However, such a small resolution $\widehat{X}$ is, in general, neither projective nor K\"ahler.

Let us fix any choice $\widehat{X}$ and denote the exceptional curves by $C_p=\rho^{-1}(p)$ for $p\in S$.
A theorem by Werner~\cite{WernerThesis} states that $X$ admits some projective small resolution if and only if the homology classes $[C_p]\in H_2(\widehat{X},\mathbb{Q})$ are non-trivial for all $p\in S$.~\footnote{For a recent english translation of~\cite{WernerThesis} see~\cite{werner2022smallresolutionsspecialthreedimensional} with the relevant result being Theorem 11.2.}
Note that, since the exceptional curves are holomorphic, the fact that one of them represents a class that is trivial in $H_2(\widehat{X},\mathbb{Q})$ also implies that there can not exist a K\"ahler form on $\widehat{X}$.

The situation where the homology classes of all of the exceptional curves are trivial in $H_2(\widehat{X},\mathbb{Q})$ corresponds to the case where $X$ is $\mathbb{Q}$-factorial, meaning that
\begin{align}
    \text{rk}\,\text{Cl}(X)/\text{Pic}(X)=0\,.
\end{align}
Here $\text{Cl}(X)$ and $\text{Pic}(X)$ are respectively the groups of Weil divisors and Cartier divisors, in each case taken modulo principal divisors.
A nice introduction to $\mathbb{Q}$-factoriality and, more generally, singularities of Calabi-Yau threefolds can be found in~\cite{Arras:2016evy,Grassi:2018rva} but the details won't be necessary for our applications.

We will be interested in the situation where $X$ is $\mathbb{Q}$-factorial but the homology classes of all of the exceptional are non-trivial in $H_2(\widehat{X},\mathbb{Z})$.
In other words, we want all of the exceptional curves to represent non-trivial torsion classes.
We therefore introduce the following definition of an \textit{almost generic} Calabi-Yau threefold:
\begin{definition}
    We say that a projective Calabi-Yau threefold $X$ is ``almost generic'' if
    \begin{enumerate}
        \item $X$ has isolated nodes $S\subset X$ as singularities and is smooth everywhere else,
        \item and given any analytic small resolution $\rho:\widehat{X}\rightarrow X$, the homology class of the exceptional curve $[\rho^{-1}(p)]\in H_2(\widehat{X},\mathbb{Z})$ is non-trivial and torsion for all $p\in S$.
    \end{enumerate}
\end{definition}
Since an almost generic Calabi-Yau threefold is $\mathbb{Q}$-factorial, it always admits a smoothing as a consequence of~\cite[Theorem 1.3]{Namikawa1995}.
In the following we will denote by $\widetilde{X}$ a smooth deformation of an almost generic Calabi-Yau threefold $X$.
The Euler characteristic of $\widetilde{X}$ is related to that of $\widehat{X}$ via
\begin{align}
    \widehat{X}=\widetilde{X}+2(\# S)\,.
\end{align}

Let us point out that the quantity $\chi(\widehat{X})$ appears to be an invariant among the MUM-points in a given moduli space.
In our first example, that is discussed in Section~\ref{sec:example1}, we have a MUM-point at $z=0$ that corresponds on the A-model side to a nodal quintic $X_1$ with $\#S=54$ nodes and the Euler characteristic of the generic smooth quintic is of course $\chi(\widetilde{X}_1)=-200$.
The moduli space contains a second MUM-point at $z=\infty$ which corresponds on the A-model side to a smooth Calabi-Yau threefold $Y_1$ with Euler characteristic $\chi(Y_1)=-200+2\cdot 54=-92$.
The same type of relations hold for our other examples as well, as can be seen in Table~\ref{tab:geometries}.

\subsection{Finding examples with conifold transitions}
\label{sec:conifoldTransitions}

To obtain the nodal quintics $X_1,X_2$ and the nodal octics $X_3,X_4$, we use conifold transitions from smooth Calabi-Yau threefolds $\widehat{X}^{\rm r}_a$ for $a=1,\ldots,4$, following the strategy laid out in~\cite{Katz:2023zan}.
In all of these cases $\widehat{X}^{\rm r}_a$ is a complete intersection in a product of projective spaces (so-called CICYs, introduced in~\cite{Candelas:1987kf,Green:1987cr}).

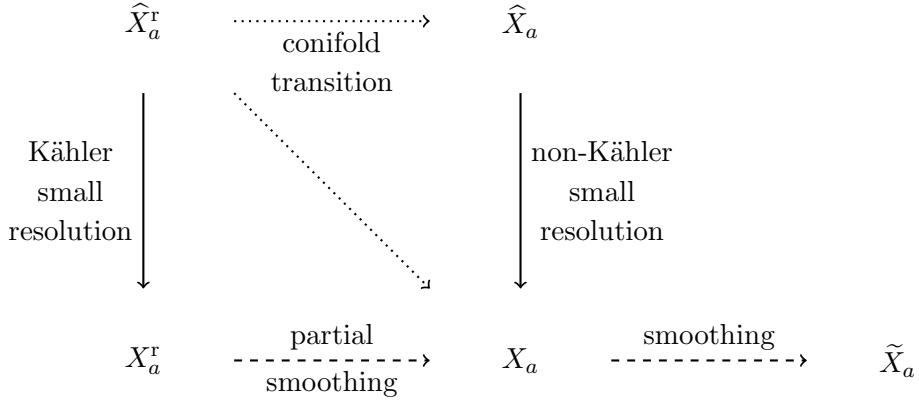
\begin{figure}[ht!]
\centering
\begin{tikzpicture}[
    box/.style={minimum width=2cm, minimum height=1.5cm, align=center},
    ghostbox/.style={minimum width=2cm, minimum height=1.5cm, align=center},
    node distance=3cm and 3cm
]

\node[box] (upperLeft) {$\widehat{X}^{\rm r}_a$};
\node[box] (upperMid) [right=of upperLeft] {$\widehat{X}_a$};
\node[ghostbox] (upperRight) [right=of upperMid] {};
\node[box] (lowerLeft) [below=of upperLeft] {$X^{\rm r}_a$};
\node[box] (lowerMid) [below=of upperMid] {$X_a$};
\node[box] (lowerRight) [below=of upperRight] {$\widetilde{X}_a$};

\coordinate (StartHorizontalLeft) at ($(upperLeft.east)+(0.2cm,0)$);
\coordinate (EndHorizontalLeft) at ($(upperMid.west)-(0.2cm,0)$);

\coordinate (StartHorizontalMid) at ($(lowerLeft.east)+(0.2cm,0)$);
\coordinate (EndHorizontalMid) at ($(lowerMid.west)-(0.2cm,0)$);

\coordinate (StartHorizontalRight) at ($(lowerMid.east)+(0.2cm,0)$);
\coordinate (EndHorizontalRight) at ($(lowerRight.west)-(0.2cm,0)$);

\coordinate (StartVerticalLeft) at ($(upperLeft.south)-(0,0.2cm)$);
\coordinate (EndVerticalLeft) at ($(lowerLeft.north)+(0,0.2cm)$);

\coordinate (StartVerticalMid) at ($(upperMid.south)-(0,0.2cm)$);
\coordinate (EndVerticalMid) at ($(lowerMid.north)+(0,0.2cm)$);

\coordinate (StartDiagonal) at ($(upperLeft.south east)+(0.2cm,-0.2cm)$);
\coordinate (EndDiagonal) at ($(lowerMid.north west)+(-0.2cm,0.2cm)$);

\draw[->,dotted,thick] (StartHorizontalLeft) -- (EndHorizontalLeft) node[midway, above] {} node[midway, below, align=center] {conifold\\transition};
\draw[->,dotted,thick] (StartDiagonal) -- (EndDiagonal);
\draw[->,dashed,thick] (StartHorizontalMid) -- (EndHorizontalMid) node[midway, above] {partial} node[midway, below] {smoothing};

\draw[->,thick] (StartVerticalLeft) -- (EndVerticalLeft) node[midway, left, align=center] {K\"ahler\\small\\resolution} ;
\draw[->,thick] (StartVerticalMid) -- (EndVerticalMid) node[midway, right, align=center] {non-K\"ahler\\small\\resolution} ;
\draw[->,dashed,thick] (StartHorizontalRight) -- (EndHorizontalRight) node[midway, above] {smoothing};

\end{tikzpicture}
\caption{The conifold transitions used to construct the geometries $X_a$ for $a=1,\ldots,4$.} \label{fig:transitions}
\end{figure}
We always proceed in the following three steps:
\begin{enumerate}
    \item We start with a smooth Calabi-Yau 3-fold $\widehat{X}^{\rm r}_a$,
    with the homology group of curves being
    \begin{align}
        H_2(\widehat{X}^{\rm r}_a,\mathbb{Z})\simeq\mathbb{Z}^m\times B(\widehat{X}^{\rm r}_a)\,,
    \end{align}
    where $B(\widehat{X}^{\rm r}_a)=\text{Tors}\,H_2(\widehat{X}^{\rm r}_a,\mathbb{Z})$ is some finite Abelian group.
    We conjecture  that $B(\widehat{X}^{\rm r}_a)$ is trivial in all of our examples.
    To simplify the exposition we will make this assumption in the rest of the paper.
    However, our results do not rely on this conjecture.
    If $B(\widehat{X}^{\rm r}_a)$ should be non-trivial, the only consequence would be that statements about the torsion of curves should be interpreted in the homology modulo $B(\widehat{X}^{\rm r}_a)$.
    
    \item We then show that $\widehat{X}^{\rm r}_a$ is a small resolution of a Calabi-Yau 3-fold $X^{\rm r}_a$ that has isolated nodal singularities $S^{\rm r}_a\subset X^{\rm r}_a$, that fall into two sets
    \begin{align}
        S^{\rm r}_a=S^{\rm r,A}_a\cup S^{\rm r,B}_a\,.
    \end{align}
    We refer to the corresponding nodes as being of Type A and Type B.
    We will denote the exceptional curve in $\widehat{X}^{\rm r}_a$ that resolves the node $p\in S^{\rm r}_a$ by $C_p\subset \widehat{X}^{\rm r}_a$.
    In all of our examples, the nodal Calabi-Yau $X^{\rm r}_a$ admits a smooth deformation $\widetilde{X}_a$.
    \item We show that $X^{\rm r}_a$ also admits a partial smoothing $X_a$ that preserves the subset $S_a^{\rm r,A}\subset S^{\rm r}_a$ of the nodes.
    Using a Mayer-Vietoris argument, analogous to the one used in the proof of~\cite[Proposition 3.3]{Katz:2023zan}, one can then show that for any analytic small resolution $\widehat{X}_a$ of $X_a$ one has
    \begin{align}
        H_2(\widehat{X}_a,\mathbb{Z})=H_2(\widehat{X}^{\rm r}_a,\mathbb{Z})/\langle\{\,[C_p],\,p\in S^{\rm r,B}_a\,\}\rangle\,.
        \label{eqn:MayerVietorisCurves}
    \end{align}
    As a result, we find that the exceptional curves in any small resolution $\widehat{X}_a$ of $X_a$ represent non-trivial torsion classes in homology and $X_a$ does not admit any K\"ahler small resolution.
\end{enumerate}
The relation between the geometries is summarized in Figure~\ref{fig:transitions}.

\paragraph{Existence of partial smoothing}
Let us briefly explain the existence of a deformation $X_a$ of $X_a^{\rm r}$ that removes only the nodes of Type B.
In all of our examples, the exceptional curves that resolve the nodes $S_a^{\rm r,B}$ satisfy a relation in homology
\begin{align}
    \sum\limits_{p\in S_a^{\rm r,B}}a_p[C_p]=0\in H_2(\widehat{X}_a^{\rm r},\mathbb{Z})\,,
    \label{eqn:homologyRelation}
\end{align}
with all coefficients $a_p\ne 0$.
We denote by $\overline{X}^{\rm r}_a$ an analytic variety that is obtained by small resolving only the nodes of Type A, $S_a^{\rm r,A}\subset X^{\rm r}_a$.
Any of the $2^{\#S_a^{\rm r,A}}$ choices will work for us.
We can then apply the results from~\cite{Friedman86,Kawamata1992UnobstructedDR,Ran1993UnobstructednessOC,tian92}, as summarized for example in the introduction of~\cite{friedman24}, to conclude that there exists a smoothing $\widehat{X}_a$ of $\overline{X}^{\rm r}_a$.
After contracting the remaining exceptional curves we then obtain the almost generic Calabi-Yau threefold $X_a$.

To be clear, this argument allows us to conclude the existence of the Calabi-Yau $X_a$ and to deduce its relevant properties.
We will not identify the explicit deformations of the defining equations that remove the nodes of Type B while preserving those of Type A, as has been possible in the case of symmetric determinantal double octics, studied in~\cite{Katz:2023zan}.
It would be very interesting to identify the explicit form of the defining equations of the geometries $X_a$ but we will leave this question for future work.

\paragraph{Torsion of partial smoothing}
Given the homology classes of the exceptional curves $C_p$ for $p\in S^{\rm r,B}_a$, it is easy to calculate the torsion of $X_a$.
For the moment, let $r=b_2(\widehat{X}^{\rm r}_a)$.
In all of our examples the lattice
\begin{align}
    \langle\{\,[C_p],\,p\in S^{\rm r,B}_a\,\}\rangle\subset H_2(\widehat{X}^{\rm r}_a,\mathbb{Z})\simeq\mathbb{Z}^{r}
\end{align}
has rank $r-1$, and is generated by elements $\vec{v}_i=(0,v_{i,1},\ldots,v_{i,r-1})$ for $i=1,\ldots,r-1$.
As a result, the homology of $\widehat{X}_a$ takes the form
\begin{align}
    H_2(\widehat{X}_a,\mathbb{Z})=\mathbb{Z}\times B(X_a)\,,\quad B(X_a)=\mathbb{Z}_N\,,
\end{align}
with the order of torsion being
\begin{align}
    N=\det\left(\begin{array}{ccc}
        v_{1,1}&\ldots&v_{1,r-1}\\
        \vdots&\ddots&\vdots\\
        v_{r-1,1}&\ldots&v_{r-1,r-1}
    \end{array}\right)\,.
    \label{eqn:Ndet}
\end{align}

\section{Topological strings and mirror symmetry}
\label{sec:topologicalStrings}
We now discuss the topological string A-model on a background $(X,\alpha)$, where $X$ is an almost generic Calabi-Yau threefold and $\alpha\in B(X)$ the topology of the flat B-field, as well as the B-model on the corresponding mirror Calabi-Yau $W_\alpha$.

We will denote by $\rho:\widehat{X}\rightarrow X$ some choice of small resolution of $X$, the choice of which doesn't matter but will be kept fixed throughout the discussion.
The set of nodes in $X$ will be denoted by $S\subset X$ and $C_p$ for $p\in S$ is the exceptional curve over the node $p$.
The smooth deformation of $X$ will be denoted by $\widetilde{X}$.
We will assume that $H_2(\widetilde{X},\mathbb{Z})=\mathbb{Z}$ and that the torsion of $X$ is
\begin{align}
    B(X)=\text{Tors}\,H^3(\widehat{X},\mathbb{Z})= \mathbb{Z}_N\,,
\end{align}
for some $N\ge 1$.
The class $\alpha$ can then be represented by an integer $k\in\{0,\ldots,N-1\}$ and we denote the equivalence class by $\alpha=[k]_N\in \mathbb{Z}_N$.
Note that the case $N=1$ corresponds to the usual situation where $X=\widetilde{X}$ is already a smooth projective Calabi-Yau threefold.

We will assume that the B-field stabilizes all of the nodes, i.e. $\text{gcd}(k,N)=1$.
The reason is that ``unstable nodes'', by which we mean nodes that are resolved by exceptional curves which have a trivial B-field holonomy, can always be removed by a complex structure deformation and do not affect the A-model topological string partition function.

\subsection{Topologically non-trivial B-fields and torsion refined GV-invariants}
Since a singular Calabi-Yau is not a manifold, the notion of the K\"ahler cone is no longer well-defined.
However, it can be naturally replaced by the cone of ample divisors.
This is because of Kleiman's criterion, which states that a line bundle on a projective scheme is ample if and only if the degree on every non-zero element in the closure of the cone of effective curves is positive.
In all of our examples, the cone of ample divisors is one-dimensional and we denote the generator by $J\in\text{Pic}(X)$.
In a slight abuse of terminology, we still refer to $\omega_X=t\,J$
as the complexified K\"ahler form on $X$ and call $t$ the complexified K\"ahler parameter.~\footnote{The corresponding metric can be obtained by performing an infinitesimal smoothing and applying the Calabi-Yau theorem.}

We denote by $Z_{\text{top.}}^X(t,[k]_N,\lambda)$ the A-model topological string partition function on $(X,[k]_N)$ with complexified K\"ahler parameter $t$.
It was observed in~\cite{Schimannek:2021pau}, and subsequently in~\cite{Katz:2022lyl,Katz:2023zan}, that the usual expansion of the topological string partition function in terms of Gopakumar-Vafa invariants~\cite{Gopakumar:1998ii,Gopakumar:1998jq} has to be modified if $k\ne 0$.
For general $k$, it takes the form
\begin{align}
    \log Z_{\text{top.}}^X(t,[k]_N,\lambda)=\sum\limits_{g\ge 0}\sum\limits_{m\ge 1}\sum\limits_{d\ge 1}\sum\limits_{p=0}^{N-1}\frac{n_g^{d,p}}{m}\left[\sin\left(\frac{m\lambda}{2}\right)\right]^{2-2g}e^{\frac{2\pi i k m p}{N}}e^{2\pi i mdt}\,,
    \label{eqn:GVex}
\end{align}
in terms of the so-called torsion refined Gopakumar-Vafa invariants $n_g^{d,p}\in\mathbb{Z}$.~\footnote{The appearance of the phase in the context of smooth Calabi-Yau threefolds with torsion was also remarked in~\cite{Dedushenko:2014nya}. At genus zero it has been pointed out for singular Calabi-Yau threefolds in~\cite{Aspinwall:1995rb} and, in the smooth case, was applied in~\cite{Braun:2007xh,Braun:2007vy}.}

The physical derivation of this modification rests on the claim that the effective supergravity associated to the M-theory compactification on $X$ exhibits a
\begin{align}
    G_{5\rm d}=\text{Hom}\left(H_2(\widehat{X},\mathbb{Z}),{\rm U}(1)\right)={\rm U}(1)\times \mathbb{Z}_N
    \label{eqn:G5d}
\end{align}
gauge symmetry and the five dimensional BPS-particles carry charges
\begin{align}
    (d,[p]_N)\in H_2(\widehat{X},\mathbb{Z})=\mathbb{Z}\times\mathbb{Z}_N\,.
\end{align}
As has been pointed out in~\cite{Katz:2023zan}, the relation of the homology groups under conifold transitions, that we use to construct the geometries $X_a$ for $a=1,\ldots,4$, directly translates to a physics proof of~\eqref{eqn:G5d}.

The original argument from~\cite{Gopakumar:1998ii,Gopakumar:1998jq} then carries over with the only modification being that the invariants carry an additional label that keeps track of the torsion charge.
We refer to~\cite{Schimannek:2021pau,Katz:2022lyl,Katz:2023zan} for more details, also on the mathematical definition of the torsion refined invariants.

It was pointed out in~\cite{Schimannek:2021pau} that the topological string partition function does not distinguish the choices $[k]_N$ and $[-k]_N$ for $\alpha$.
This is related to the phenomenon, further discussed in~\cite{Katz:2022lyl,Katz:2023zan}, that M-theory on $X$ appears to ``see'' all of the small resolutions simultaneously.
This has already been observed for the non-compact conifold in~\cite{Szendroi:2007nu}.
It is useful to keep this in mind since it will also explain several factors of two that appear later on in our discussion.

\subsection{The B-model at genus zero}
\label{sec:BmodelGenus0}
Let us now assume that the Calabi-Yau background $(X,[k]_N)$ is mirror to a Calabi-Yau background $(W_k,\beta_k)$, where $W_k$ is a projective Calabi-Yau threefold and $\beta_k$ represents the B-field topology on $W_k$ and takes values in some suitable group.
The question of whether or not $W_k$ is smooth and whether $\beta_k$ is non-trivial are of course very interesting but we will assume that this does not affect the calculation of the periods.~\footnote{There is certainly a lot to be understood here, for example what happens if the B-field topology on $W_k$ obstructs certain deformations. This is beyond the scope of our paper but we hope to come back to this question in future work.}
We will therefore forget about $\beta_k$ and will sometimes just talk about $W_k$ as the mirror of $(X,[k]_N)$.

The K\"ahler potential that is associated to the special K\"ahler structure on the complex structure moduli space $\mathcal{M}_{\text{cs}}$ of $W_k$ takes the form
\begin{align}
	e^{-K(z,\bar{z})}={\rm i}\int_{W_k}\Omega\wedge\bar{\Omega}\,,
\end{align}
in terms of the holomorphic $3$-form $\Omega$, which is a section of the K\"ahler line bundle $\mathcal{L}$.
A rescaling $\Omega\rightarrow e^{f(z)}\Omega$, in terms of a holomorphic function $f(z)$ on $\mathcal{M}_{\text{cs}}$, therefore corresponds to a K\"ahler transformation
\begin{align}
	K(z,\bar{z})\rightarrow K(z,\bar{z})-f(z)-\bar{f}(\bar{z})\,.
\end{align}
The Weil-Petersson metric on the moduli space is then given by $G_{z\bar{z}}=\partial_z\partial_{\bar{z}}K$.

We will now assume that the limit $z\rightarrow 0$ corresponds to a point of maximally unipotent monodromy (MUM-point) that is mirror to the large volume limit $t\rightarrow {\rm i}\infty$ on $(X,[k]_N)$.
One can then choose a basis of rational 3-cycles on $W_k$ such that the corresponding  periods of $\Omega(z)$ take the form
\begin{align}
    \Pi'(z)=\left(\begin{array}{c}
        \varpi_0(z)\\
        \varpi_0(z)\log(z)+\varpi_1(z)\\
        \varpi_0(z)\log(z)^2+2\varpi_1(z)\log(z)+\varpi_2(z)\\
        \varpi_0(z)\log(z)^3+3\varpi_1(z)\log(z)^2+3\varpi_2(z)\log(z)+\varpi_3(z)\\
    \end{array}\right)\,,
    \label{eqn:Qbasis}
\end{align}
in terms of locally analytic functions $\varpi_0(z)=1+\mathcal{O}(z)$ and $\varpi_i(z)=\mathcal{O}(z)$ for $i=1,2,3$.
We refer to $\varpi_0(z)$ as the \textit{fundamental period} and it is a section of $\mathcal{L}$.

As a consequence of Griffiths transversality, the holomorphic 3-form -- and therefore also the periods -- are annihilated by the so-called Picard-Fuchs operator
\begin{align}
    \mathcal{D}=\sum\limits_{i=0}^4p_i(z)\theta^i\,,\quad \theta=z\partial_z\,.
\end{align}
The coefficients $p_i(z)$, $i=0,\ldots,4$ are polynomials in $z$.
The Picard-Fuchs operator can be reconstructed from the fundamental period $\varpi_0(z)$ by making an ansatz for $\mathcal{D}$ and solving $\mathcal{D}\varpi_0(z)=0$.

For a suitable choice of coordinate $z$, the mirror map that relates this to the complexified K\"ahler parameter $t$ of $(X,[k]_N)$ is given by
\begin{align}
    t(z)=\frac{1}{2\pi{\rm i}}\left(\log(z)+\frac{\varpi_1(z)}{\varpi_0(z)}\right)\,.
    \label{eqn:flatCoordinate}
\end{align}
The mirror map can be inverted around $z=0$ to obtain the inverse mirror map $z(t)$.

One can then choose an integral symplectic basis of 3-cycles
\begin{align}
    A^I,B_I\in H_3(W_k,\mathbb{Z})\,,\quad A^I\cdot B_J=-B_J\cdot A^I=\delta^I_J\,,\quad I,J=0,1\,,
    \label{eqn:ABbasis}
\end{align}
such that the corresponding vector of periods takes the form
\begin{align}
    \vec{\Pi}=\left(\int_{B_I}\Omega(z),\,\int_{A^I}\Omega(z)\right)=f_0(z)\left(N\left[2F(t)-t\partial_tF(t)\right]\,,
            \partial_tF(t),\,
            \frac{1}{N},\,
            t\right)\,,
    \label{eqn:integralBasis}
\end{align}
in terms of a prepotential $F(t)$.
The appearance of the factor of $N$ has first been observed in~\cite{Katz:2022lyl,Katz:2023zan}, for $N=2$, and we generalize the results here for arbitrary $N$.

As was also discussed in~\cite{Katz:2022lyl,Katz:2023zan}, generalizing the well known results for $N=1$ from~\cite{CANDELAS199121,Hosono:1993qy,Hosono:1994ax}, the prepotential $F(t)$ takes the form
\begin{align}
    F(t)=-\frac{\kappa}{6}t^3+\frac{\sigma}{2} t^2+\frac{b}{24N}\, t+\frac12\frac{1}{(2\pi {\rm i})^3}\epsilon+\mathcal{O}(e^{2\pi{\rm i}t})\,.
    \label{eqn:prepotential}
\end{align}
Here $\kappa$ is the triple intersection number of the smooth deformation $\widetilde{X}$ of $X$ and the coefficient of the constant term takes the form
\begin{align}
    \epsilon=\chi(\widehat{X})\zeta(3)-\sum\limits_{q=1}^{\lfloor\frac{N}{2}\rfloor}n_q\left[\text{Li}_3\left(e^{\frac{2\pi{\rm i} k q}{N}}\right)+\text{Li}_3\left(e^{-\frac{2\pi{\rm i} k q}{N}}\right)\right]\,.
    \label{eqn:zeta3term}
\end{align}
Here we define $n_q$ for $q=1,\ldots,\lfloor N/2\rfloor$ as
\begin{align}
    n_q=\#\left\{\,p\in S\,\vert\,[C_p]\in\{[q]_N,\,[-q]_N\}\,\right\}\,.
\end{align}

The geometric interpretation of the coefficient $b$ is not yet understood, except when $X$ is smooth, i.e. $N=1$, in which case $b=J\cdot c_2(TX)$.
More generally, we observe that in all of the examples that have been considered in~\cite{Katz:2022lyl,Katz:2023zan} and in this paper it takes values $b\in\mathbb{Z}$.
The coefficient $\sigma\in\frac12\mathbb{Z}$ can often be fixed from integrality of the monodromy matrices.~\footnote{As for some of our examples, it can happen that there is a second MUM-point in the moduli space that is related via mirror symmetry to an ordinary smooth Calabi-Yau threefold. In those cases the Gamma-class formula, see e.g.~\cite{Halverson:2013qca}, can be used to determine an integral basis of periods around that point and numerical analytic continuation yields a corresponding basis around $z=0$.}

The Yukawa coupling $C_{zzz}=\int_W\Omega\wedge\partial_z^3\Omega$ is related to the prepotential via
\begin{align}
	C_{zzz}=-\varpi_0(z)^2\partial_{t}^3F(t)\,.
\end{align}
It is a rational function in $z$ and a section of $\mathcal{L}^2\otimes \text{Sym}^3\left(T^*\mathcal{M}_{\text{cs}}\right)^{1,0}$.

\subsection{Higher genus free energies and holomorphic anomaly equations}
The B-model genus $g$ topological string free energy $\mathcal{F}_g(z,\bar{z})$ is a section of $\mathcal{L}^{2-2g}$.
To be independent of the choice of K\"ahler gauge, one can introduce $F_g(t,\bar{t})=\varpi_0(z)^{2g-2}\mathcal{F}_g(z,\bar{z})\big\vert_{z\rightarrow z(t)}$.
We refer to the quantities $F_g(t)$ that one obtains in the holomorphic limit
\begin{align}
    F_g(t)=\lim\limits_{\bar{t}\rightarrow {\rm i}\infty}F_g(t,\bar{t})\,,
\end{align}
as the A-model topological string free energies.
They are related to the topological string partition function in~\eqref{eqn:GVex} via
\begin{align}
    \log Z_{\text{top.}}^X(t,[k]_N,\lambda)=\sum\limits_{g\ge 0}\lambda^{2g-2}F_g(t)\,.
\end{align}
The genus zero free energies are already holomorphic and the prepotential~\eqref{eqn:prepotential} can be identified with $F(t)=-F_0(t)$.
In the holomorphic limit, the derivative of the K\"ahler potential and the Weil-Petersson metric respectively become
\begin{align}
    K_z\rightarrow-\partial_z\log\varpi_0(z)\,,\quad G_{z\bar{z}}\rightarrow \partial_z t\,.
\end{align}

The genus one free energy satisfies the holomorphic anomaly equation~\cite{Bershadsky:1993ta}
\begin{align}
	\partial_z\partial_{\bar{z}}\mathcal{F}_1(z,\bar{z})=\frac12C_{zzz}C^{zz}_{\bar{z}}-\left(\frac{\chi(\widehat{X})}{24}-1\right)G_{z\bar{z}}\,,
    \label{eqn:holAnomaly1}
\end{align}
in terms of $C^{zz}_{\bar{z}}=e^{2K}\overline{C}_{\bar{z}\bar{z}\bar{z}}G^{z\bar{z}}G^{z\bar{z}}$.
Recall from Section~\ref{sec:almostGeneric} that the Euler characteristic of the small resolution can also be expressed as $\chi(\widehat{X})=\chi(\widetilde{X})+2(\#S)$.
The necessary modification for the number of nodes $\#S\ne 0$ has been first observed in~\cite{Schimannek:2021pau}.

Integrating~\eqref{eqn:holAnomaly1}, and using the boundary behaviour of $\mathcal{F}_1$ gives the ansatz 
\begin{align}
	\begin{split}
		\mathcal{F}_1(z,\bar{z})=&-\frac12\left(4-\frac{\chi(\widehat{X})}{12}\right)K-\frac12\log\,\det\,G_{z\bar{z}}+\log\vert f_1(z)\vert^2\,,
	\end{split}
	\label{eqn:f1ansatz}
\end{align}
in terms of the ambiguity
\begin{align}
\begin{split}
    f_1(z)=&z^{-\frac{1}{24}(\tilde{b}+12)}\prod_i\Delta_i^{-\frac{c_i}{12}}\,.
    \label{eqn:genus1ambiguity}
    \end{split}
\end{align}
Here $\widetilde{b}=\int_{\widetilde{X}}\widetilde{J}\cdot c_2(T\widetilde{X})$, in terms of a primitive ample divisor $\widetilde{J}$ on $\widetilde{X}$, and the coefficients $c_i\in\mathbb{Z}$ depend on the number of vector- and hypermultiplets that become massless at the sublocus in moduli space $\{\Delta_i=0\}\subset\mathcal{M}_{\text{cs}}$~\cite{Vafa:1995ta}.

The free energies at genus $g\ge 2$ satisfy similar holomorphic anomaly equations~\cite{Bershadsky:1993cx}
\begin{align}
	\partial_{\bar{z}}\mathcal{F}_g(z,\bar{z})=\frac12 C^{zz}_{\bar{z}}\left(D_zD_z\mathcal{F}_{g-1}(z,\bar{z})+\sum\limits_{h=1}^{g-1}D_z\mathcal{F}_h(z,\bar{z})D_z\mathcal{F}_{g-h}(z,\bar{z})\right)\,.
    \label{eqn:holAnomaly2}
\end{align}
The integration of these equations using a set of propagators, the BCOV ring and the polynomial structure of the topological string free energies is described in Appendix~\ref{sec:BCOVring}.

\subsection{Boundary conditions}
\label{sec:boundaryConditions}
Integrating the holomorphic anomaly equations~\eqref{eqn:holAnomaly2} determines $\mathcal{F}_g$ in terms of the free energies at lower genus $\mathcal{F}_{g'}$, $g'<g$ up to an integration constant $f_g(z)$ that is referred to as the holomorphic ambiguity.

In all of our examples, the discriminant polynomial takes the form
\begin{align}
    \Delta=\prod\limits_{i=1}^{n_{\rm C}}\Delta_i^{\rm C} \cdot \prod\limits_{i=1}^{n_{\rm HC}}\Delta_i^{\rm HC}\,,
\end{align}
where $n_{\rm C}+n_{\rm HC}$ is the degree of $\Delta$ and the loci $\{\,\Delta^{\rm C}_i=0\,\}$ for $i=1,\ldots,n_{\rm C}$ are all conifold points while the loci $\{\,\Delta^{\rm HC}_i=0\,\}$ for $i=1,\ldots,n_{\rm HC}$ are all hyperconifold points.
In fact, the only examples where $n_{\rm HC}\ne 0$ are $(X_3,[\pm 1]_3)$ and the dual Calabi-Yau $Y_3$.
In that case $n_{\rm HC}=1$ and we find one hyperconifold point in the moduli space where an $S^3/\mathbb{Z}_2$ appears to shrink inside the mirror.

Also, in our examples the point at $z=\infty$ is always another MUM-point.
The corresponding local expansion is related to that around $z=0$ by a coordinate and K\"ahler transformation
\begin{align}
    z \propto\frac{1}{v^{e_1}}\,,\quad \varpi_0\rightarrow f_K \varpi_0\,,\quad f_K\propto\frac{1}{v^{e_2}}\,,
\end{align}
for some $e_1,e_2\in\mathbb{N}$.
For reasons that will become clear in a moment, we call $\rho=e_1/e_2$ the regulator.
The holomorphic ambiguity then takes the form
\begin{align}
    f_g(z)=\sum\limits_{i=1}^{n_{\rm C}}\frac{p^{\rm C}_i(z)}{(\Delta_i^{\rm C})^{2g-2}}+\sum\limits_{i=1}^{n_{\rm HC}}\frac{p^{\rm HC}_i(z)}{(\Delta_i^{\rm HC})^{g-1}}+\sum\limits_{n=0}^{\lfloor\frac{2(g-1)}{\rho}\rfloor}b_nz^n\,.
\end{align}
The lower and the upper bound of the last sum are respectively determined by the regularity of the genus $g\ge 2$ free energies in a large volume limit.
The numerator $p_i^{\rm C}$ for $i=1,\ldots,n_{\rm C}$ is a polynomial of degree $(\text{deg}\,\Delta_i^{\rm C})^{2g-2}-1$
and is fixed by the gap condition at $\Delta^{\rm C}_i=0$ that we will discuss in a moment.
On the other hand, the numerator $p_i^{\rm HC}$ for $i=1,\ldots,n_{\rm HC}$ is a polynomial of degree $(\text{deg}\,\Delta_i^{\rm HC})^{g-1}-1$.
There does not appear to be an ordinary gap at the hyperconifold points and therefore we fix the $p_i^{\rm HC}$, just like the coefficients $b_n$, by using additional boundary conditions and the Castelnuovo vanishing condition.

Before we discuss the gap condition, let us briefly recall that the monodromy around a conifold point always takes the form
\begin{align}
    M_{\rm C}=\text{Id}_{4\times 4}-m\Sigma^T\vec{v}\,\vec{v}^{\,T}\,,\quad m\in\mathbb{N}\,,\quad \vec{v}\in\mathbb{Z}^4\,,
    \label{eqn:conifoldMonodromy}
\end{align}
where $\vec{v}$ is a primitive 3-cycle in the mirror $W_k$ that shrinks at the conifold point, in the basis $A^I,B_J$ from~\eqref{eqn:ABbasis}, and the symplectic intersection form $\Sigma$ is given by
\begin{align}
    \Sigma=\left(
\begin{array}{cccc}
 0 & 0 & 1 & 0 \\
 0 & 0 & 0 & 1 \\
 -1 & 0 & 0 & 0 \\
 0 & -1 & 0 & 0 \\
\end{array}
\right)\,.
\end{align}
The coefficient $m$ here is identical to the exponent $c_i$ that appears in~\eqref{eqn:genus1ambiguity} for the corresponding discriminant component.

The so-called gap condition, first described in~\cite{Huang:2006si,Huang:2006hq}, amounts to the conjecture that
\begin{align}
    F_g(t_{\rm c})=\frac{(-1)^{g-1}B_{2g}}{2g(2g-2)}\frac{m}{t_{\rm c}^{2g-2}}+\mathcal{O}(t_{\rm c}^0)\,,
    \label{eqn:gap}
\end{align}
where $t_{\rm c}$ is a local flat coordinate around this conifold point.
The coefficient in~\eqref{eqn:gap}, for $m=1$, has been determined in~\cite{Ghoshal:1995wm}.
A correct choice for $t_{\rm c}$ is the analytic continuation of the period $\vec{v}\cdot\vec{\Pi}$, normalized by a regular period that takes value $1$ at the conifold point.
More details on the choice of normalization can be found in~\cite{Huang:2006hq,EmanuelWIP}.

One additional boundary condition (in fact two, given that we have two MUM-points) is provided by the generic expression for the constant term of $F_g(t)$,
\begin{align}
    \begin{split}
	F_g(t)\big\vert_{\text{const.}}=&\frac{(-1)^{g-1}}{2}\frac{B_{2g}}{2g[(2g-2)!]}\left(\frac{B_{2g-2}}{2g-2}\chi(\widehat{X})\right.\\
    &\left.+\sum\limits_{q=1}^{\lfloor\frac{N}{2}\rfloor}n_q\left[\text{Li}_{3-2g}\left(e^{\frac{2\pi{\rm i} k q}{N}}\right)+\text{Li}_{3-2g}\left(e^{-\frac{2\pi{\rm i} k q}{N}}\right)\right]\right)\,,
    \end{split}
\end{align}
The necessary modification in the case $N=2$ has first been explained in~\cite{Katz:2022lyl} and the argument directly generalizes to arbitrary $N$.

Finally, we can use the Castelnuovo vanishing property, which states that for a given degree $d\in\mathbb{N}$ there exists a maximal genus $g_{\text{max}}(d)$ such that $n^{d,p}_g=0$ if $g>g_{\text{max}}(d)$.
The existence of such a bound in the unrefined case has been proven, using the relation with Gromov-Witten invariants, in~\cite{doan2021gopakumarvafafinitenessconjecture}.
On the other hand, concrete bounds for some geometries, again in the unrefined case, have been proven in~\cite{Liu:2022agh,Alexandrov:2023zjb}.
Instead of using a concrete bound, we will make the common simplifying assumption that if for some $(d,p,g)$ we find that $n^{d,p}_g=0$, then $n^{d,p}_{g'}=0$ for all $g'>g$.

\section{Almost generic nodal quintics with $\mathbb{Z}_2$-torsion}
\label{sec:example1}

In this section we use conifold transitions, as described in Section~\ref{sec:conifoldTransitions}, to construct almost generic quintic Calabi-Yau threefolds $X_1$ and $X_2$ that respectively have $54$ and $48$ isolated nodal singularities.
In both cases the torsion is $B(X_1)=B(X_2)=\mathbb{Z}_2$.

The corresponding geometries $\widehat{X}^{\rm r}_1$ and $\widehat{X}^{\rm r}_2$, which are the starting points of the conifold transitions, are CICY 6675 and CICY 6989 from the classification~\cite{Candelas:1987kf,Green:1987cr}.
The corresponding configuration matrices are
\begin{align}
    \widehat{X}^{\rm r}_1=\left[\begin{array}{c|cccccc}
        \mathbb{P}^4&1&1&1&1&1&0\\
        \mathbb{P}^2&1&1&0&0&0&1\\
        \mathbb{P}^2&0&0&1&1&0&1\\
        \mathbb{P}^1&0&0&0&0&1&1
    \end{array}\right]^{4,35}_{-62}\,,\qquad \widehat{X}^{\rm r}_2=\left[\begin{array}{c|cccccc}
        \mathbb{P}^4&1&1&1&1&0\\
        \mathbb{P}^2&1&1&0&0&1\\
        \mathbb{P}^1&0&0&2&0&1\\
        \mathbb{P}^1&0&0&0&1&1
    \end{array}\right]^{4,37}_{-66}\,,
\end{align}
where the superscript indicates $h^{1,1}$ and $h^{2,1}$ of the geometry, while the subscript is the Euler characteristic.

\subsection{Porteous' formula}
\label{sec:porteous}
Before discussing the geometries, let us briefly recall some properties of the degeneracy loci of bundle maps and Porteous' formula.

Consider vector bundles $\mathcal{E},\mathcal{F}$ on a variety $V$ of respective rank $e$ and $f$ and a map $\phi:\mathcal{E}\rightarrow\mathcal{F}$.
We denote the vanishing locus of $(k+1)\times(k+1)$ minors of a matrix representation of $\phi$ by $M_k(\phi)$.
The expected codimension of $M_k(\phi)$ in $V$ is $(e-k)(f-k)$ and if the actual codimension of $M_k(\phi)$ matches the expected codimension then the cohomology class $[M_k(\phi)]$ is given by Porteous' formula.
This is in particular the case if all of the entries of the matrix representation of $\phi$ are sections of globally generated line bundles on $V$.

Denoting by $c_0,c_1,\ldots$ the Chern classes of $\mathcal{F}/\mathcal{E}$, Porteous' formula states that
\begin{align}
	[M_k(\phi)]=\det\left(\begin{array}{cccc}
		c_{f-k}&c_{f-k+1}&\cdots&c_{e+f-2k-1}\\
		\vdots&\vdots&\ddots&\vdots\\
		\vdots&\vdots&\ddots&\vdots\\
		c_{f-e+1}&c_{f-e+2}&\cdots&c_{f-k}
	\end{array}\right)\,.
\end{align}

To illustrate this, let us consider an $f\times e$ matrix $B$ such that the entries of the $i$-th row are sections of a line bundle $L_i$ on some space $W$.
We can interpret this as a bundle map
\begin{align}
    B:\,\mathcal{O}_W^{\oplus e}\rightarrow \bigoplus\limits_{i=1}^fL_i\,,
\end{align}
that we denote by the same symbol in a slight abuse of notation.
Let us assume for example that $f=2$ and $e=3$.
Then the degeneracy locus $M_1(B)=\{\,\text{corank}\,B=1\,\}$ has expected codimension $2$ and Porteous' formula implies that it represents the cohomology class
\begin{align}
    [M_1(B)]=c_1(L_1)^2+c_1(L_1)c_1(L_2)+c_1(L_2)^2\,.
    \label{eqn:porteous2x3}
\end{align}

\subsection{The geometry of $\widehat{X}^{\rm r}$ and $X^{\rm r}$}
In order to understand the geometry of $\widehat{X}^{\rm r}_1$ and $\widehat{X}^{\rm r}_2$ it will be useful to interpret them as a special case of a slightly more general construction and consider instead the complete intersection inside $V=\mathbb{P}^4\times\mathbb{P}^2\times\mathbb{P}^2\times\mathbb{P}^2$ that is associated to the configuration matrix
\begin{align}
    \left[\begin{array}{c|ccccccc}
        \mathbb{P}^4&a_{1,1}&a_{1,2}&a_{2,1}&a_{2,2}&a_{3,1}&a_{3,2}&0\\
        \mathbb{P}^2&1&1&0&0&0&0&1\\
        \mathbb{P}^2&0&0&1&1&0&0&1\\
        \mathbb{P}^2&0&0&0&0&1&1&1
    \end{array}\right]\,,
\end{align}
with $a_{i,j}\in\mathbb{N}$.
The corresponding complete intersection will be a Calabi-Yau threefold if the sum of all $a_{i,j}$ equals $5$.
One can easily check, for example using CohomCalg~\cite{Blumenhagen:2010pv,cohomCalg:Implementation}, that only two cases lead to a Calabi-Yau threefold with $h^{1,1}=4$, namely
\begin{align}
    (a_{1,1},a_{1,2},a_{2,1},a_{2,2},a_{3,1},a_{3,2})=\left\{\begin{array}{cc}
        (1,1,1,1,1,0)&\,\,\text{for}\,\,\,\widehat{X}^{\rm r}_1\,,\\
        (1,1,2,0,1,0)&\,\,\text{for}\,\,\,\widehat{X}^{\rm r}_2\,.
        \end{array}\right.
\end{align}
For now, let us leave the choice for $a_{i,j}$ open and only assume that $H_2(\widehat{X}^{\rm r},\mathbb{Z})=\mathbb{Z}^4$.~\footnote{As already mentioned in Section~\ref{sec:almostGeneric}, we conjecture that $H_2(\widehat{X}^{\rm r},\mathbb{Z})$ is torsion free. This mostly allows us to simplify our notation without affecting our results in any essential way.}

We introduce homogeneous coordinates
\begin{align}
    [x_1:\ldots:x_5]\in\mathbb{P}^4\,,\quad [u_1:u_2:u_3]\in\mathbb{P}^2\,,\quad [v_1:v_2:v_3]\in\mathbb{P}^2\,,\quad [w_1:w_2:w_3]\in \mathbb{P}^2\,,
\end{align}
and write the defining equations as 
	\begin{align}
		\begin{split}
			\widehat{X}^{\rm r}=&\left\{\,B_1(x_1,\ldots,x_5)(u_1,u_2,u_3)^{T}=B_2(x_1,\ldots,x_5)(v_1,v_2,v_3)^{T}\right.\\
			&\left.=B_3(x_1,\ldots,x_5)(w_1,w_2,w_3)^{T}=\vec{0}\,,\quad \sum\limits_{i,j,k=1}^3A_{i,j,k}u_iv_jw_k=0\,\right\}\subset V\,.
		\end{split}
	\end{align}
Here $B_1,B_2,B_3$ are $2\times 3$ matrices with entries that are generic homogeneous polynomials in $[x_1:\ldots:x_5]$, with the degree of the entries in the $j$-th row of $B_i$ being $a_{i,j}$, while $A_{i,j,k}\in\mathbb{C}$ are generic complex coefficients.

We denote by $\pi_i:V\rightarrow \mathbb{P}^{r_i}$, for $i=1,\ldots,4$ and $\vec{r}=(4,2,2,2)$, the projection to the $i$-th factor of $V$ and write
\begin{align}
    J_i=\pi_i^*(H_i)\big\vert_{\widehat{X}^{\rm r}}\,,\quad i=1,\ldots, 4\,,
\end{align}
where $H_i$ is a generic hyperplane divisor in $\mathbb{P}^{r_i}$.
We define the degree of a curve $C\subset \widehat{X}^{\rm r}$ as $(d_1,\ldots,d_4)=C\cap (J_1,\ldots,J_4)$.

Let us denote by $B_{i}^{[j]}$ the matrix that is obtained from $B_i$ by omitting the $j$-th column and define the singular quintic
\begin{align}
	X^{\rm r}=\left\{\,\sum\limits_{i,j,k=1}^3(-1)^{i+j+k}A_{i,j,k}\det B_{1}^{[i]}(x)\det B_{2}^{[j]}(x)\det B_{3}^{[k]}(x)=0\,\right\}\subset\mathbb{P}^4\,.
    \label{eqn:detQuintic}
\end{align}
We also define the surfaces $S_i=M_1(B_i)=\{\,\text{corank}\,B_i(x)=1\,\}\subset\mathbb{P}^4$,
which by~\eqref{eqn:porteous2x3} have respective degrees
\begin{align}
    \text{deg}\,S_i=a_{i,1}^2+a_{i,1}a_{i,2}+a_{i,2}^2\,.
\end{align}
We then see that~\eqref{eqn:detQuintic} is the special quintic that contains all three surfaces $S_i$, $i=1,2,3$.
Note that if for some $i\in\{1,2,3\}$ one has $a_{i,1}=a_{i,2}=0$ then $S_i=\emptyset$ but this will not be the case of interest to us.
On the other hand, we can safely assume that $M_0(B_i)=\emptyset$.

The restriction of $\pi_1$ to $\widehat{X}^{\rm r}$ induces a birational morphism
\begin{align}
        \pi:\,\widehat{X}^{\rm r}\rightarrow X^{\rm r}\,.
\end{align}
To see this, consider first $p\in\widehat{X}^{\rm r}$, corresponding to coordinates $\vec{x},\vec{u},\vec{v},\vec{w}$, and assume that $p\ne S_1\cup S_2\cup S_3$.
It is easy to check that if, for example, the matrix $B_1$ has full rank one has
\begin{align}
    B_1\vec{u}=0\qquad \Leftrightarrow \qquad [u_1:u_2:u_3]=\left[\det B_1^{[1]}:-\det B_1^{[2]}:\det B_1^{[3]}\right]\,.
\end{align}
This implies that $\vec{x}\in X^{\rm r}$ and that $\pi$ is an isomorphism away from $S_1\cup S_2\cup S_3$.
On the other hand, if $p\in S_1\cup S_2\cup S_3$ it is clear that again $\vec{x}\in X^{\rm r}$.

The threefold $X^{\rm r}$ has isolated nodal singularities that are resolved by the map $\pi:\,\widehat{X}^{\rm r}\rightarrow X^{\rm r}$.
There are two different types of nodes in $X^{\rm r}$ and for each of the two one can further distinguish three situations.

\paragraph{Nodes of Type B}
The nodes that we denote of Type ${\rm B}_i$ for $i=1,2,3$ arise at the intersection $S_j\cap S_k$ for $i\ne j\ne k$.
We denote the set of these nodes by
\begin{align}
    S^{{\rm r},{\rm B}_i}=\{\,\text{corank}\,B_j=\text{corank}\,B_k=1\,\}\,,\quad i,j,k\in\{1,2,3\}\,,\quad  i\ne j\ne k\,,
\end{align}
and the number of nodes $S^{{\rm r},{\rm B}_i}$ is
\begin{align}
    n_{{\rm B}_i}=&(\text{deg}\,S_j)(\text{deg}\,S_k)=\left(a_{j,1}^2+a_{j,1}a_{j,2}+a_{j,2}^2\right)\left(a_{k,1}^2+a_{k,1}a_{k,2}+a_{k,2}^2\right)\,.
    \label{eqn:ex1numBnodes}
\end{align}
Since these loci are determined by the simultaneous vanishing of the minors $\det B_j^{[l]}$ and $\det B_k^{[l]}$ for $l=1,2,3$ it is clear that these points lie on $X^{\rm r}$.
The degrees of the exceptional curves $C_p$ in $\widehat{X}^{\rm r}$ that resolve these nodes are
\begin{align}
    (J_1,\ldots,J_4)\cap C_p=\left\{\begin{array}{cc}
        (0,0,1,1)&\text{ for }p\in S^{{\rm r},{\rm B}_1}\,,\\
        (0,1,0,1)&\text{ for }p\in S^{{\rm r},{\rm B}_2}\,,\\
        (0,1,1,0)&\text{ for }p\in S^{{\rm r},{\rm B}_3}\,.
    \end{array}\right.
\end{align}

\paragraph{Nodes of Type A}
On the other hand, the nodes that we call of Type ${\rm A}_i$ for $i=1,2,3$ are respectively located on the surface $S_i$, away from the intersections.
We will first describe the nodes $p\in S^{{\rm r},{\rm A}_1}$ of Type ${\rm A}_1$ such that $p\in S_1$ and $p\ne S_2\cup S_3$.
Since the matrices $B_2$ and $B_3$ have full rank over such a point, there are unique points $[v_1:v_2:v_3]$ and $[w_1:w_2:w_3]$ over $p$ such that
\begin{align}
    B_2(p)\vec{v}=B_3(p)\vec{w}=\vec{0}\,.
\end{align}
If $p$ is a generic point on $S_1$ then
\begin{align}
    B_1(p)\vec{u}=\sum\limits_{i,j,k=1}^3A_{i,j,k}u_iv_jw_k=0\,,
    \label{eqn:linearConditions}
\end{align}
determines a unique point $[u_1:u_2:u_3]\in \mathbb{P}^2$.
However, if the linear conditions~\eqref{eqn:linearConditions} on $\vec{u}$ are not independent, which is the case when the rank of the matrix
\begin{align}
    \widetilde{B}_1=\left(\begin{array}{ccc}
	B_{1,1,1}&B_{1,1,2}&B_{1,1,3}\\
	B_{1,2,1}&B_{1,2,2}&B_{1,2,3}\\
    \sum\limits_{j,k=1}^3A_{1,j,k}v_jw_k&\sum\limits_{j,k=1}^3A_{2,j,k}v_jw_k&\sum\limits_{j,k=1}^3A_{3,j,k}v_jw_k
    \end{array}\right)\,,
\end{align}
drops by two, the solutions to~\eqref{eqn:linearConditions} span a rational curve inside $\mathbb{P}^2$.
The multiplicity of these points, as well as the corresponding expressions for the nodes of Type A$_2$ and A$_3$, are listed in Table~\ref{tab:ex1nodesTypeA}.

\begin{table}[ht!]
\centering
    \begin{tabular}{|p{14cm}|}\hline
    \textbf{Type A$_1$}, resolved by curves of degree $(0,1,0,0)\in H_2(\widehat{X}^{\rm r},\mathbb{Z})\simeq \mathbb{Z}^4$\\\hline
    {\begin{align*}
        	\begin{split}
				S^{{\rm r},{\rm A}_1}=&\pi(\{\,\text{corank}\,\widetilde{B}_1=2\,\}\cap \{\,B_2\cdot(v_1,v_2,v_3)^{T}=B_3\cdot(w_1,w_2,w_3)^{T}=\vec{0}\,\})\,,\\[.3cm]
			\widetilde{B}_1=&\left(\begin{array}{ccc}
				B_{1,1,1}&B_{1,1,2}&B_{1,1,3}\\
				B_{1,2,1}&B_{1,2,2}&B_{1,2,3}\\
				\sum\limits_{j,k=1}^3A_{1,j,k}v_jw_k&\sum\limits_{j,k=1}^3A_{2,j,k}v_jw_k&\sum\limits_{j,k=1}^3A_{3,j,k}v_jw_k
			\end{array}\right)\,,\\[.3cm]
				n_{{\rm A}_1}=&\left(a_{1,1}^2+a_{1,1} a_{1,2}+a_{1,2}^2\right) \left(a_{2,1} a_{2,2}+a_{3,1} a_{3,2}+2 \left(a_{2,1}+a_{2,2}\right) \left(a_{3,1}+a_{3,2}\right)\right)\\
			 &+a_{1,1} a_{1,2} \left(a_{1,1}+a_{1,2}\right) \left(a_{2,1}+a_{2,2}+a_{3,1}+a_{3,2}\right)+a_{1,1}^2 a_{1,2}^2 \,.
   			\end{split}
    \end{align*}}\\\hline\hline
    \textbf{Type A$_2$}, resolved by curves of degree $(0,0,1,0)\in H_2(\widehat{X}^{\rm r},\mathbb{Z})\simeq \mathbb{Z}^4$\\\hline
    {\begin{align*}
        	\begin{split}
				S^{{\rm r},{\rm A}_2}=&\pi(\{\,\text{corank}\,\widetilde{B}_2=2\,\}\cap \{\,B_1\cdot(u_1,u_2,u_3)^{T}=B_3\cdot(w_1,w_2,w_3)^{T}=\vec{0}\,\})\,,\\[.3cm]
			\widetilde{B}_2=&\left(\begin{array}{ccc}
				B_{2,1,1}&B_{2,1,2}&B_{2,1,3}\\
				B_{2,2,1}&B_{2,2,2}&B_{2,2,3}\\
				\sum\limits_{i,k=1}^3A_{i,1,k}u_iw_k&\sum\limits_{i,k=1}^3A_{i,2,k}u_iw_k&\sum\limits_{i,k=1}^3A_{i,3,k}u_iw_k
			\end{array}\right)\,,\\[.3cm]
				n_{{\rm A}_2}=&\left(a_{2,1}^2+a_{2,1} a_{2,2}+a_{2,2}^2\right) \left(a_{1,1} a_{1,2}+a_{3,1} a_{3,2}+2 \left(a_{1,1}+a_{1,2}\right) \left(a_{3,1}+a_{3,2}\right)\right)\\
				&+a_{2,1} a_{2,2} \left(a_{2,1}+a_{2,2}\right) \left(a_{1,1}+a_{1,2}+a_{3,1}+a_{3,2}\right)+a_{2,1}^2 a_{2,2}^2\,.
   			\end{split}
    \end{align*}}\\\hline\hline
    \textbf{Type A$_3$}, resolved by curves of degree $(0,0,0,1)\in H_2(\widehat{X}^{\rm r},\mathbb{Z})\simeq \mathbb{Z}^4$\\\hline
    {\begin{align*}
        	\begin{split}
				S^{{\rm r},{\rm A}_3}=&\pi(\{\,\text{corank}\,\widetilde{B}_3=2\,\}\cap \{\,B_1\cdot(u_1,u_2,u_3)^{T}=B_2\cdot(v_1,v_2,v_3)^{T}=\vec{0}\,\})\,,\\[.3cm]
			\widetilde{B}_3=&\left(\begin{array}{ccc}
				B_{3,1,1}&B_{3,1,2}&B_{3,1,3}\\
				B_{3,2,1}&B_{3,2,2}&B_{3,2,3}\\
				\sum\limits_{i,j=1}^3A_{i,j,1}u_iv_j&\sum\limits_{i,j=1}^3A_{i,j,2}u_iv_j&\sum\limits_{i,j=1}^3A_{i,j,3}u_iv_j
			\end{array}\right)\,,\\[.3cm]
				n_{{\rm A}_3}=&\left(a_{3,1}^2+a_{3,1} a_{3,2}+a_{3,2}^2\right) \left(a_{1,1} a_{1,2}+a_{2,1} a_{2,2}+2 \left(a_{1,1}+a_{1,2}\right) \left(a_{2,1}+a_{2,2}\right)\right)\\
				&+a_{3,1} a_{3,2} \left(a_{3,1}+a_{3,2}\right) \left(a_{1,1}+a_{1,2}+a_{2,1}+a_{2,2}\right)+a_{3,1}^2 a_{3,2}^2\,.
   			\end{split}
    \end{align*}}\\\hline
    \end{tabular}
    \caption{The nodes of Type A$_i$ in $X^{\rm r}$, their locations $S^{{\rm r},{\rm A}_i}$ and multiplicities $n_{{\rm A}_i}$ for $i=1,2,3$.}
    \label{tab:ex1nodesTypeA}
\end{table}

\subsection{Almost generic quintic threefolds $X_1$ and $X_2$}
In order to obtain the geometries $X_1$ and $X_2$, we start with the corresponding quintic threefolds $X^{\rm r}_a$ from~\eqref{eqn:detQuintic} and deform the equation such that we preserve only the nodes of Type A while smoothing the nodes of Type B.
Using~\eqref{eqn:Ndet}, and
\begin{align}
    \det\left(\begin{array}{ccc}
        0&1&1\\
        1&0&1\\
        1&1&0
    \end{array}\right)=2\,,
\end{align}
we see that the resulting torsion is $B(X_1)=B(X_2)=\mathbb{Z}_2$.
The existence of such a deformation has been explained in Section~\ref{sec:conifoldTransitions} and relies on the fact that in both cases $n_{{\rm B},i}>1$ for all $i=1,2,3$.
This implies that the homology classes of the corresponding exceptional curves satisfy a relation~\eqref{eqn:homologyRelation} with non-zero coefficients.

\paragraph{The quintic $X_1$ with 54 nodes}
Let us first specialize to the values
\begin{align}
    (a_{1,1},a_{1,2},a_{2,1},a_{2,2},a_{3,1},a_{3,2})=(1,1,1,1,1,0)\,.
\end{align}
With standard methods one can calculate
\begin{align}
	h^{1,1}(\widehat{X}^{\rm r}_1)=4\,,\quad h^{2,1}(\widehat{X}^{\rm r}_1)=35\,,\quad \chi(\widehat{X}^{\rm r}_1)=-62\,,
\end{align}
as well as $c_2(T\widehat{X}^{\rm r}_1)\cdot (J_1,J_2,J_3,J_4)=(50,36,36,24)$.
Using~\eqref{eqn:ex1numBnodes} and Table~\ref{tab:ex1nodesTypeA}, we find that the numbers of nodes in $X^{\rm r}_1$ are
\begin{align}
	n_{{\rm A}_1}=22\,,\quad n_{{\rm A}_2}=22\,,\quad n_{{\rm A}_3}=10\,,\quad n_{{\rm B}_1}=&9\,,\quad n_{{\rm B}_2}=3\,,\quad n_{{\rm B}_3}=3\,.
\end{align}
After the transition, we find that the exceptional curves resolving the remaining $\#S=54$ nodes all represent the same non-trivial 2-torsion class in $H_2(\widehat{X}_1,\mathbb{Z})=\mathbb{Z}\times\mathbb{Z}_2$.

\paragraph{The quintic $X_2$ with 48 nodes}
Our second examples corresponds to the choice
\begin{align}
    (a_{1,1},a_{1,2},a_{2,1},a_{2,2},a_{3,1},a_{3,2})=(1,1,2,0,1,0)\,,
\end{align}
corresponding to the topological invariants
\begin{align}
	h^{1,1}(\widehat{X}^{\rm r}_2)=4\,,\quad h^{2,1}(\widehat{X}^{\rm r}_2)=37\,,\quad \chi(\widehat{X}^{\rm r}_2)=-66\,,
\end{align}
as well as $c_2(T\widehat{X}^{\rm r}_2)\cdot (J_1,J_2,J_3,J_4)=(50,36,24,24)$.
Using again~\eqref{eqn:ex1numBnodes} and Table~\ref{tab:ex1nodesTypeA}, we find that the numbers of nodes in $X^{\rm r}_2$ are
\begin{align}
	n_{{\rm A}_1}=19\,,\quad n_{{\rm A}_2}=20\,,\quad n_{{\rm A}_3}=9\,,\quad n_{{\rm B}_1}=&12\,,\quad n_{{\rm B}_2}=3\,,\quad n_{{\rm B}_3}=4\,.
\end{align}
After the transition, we find that the exceptional curves resolving the remaining $\#S=48$ nodes all represent the same non-trivial 2-torsion class in $H_2(\widehat{X}_2,\mathbb{Z})=\mathbb{Z}\times\mathbb{Z}_2$.

\subsection{Mirror symmetry and the conifold transitions $\widehat{X}^{\rm r}\rightarrow (X,[1]_2)$}
\label{sec:ex1conifold}
We now want to find the fundamental periods of the mirrors of the Calabi-Yau backgrounds $(X_a,[1]_2)$ for $a=1,2$.
To this end, we find a limit that is mirror to the transition $\widehat{X}^{\rm r}_a\rightarrow (X_a,[1]_2)$.
The behaviour of the flat coordinates in this limit allows us to deduce the presence of the flat but topologically non-trvial B-field.
We will first discuss the transitition $\widehat{X}^{\rm r}_1\rightarrow (X_1,[1]_2)$ in detail and then provide the corresponding expressions for $(X_2,[1]_2)$.
\paragraph{The transition $\widehat{X}^{\rm r}_1\rightarrow (X_1,[1]_2)$}
Using the techniques from~\cite{Batyrev:1993oya,Hosono:1993qy,Hosono:1994ax} we can easily write down the fundamental period of the mirror of $\widehat{X}^{\rm r}_1$.
It takes the form
\begin{align}
\begin{split}
    \varpi_0^{\widehat{X}^{\rm r}_1}(v,z_1,z_2,z_3)=\sum\limits_{n,l_1,l_2,l_3\ge 0}\binom{n+l_1}{l_1}^2\binom{n+l_2}{l_2}^2\binom{n+l_3}{l_3}\frac{(l_1+l_2+l_3)!}{l_1!l_2!l_3!}v^nz_1^{l_1}z_2^{l_2}z_3^{l_3}\,.
    \end{split}
\end{align}

Let us first find the values for $z_1,z_2,z_3$ that are mirror to points at which we can perform a conifold transition from $\widehat{X}^{\rm r}_1$ to $X_1$.
We can focus on the slice of the moduli space where $z_1=z_2=z_3=z$.
Taking also the limit $v\rightarrow 0$ and using the Frobenius method, we obtain the relevant logarithmic period
\begin{align}
    f(z)=\lim\limits_{\rho\rightarrow 0}\sum\limits_{l_1,l_2,l_3\ge 0}\partial_\rho\left[\frac{(l_1+l_2+l_3+\rho)!}{(l_1+\rho)!l_2!l_3!}z^{l_1+l_2+l_3+\rho}\right]=\frac{1}{1-3z}\log\left(\frac{z}{1-2z}\right)\,.
\end{align}
From this we can conclude that, in the limit $v\rightarrow 0$, the complexified volume of the exceptional curves $C_p$ that resolve the nodes $p\in X^{\rm r}$ is given by
\begin{align}
    \text{Vol}_{\mathbb{C}}(C_p)=\left\{\begin{array}{cc}
        t_e&\text{ for }\,p\in S^{{\rm r},{\rm A}}_1\\
        2t_e&\text{ for }\,p\in S^{{\rm r},{\rm B}}_1
    \end{array}\right.\,,\quad t_e(z)=\frac{1}{2\pi i}\log\left(\frac{z}{1-2z}\right)\,.
\end{align}

Recall that we want to smoothen the nodes of Type B and therefore we want the complexified volume of the curves $C_p$ for $p\in S^{\rm B}_1$ to be trivial.
Since the B-field holonomy is only defined up to shifts by integers, this amounts to the condition $2t_e\in\mathbb{Z}$.
We thus find two points that allow for the desired conifold transition,
\begin{align}
    \lim\limits_{z\rightarrow 1/3}t_e(z)=0\,,\quad \lim\limits_{z\rightarrow 1}t_e(z)=\frac12\,.
\end{align}

In the limit $z\rightarrow 1/3$ the complexified volume of all of the exceptional curves vanishes and therefore we can perform a conifold transition from $\widehat{X}^{\rm r}_1$ to the generic smooth quintic Calabi-Yau $\widetilde{X}_1$.
Indeed, we can perform a suitably regularized limit of the mirror fundamental period and find
\begin{align}
    \begin{split}
    &\lim\limits_{z\rightarrow 1/3}(1-3z)\varpi_0^{\widehat{X}^{\rm r}_1}\left(\left(\frac{1-3z}{z}\right)^5w,z,z,z\right)\\
    =&\sum\limits_{n\ge 0}\frac{(5n)!}{(n!)^5}w^n=1+120w+113400w^2+\mathcal{O}(w^3)\,.
    \end{split}
\end{align}
This is of course nothing but the fundamental period of the mirror quintic~\cite{CANDELAS199121}.

In the second limit, $z\rightarrow 1$, we observe that the complexified volume of the exceptional curves that resolve nodes of Type B is again trivial.
However, the curves $C_p$ for $p\in S^{{\rm r},{\rm A}}_1$, that resolve nodes of Type A, measure a non-trivial B-field holonomy $1/2$ that prevents us from deforming those nodes as well.
Note that the holonomy of $1/2$ is precisely compatible with the fact that, after the transition to $\widehat{X}$, the homology class of the curves becomes 2-torsion.

Taking again the regularized limit we find the corresponding mirror fundamental period
\begin{align}
    \begin{split}
    &\lim\limits_{z\rightarrow 1}(1-3z)\varpi_0^{\widehat{X}^{\rm r}_1}\left(\left(\frac{1-3z}{z}\right)^5w,z,z,z\right)\\
    =&1+88 w^2+1728 w^3+99576 w^4+4104000 w^5+\mathcal{O}(w^6)\,.
    \end{split}
    \label{eqn:AESZ203fundamental}
\end{align}
This is annihilated by the Picard-Fuchs operator AESZ~203~\eqref{eqn:AESZ203}.
We can therefore interpret this operator as annihilating the periods of the mirror of the almost generic nodal quintic $X_1$ that is equipped with a flat but topologically non-trivial B-field $[1]_2\in B(X_1)$.

\paragraph{The transition $\widehat{X}^{\rm r}_2\rightarrow (X_2,[1]_2)$}
The mirror transition of the conifold transition $\widehat{X}^{\rm r}_2\rightarrow (X_2,[1]_2)$ can be understood in a completely analogous fashion.
The fundamental period of the mirror of $\widehat{X}^{\rm r}_2$ takes the form
\begin{align}
    \begin{split}
    &\varpi_0^{\widehat{X}^{\rm r}_2}(v,z_1,z_2,z_3)\\
    =&\sum\limits_{n,l_1,l_2,l_3\ge 0}\binom{n+l_1}{l_1}^2 \binom{2 n+l_2}{l_2} \binom{n+l_3}{l_3} \binom{2 n}{n}\frac{\left(l_1+l_2+l_3\right)!}{l_1! l_2! l_3!}v^n z_1^{l_1}z_2^{l_2}z_3^{l_3}\,.
    \end{split}
\end{align}
Again we find that in the limit $z\rightarrow 1/3$ this can be regularized to yield the fundamental period of the generic mirror quintic
\begin{align}
    \begin{split}
    &\lim\limits_{z\rightarrow 1/3}(1-3z)\varpi_0^{\widehat{X}^{\rm r}_2}\left(\left(\frac{1-3z}{z}\right)^5w,z,z,z\right)=1+120w+113400w^2+\mathcal{O}(w^3)\,.
    \end{split}
\end{align}
However, in the limit $z\rightarrow 1$ we obtain the fundamental period
\begin{align}
    \begin{split}
    &\lim\limits_{z\rightarrow 1}(1-3z)\varpi_0^{\widehat{X}^{\rm r}_1}\left(-\left(\frac{1-3z}{z}\right)^5w,z,z,z\right)\\
    =&1+8 w+504 w^2+36800 w^3+3518200 w^4+365275008 w^5+\mathcal{O}(w^6)\,,
    \end{split}
    \label{eqn:AESZ222fundamental}
\end{align}
that is annihilated by the Picard-Fuchs operator AESZ~222~\eqref{eqn:AESZ222}.
We can therefore interpret this as corresponding to the mirror of the Calabi-Yau background $(X_2,[1]_2)$.~\footnote{We have introduced an additional sign for $w$ in the transition in order to directly reproduce AESZ~222. At the level of the torsion refined GV-invariants, the only consequence is that the $\mathbb{Z}_2$-charges of curves of odd degree in $\widehat{X}_2$ are flipped.}

\section{Almost generic nodal octics with $\mathbb{Z}_3$-torsion}
\label{sec:example2}

Our second type of examples are almost generic octic hypersurfaces $X_3$ and $X_4$ in $\mathbb{P}^4_{1,1,1,1,4}$.
They respectively have 104 and 100 isolated nodes and in both cases the torsion is
\begin{align}
    B(X_3)=B(X_4)=\mathbb{Z}_3\,.
\end{align}
The geometries are again obtained by using conifold transitions from certain CICY Calabi-Yau threefolds $\widehat{X}^{\rm r}_a$, $a=3,4$ as described in Section~\ref{sec:conifoldTransitions}.

Now the geometries $\widehat{X}^{\rm r}_3$ and $\widehat{X}^{\rm r}_4$ are CICY 7242 and CICY 7237 from the classification~\cite{Candelas:1987kf,Green:1987cr}.
The corresponding configuration matrices take the form
\begin{align}
    \widehat{X}^{\rm r}_a=\left[\begin{array}{c|cccccc}
        \mathbb{P}^3&1&1&0&0&1&1\\
        \mathbb{P}^3&0&0&1&1&1&1\\
        \mathbb{P}^3&1&1&a_1&a_2&0&0\\
    \end{array}\right]^{3,39}_{-72}\,,\quad (a_1,a_2)=\left\{\begin{array}{cc}
        (1,1)&\,\,\text{for}\,\,\,a=3\,,\\
        (2,0)&\,\,\text{for}\,\,\,a=4\,.
    \end{array}\right.
    \label{eqn:ex2CICYs}
\end{align}
In order to understand the geometry behind the conifold transition, it will be useful to first familiarize ourself with the concept of hyperdeterminants.

\subsection{Hyperdeterminants from projective duals}
\label{sec:hyperdeterminants}
We start by considering the Veronese embedding of $X=\mathbb{P}^1$ in $\mathbb{P}^2$, given by
\begin{align}
	\nu_2:\,\mathbb{P}^1\hookrightarrow\mathbb{P}^2\,,\quad \nu_2:\,[x_1:x_2]\mapsto[x_1^2:x_1x_2:x_2^2]\,.
\end{align}
The projective dual variety $X^{\vee}$ is the subvariety in the dual projective space that consists of the hyperplanes that are tangent to a smooth point of $X$.
Concretely, we denote the homogeneous coordinates on the dual projective space by $[a_1:a_2:a_3]\in\mathbb{P}^2$ such that the intersection of the corresponding hyperplane with $X$ is
\begin{align}
	I=\{\,a_1x_1^2+a_2x_1x_2+a_3x_2^2=0\,\}\subset X\,.
	\label{eqn:quadric1}
\end{align}
The hyperplane is tangent to $X$ if the quadric is degenerate, i.e. the intersection~\eqref{eqn:quadric1} consists of a single point or $X$ itself.
After rewriting
\begin{align}
	I=\{\,\vec{x}^{\,T}A\vec{x}=0\}\subset X\,,\quad A=\left(\begin{array}{cc}a_1&a_2/2\\a_2/2&a_3\end{array}\right)\,,
\end{align}
we can identify $X^{\vee}$ as the symmetric determinantal variety
\begin{align}
	X^{\vee}=\{\,\det A=0\,\}\subset\mathbb{P}^2\,.
\end{align}
The construction generalizes and gives a definition of the determinant of a symmetric $(n+1)\times (n+1)$ matrices as the defining equation of the projective dual variety of the Veronese embedding of $\mathbb{P}^{n}$ in $\mathbb{P}^{\frac{(n+1)(n+2)}{2}-1}$.

The determinant of a general $(n+1)\times (n+1)$ matrix can similarly be defined in terms of the Segre embedding of $\mathbb{P}^n\times\mathbb{P}^n$ into $\mathbb{P}^{(n+1)^2-1}$.
For example, let $X=\mathbb{P}^1\times\mathbb{P}^1$ with respective homogeneous coordinates $[x_1:x_2]$ and $[y_1:y_2]$ and consider the Segre embedding
\begin{align}
	\sigma:\,[x_1:x_2]\times[y_1:y_2]\mapsto [x_1y_1:x_1y_2:x_2y_1:x_2y_2]\in\mathbb{P}^3\,.
\end{align}
Denoting the homogeneous coordinates on the dual $\mathbb{P}^3$ by $[a_1:a_2:a_3:a_4]$ the intersection of the corresponding hyperplane with $X$ is given by
\begin{align}
	I=\{\,\vec{y}^{\,T}A\vec{x}=0\,\}\subset X\,,\quad A=\left(\begin{array}{cc}a_1&a_3\\a_2&a_4\end{array}\right)\,,
\end{align}
and the projective dual variety again takes the form $X^{\vee}=\{\,\det A=0\,\}\subset\mathbb{P}^3$.

If $A$ has full rank then $I$ is just a rational curve since the projection to any of the $\mathbb{P}^1$ factors of $X$ is a bijection.
On the other hand, if the rank of $A$ drops to one, then $I$ consists of two rational curves that intersect transversely in a point.
To see this, note that when $\vec{x}$ is the up to scale unique (right) zero-eigenvector of $A$ then the condition on $\vec{y}$ is trivial while for generic $\vec{x}$ we again have a bijective projection to the first $\mathbb{P}^1$ factor of $X$.

\paragraph{Hypermatrices and hyperdeterminants}
In the case of an $(n+1)\times (m+1)$ matrix, with $n\ne m$, the analogous construction using the Segre embedding of $\mathbb{P}^n\times\mathbb{P}^m$ leads to a projective dual variety that is not a hypersurface and therefore the determinant of a non-square matrix can be defined to be $1$.
However, when applied to the Segre embedding of more general products $\mathbb{P}^{k_1}\times\ldots\times\mathbb{P}^{k_r}$, the defining equation of the projective dual variety leads to the concept of hyperdeterminants of $(k_1+1)\times\ldots\times (k_r+1)$ hypermatrices.

One refers to the dimensions $(k_1+1)\times\ldots\times (k_r+1)$ as the format of the corresponding hypermatrix.
In the following we will denote by $\text{Det}\,A$ the hyperdeterminant of a hypermatrix $A$ and reserve $\det A$ for the case when $A$ is of format $m\times m$.

Hyperdeterminants of format $2\times 2\times 2$ have first appeared in the work of Cayley~\cite{cayley1845} and for general format they have been systematically introduced over a century later in~\cite{Gelfand1994} (for a nice review see also~\cite{ottaviani2013introduction}).
While the determinant of an $n\times m$ matrix is non-trivial only if $n=m$,~\cite[Theorem 1.3]{Gelfand1994} states that more generally the hyperdeterminant of format $(k_1\times 1)\times\ldots\times (k_r+1)$  is non-trivial if and only if
\begin{align}
	k_j\le\sum\limits_{i\ne j}k_i\,,
\end{align}
for all $j=1,\ldots,r$.
The degree of the hyperdeterminant of format $2\times n\times n$ as a homogeneous polynomial in the entries of the hypermatrix is given by $2n(n-1)$.

A hypermatrix $A$ of format $(k_1+1)\times\ldots\times(k_r+1)$ can be interpreted as a multilinear map $f_A:\,\mathbb{C}^{k_1+1}\times\ldots\times\mathbb{C}^{k_r+1}\rightarrow\mathbb{C}$.
It is called degenerate if there exist non-zero vectors $\vec{x}^{(i)}\in\mathbb{C}^{k_i+1},\,i=1,\ldots,r$ such that 
\begin{align}
	f_A\left(\vec{x}^{(1)},\ldots,\vec{x}^{(i-1)},\vec{y},\vec{x}^{(i+1)},\ldots,\vec{x}^{(r)}\right)=0\,,
\end{align}
for all $i=1,\ldots,r$ and $\vec{y}=\mathbb{C}^{k_i+1}$.
The set of all such tuples $\vec{x}^{(i)},\,i=1,\ldots,r$ in $\mathbb{P}^{k_1}\times\ldots\times\mathbb{P}^{k_r}$ is called the kernel $\mathcal{K}(A)$ and~\cite[Proposition 1.1]{Gelfand1994} states that $\mathcal{K}(A)$ is non-empty if and only if the hyperdeterminant $\text{Det}\,A$ of $A$ vanishes.
The hypermatrix $A[n,m]$ of format $(k_1+1)\times\ldots\times(k_{n-1}+1)\times(k_{n+1}+1)\times\ldots\times (k_r+1)$ with entries
\begin{align}
	A[n,m]_{i_1,\ldots,i_{r-1}}=A_{i_1,\ldots,i_{n},m,i_{n+1},\ldots,i_{r-1}}\,,
\end{align}
is called the $m$-th slice in the $n$-th direction of $A$.

While a general method to calculate hyperdeterminants in terms of a so-called Cayley-Koszul complex has been developed in~\cite{Gelfand1994}, again building on earlier work by Cayley, it is sometimes possible to calculate hyperdeterminants using the so-called Schl\"afli trick.
In the case of a hypermatrix $A$ of format $2\times n\times n$, one introduces an auxiliary variable $X$ and defines the $n\times n$ matrix
\begin{align}
	\widetilde{A}=x A[1,1]+A[1,2]\,.
\end{align}
The hyperdeterminant of $A$ is then the discriminant of the polynomial $\det \widetilde{A}$ in $x$.

\paragraph{Hyperdeterminants of format $2\times 2\times 2$}
As an example, consider the Segre embedding $\sigma:\,\mathbb{P}^1\times\mathbb{P}^1\times\mathbb{P}^1\hookrightarrow\mathbb{P}^7$ corresponding to a hypermatrix $A$ of format $2\times2\times 2$.
We denote the homogeneous coordinates on the factors of $X=\mathbb{P}^1\times\mathbb{P}^1\times\mathbb{P}^1$ respectively by $[p_1:p_2]$, $[x_1:x_2]$ and $[y_1:y_2]$.
The entries $A_{s,i,j}$ of $A$ provide homogeneous coordinates on the dual projective space and the intersection of the corresponding hyperplane with $X$ takes the form
\begin{align}
	I=\left\{\,f_A(\vec{p},\vec{x},\vec{y})=0\,\right\}\subset X\,,\quad f_A(\vec{p},\vec{x},\vec{y})=\sum\limits_{s,i,j}A_{s,i,j}p_sx_iy_j\,.
\end{align}
The Schl\"afli trick for the case $n=2$ gives the hyperdeterminant
\begin{align}
	\begin{split}
	\text{Det}\,A=&A_{1,2,2}^2 A_{2,1,1}^2-2 A_{1,2,1} A_{1,2,2} A_{2,1,2} A_{2,1,1}-2 A_{1,1,2} A_{1,2,2} A_{2,2,1} A_{2,1,1}\\
		&+4 A_{1,1,2} A_{1,2,1} A_{2,2,2} A_{2,1,1}-2 A_{1,1,1} A_{1,2,2} A_{2,2,2} A_{2,1,1}+A_{1,2,1}^2 A_{2,1,2}^2\\
		&+A_{1,1,2}^2 A_{2,2,1}^2+A_{1,1,1}^2 A_{2,2,2}^2-2 A_{1,1,2} A_{1,2,1} A_{2,1,2} A_{2,2,1}\\
		&+4 A_{1,1,1} A_{1,2,2} A_{2,1,2} A_{2,2,1}-2 A_{1,1,1} A_{1,2,1} A_{2,1,2} A_{2,2,2}\\
		&-2 A_{1,1,1} A_{1,1,2} A_{2,2,1} A_{2,2,2}\,.
	\end{split}
	\label{eqn:hyper222}
\end{align}
We observe, that this is the determinant of the symmetric $2\times 2$ matrix $\rm A$, with entries
\begin{align}
    {\rm A}_{i,j}=\sum\limits_{\sigma_1,\sigma_2\in S_2}\text{sgn}(\sigma_1)\text{sgn}(\sigma_2)A_{i,\sigma_1(1),\sigma_2(1)}A_{j,\sigma_1(2),\sigma_2(2)}\,,
\end{align}
where $S_2$ is the group of permutations of \{1,2\}, such that
\begin{align}
    \text{Det}\,A=\sum\limits_{\sigma_1\in S_{2}}\prod_{i=1}^2\text{sgn}(\sigma_1)\sum\limits_{\sigma_2,\sigma_3\in S_{2}}\text{sgn}(\sigma_2)\text{sgn}(\sigma_3)A_{i,\sigma_2(1),\sigma_3(1)}A_{\sigma_1(i),\sigma_2(2),\sigma_3(2)}\,.
\end{align}

We can equally interpret $\text{Det}\,A=0$ as the condition for the intersection
\begin{align}
	I_A=\left\{\,\sum\limits_{i,j}A_{1,i,j}x_iy_j=\sum\limits_{i,j}A_{2,i,j}x_iy_j=0\,\right\}\subset\mathbb{P}^1\times\mathbb{P}^1\,,
    \label{eqn:cint11}
\end{align}
to degenerate.
To see this, let us define
\begin{align}
    f_k(\vec{x},\vec{y})=\sum\limits_{i,j}A_{k,i,j}x_iy_j\,,\quad k=1,2\,.
\end{align}
The variety $I_A$ is singular at a point if and only if $f_1=f_2=0$ and
\begin{align}
    \text{rank}\,\left(\begin{array}{cccc}
        \partial_{x_1}f_1&\partial_{x_2}f_1&\partial_{y_1}f_1&\partial_{y_2}f_1\\
        \partial_{x_1}f_2&\partial_{x_2}f_2&\partial_{y_1}f_2&\partial_{y_2}f_2
    \end{array}\right)\le 1\,.
\end{align}
The latter condition just means that there exists $\vec{p}\in\mathbb{P}^1$ such that
\begin{align}
    \sum\limits_{k,j}A_{k,a,j}p_ky_j=0\,,\quad \sum\limits_{k,i}A_{k,i,b}p_kx_i=0\,,\quad a,b=1,2\,.
\end{align}
Therefore a point in $\mathbb{P}^1\times\mathbb{P}^1$ is a singular point on $I_A$ if and only if the determinant of the hypermatrix $A_{k,i,j}$ vanishes.

\paragraph{Degeneration loci in the space of $2\times 2\times 2$ hypermatrices}
As was discussed for example in~\cite{Weyman1996}, the space $M$ of $2\times 2\times 2$ hypermatrices contains subspaces $\nabla,\nabla_{\text{cusp}}$ of respective codimensions 1 and 3 such that
\begin{align}
    0\in \nabla_{\text{cusp}}\subset \nabla\subset M\,.
\end{align}
The subset $\nabla\subset M$ consists of the matrices with vanishing hyperdeterminant while $\nabla_{\text{cusp}}$ is the set of singular points of $\nabla$.
Depending on the choice for $A$, the variety $I_A$ then takes the following form
\begin{align}
    I_A=\left\{\begin{array}{cl}
        \text{2 points}&\text{ for }A\in M\backslash\nabla\,,\\
        \text{1 point}&\text{ for }A\in \nabla\backslash\nabla_{\text{cusp}}\,,\\
        \mathbb{P}^1&\text{ for }A\in \nabla_{\text{cusp}}\backslash\{0\}\,,\\
        \mathbb{P}^1\times\mathbb{P}^1&\text{ for }A=0\,.
    \end{array}\right.
\end{align}

Let us introduce the divisors $J_1=\{\,x_1=0\,\}$ and $J_2=\{\,y_1=0\,\}$ on $\mathbb{P}^1\times\mathbb{P}^1$ and denote the degree of a curve $C$ by $(d_1,d_2)=(J_1,J_2)\cdot C$.
For hypermatrices $A\in \nabla_{\text{cusp}}\backslash\{0\}$ we can then further distinguish the following three cases, that we refer to as Type A$_i$, $i=1,2,3$:

\begin{enumerate}
	\item The degeneration of Type A$_1$ corresponds to
		\begin{align}
                \text{rank}\,\left(\begin{array}{cc}
                    A[1,1]&A[1,2]
                \end{array}\right)=
			\text{rank}\,\left(\begin{array}{cccc}
				A_{1,1,1}&A_{1,1,2}&A_{2,1,1}&A_{2,1,2}\\
				A_{1,2,1}&A_{1,2,2}&A_{2,2,1}&A_{2,2,2}
			\end{array}\right)=1\,.
		\end{align}
		Then there exists a unique point $\vec{x}\in\mathbb{P}^1$ for which the conditions on $\vec{y}\in\mathbb{P}^1$ vanish identically.
		It is also clear that this is the only point $\vec{x}$ for which a solution $\vec{y}$ exists and one finds that $I_A\simeq\mathbb{P}^1$ is a rational curve of degree $(0,1)$.
	\item Similarly, the degeneration of Type A$_2$ corresponds to
		\begin{align}
                \text{rank}\,\left(\begin{array}{cc}
                    A[1,1]^T&A[1,2]^T
                \end{array}\right)=
			\text{rank}\,\left(\begin{array}{cccc}
				A_{1,1,1}&A_{1,2,1}&A_{2,1,1}&A_{2,2,1}\\
				A_{1,1,2}&A_{1,2,2}&A_{2,1,2}&A_{2,2,2}
			\end{array}\right)=1\,,
		\end{align}
		and we find that $I_A\simeq\mathbb{P}^1$ is a rational curve of degree $(1,0)$.
	\item Type A$_3$ corresponds to
		\begin{align}
                \text{rank}\,\left(\begin{array}{cc}
                    A[2,1]&A[2,2]
                \end{array}\right)=
			\text{rank}\,\left(\begin{array}{cccc}
				A_{1,1,1}&A_{1,1,2}&A_{1,2,1}&A_{1,2,2}\\
				A_{2,1,1}&A_{2,1,2}&A_{2,2,1}&A_{2,2,2}\\
			\end{array}\right)=1\,,
		\end{align}
		such that there exists a linear combination of the two defining equations of $I_A$ that vanishes identically.
		Without loss of generality we can assume that neither of the equations vanishes itself identically and in that case $I_A$ is a rational curve 
		\begin{align}
			I_A=\left\{\,\sum\limits_{i,j}A_{1,i,j}x_iy_j=0\,\right\}=\left\{\,\sum\limits_{i,j}A_{2,i,j}x_iy_j=0\,\right\}\subset\mathbb{P}^1\times\mathbb{P}^1\,,
		\end{align}
		of degree $(1,1)$.
\end{enumerate}

\paragraph{Restricting to subspaces}
As our notation suggests, we have already understood conceptually how the nodes of Type A will arise.
However, in order to get the nodes of Type B we have to go one step further.

Let $m\ge 2$ and consider a hypermatrix $A$ of format $2\times (m+1)\times(m+1)$ as well as two ordinary matrices $B$ and $C$ of respective format $(m+1)\times (m-1)$ and $(m-1)\times(m+1)$.
We can then define a complete intersection
\begin{align}
	\begin{split}
		I=&\left\{\,\sum\limits_{i,j=1}^{m+1} A_{1,i,j}x_iy_j=\sum\limits_{i,j=1}^{m+1} A_{2,i,j}x_iy_j=0\,,\,\,\vec{x}\,B=C\,\vec{y}=\vec{0}\,\right\}\subset\mathbb{P}^{m}\times\mathbb{P}^m\,,
	\end{split}
    \label{eqn:restriction}
\end{align}
where $\vec{x}$ and $\vec{y}$ are respective homogeneous coordinates on the first and second factor of $\mathbb{P}^m\times \mathbb{P}^m$.

If the entries of $A$, $B$ and $C$ are chosen to be sufficiently generic, then the set $I$ will consist of two points.
On the other hand, if the matrices $B$ and $C$ have full rank, then we can solve the equations
\begin{align}
    \{\,\vec{x}\,B=C\,\vec{y}=\vec{0}\,\}\simeq \mathbb{P}^1\times\mathbb{P}^1\,,
\end{align}
and translate the problem again into the form~\eqref{eqn:cint11} for some $2\times 2\times 2$ hypermatrix $\widetilde{A}$ that depends on the entries of $A$, $B$ and $C$.
In this way, after staring at the resulting expression for a sufficient amount of time, we find that in the space $M$ of matrices $A,B,C$, the locus where $I$ degenerates is given by
\begin{align}
    \{\,\det {\rm A}=0\,\}\subset M\,,
\end{align}
where ${\rm A}$ is now the $2\times 2$ matrix
\begin{align}\tiny
        {\rm A}_{i,j}=\sum\limits_{\sigma_1,\sigma_2\in S_{m+1}}\text{sgn}(\sigma_1)\text{sgn}(\sigma_2)A_{i,\sigma_1(1),\sigma_2(1)}A_{j,\sigma_1(2),\sigma_2(2)}\prod\limits_{r=1}^{m-1}B_{\sigma_1(2+r),r}(z)C_{r,\sigma_2(2+r)}(z)\,.
\end{align}

We say that $(A,B,C)\in M$ is a degeneration of Type A$_i$ if $B$ and $C$ have full rank but $\widetilde{A}$ is degenerate of Type A$_i$.
Let us denote the loci where the rank of $B$ and $C$ drops respectively by $M_B$ and $M_C$ in $M$.
From Section~\ref{sec:porteous} we recall that the codimension of both $M_B$ and $M_C$ is three.
If we keep $A$ fixed with generic entries, then we can assume that both $M_B$ and $M_C$ intersect the locus where Type A degenerations occur only in codimension $\ge 4$.

We say that a degeneration of $I$ is of Type B$_1$ if $A,B,C$ correspond to a generic point of $M_B$, such that the rank of $B$ drops to $m-2$.
Similarly, we say that a degeneration is of Type B$_2$ if $A,B,C$ correspond to a generic point of $M_C$, such that the rank of $C$ drops to $m-2$.

Let us introduce again divisors $J_1=\{\,x_1=0\,\}$ and $J_2=\{\,y_1=0\,\}$ on $\mathbb{P}^m\times\mathbb{P}^m$ and denote the degree of a curve $C$ by $(d_1,d_2)=(J_1,J_2)\cdot C$.
For degenerations of Type B we then find that $I$ is a curve of the following degree:
\begin{align}
    (J_1,J_2)\cdot I=\left\{\begin{array}{cc}
        (2,1)&\text{for $I$ of Type B}_1\\
        (1,2)&\text{for $I$ of Type B}_2
    \end{array}\right.
\end{align}

\subsection{The geometry of $\widehat{X}^{\rm r}$ and $X^{\rm r}$}
We will now apply the discussion from Section~\ref{sec:hyperdeterminants} to the CICY Calabi-Yau threefolds $\widehat{X}^{\rm r}_a$ with $a=3,4$ from~\eqref{eqn:ex2CICYs}.
Denote by $\vec{x},\vec{y},\vec{z}$ the homogeneous coordinates on $\mathbb{P}^3\times \mathbb{P}^3\times \mathbb{P}^3$.
In both cases the defining equations can then be written as
\begin{align}
	\begin{split}
		\widehat{X}^{\rm r}=&\left\{\,\sum\limits_{i,j=1}^4 A_{1,i,j}x_iy_j=\sum\limits_{i,j=1}^4 A_{2,i,j}x_iy_j=0\,,\,\,\vec{x}B(z)=C(z)\vec{y}=\vec{0}\,\right\}\subset\left(\mathbb{P}^3\right)^3\,,
	\end{split}
    \label{eqn:ex2Xrhat}
\end{align}
where $A_{k,i,j}\in\mathbb{C}$ are the entries of a generic $2\times4\times 4$ hypermatrix $A$, $B(z)$ is a $4\times 2$ matrix and $C(z)$ is a $2\times 4$ matrix.
The entries of the $k$-th column of $B(z)$ are homogeneous polynomials of degree $a_k$ in $z$, with $a_k$, $k=1,2$ as in~\eqref{eqn:ex2CICYs}, while the entries of $C(z)$ are all linear homogeneous polynomials in $z$.

From the discussion at the end of Section~\ref{sec:hyperdeterminants}, we know that there exists a morphism $\rho:\widehat{X}^{\rm r}\rightarrow X^{\rm r}$, where $X^{\rm r}$ is a double cover of $\mathbb{P}^3$ that is ramified over the octic surface $\{\,\det{\rm A}=0\,\}\subset\mathbb{P}^3$ in terms of the symmetric matrix ${\rm A}_{i,j}$ with entries
\begin{align}
        {\rm A}_{i,j}=\sum\limits_{\sigma_1,\sigma_2\in S_{4}}\text{sgn}(\sigma_1)\text{sgn}(\sigma_2)A_{i,\sigma_1(1),\sigma_2(1)}A_{j,\sigma_1(2),\sigma_2(2)}\prod\limits_{r=1}^{2}B_{\sigma_1(2+r),r}(z)C_{r,\sigma_2(2+r)}(z)\,.
\end{align}
Concretely, we can construct $X^{\rm r}$ as a hypersurface
\begin{align}
    X^{\rm r}=\{\,\det{\rm A}=0\,\}\subset\mathbb{P}^4_{1,1,1,1,4}\,.
\end{align}

The degenerations of Type A$_i$ and Type B$_j$ discussed in Section~\ref{sec:hyperdeterminants} correspond to nodes of the surface $\{\,\det{\rm A}=0\,\}$ and therefore also to nodes of the double cover $X^{\rm r}$.
In order to deduce the number of the nodes of each type, we will take a shortcut and use the Gopakumar-Vafa invariants of $\widehat{X}^{\rm r}_a$ for $a=3,4$ that can easily be calculated using standard techniques e.g. from~\cite{Hosono:1994ax}.

We denote the divisors on $\widehat{X}^{\rm r}$ that are inherited from the hyperplane classes of the three $\mathbb{P}^3$ factors respectively by $J_1,J_2,J_3$, and the degree of a curve $C\subset \widehat{X}^{\rm r}$ by $(d_1,d_2,d_3)=C\cdot  (J_1,J_2,J_3)$.
The genus zero Gopakumar-Vafa invariants for $d_3=0$ are then as follows:
\begin{align}
	\widehat{X}^{\rm r}_3:\quad\begin{array}{|c|ccc|}\hline
		n^{d_1,d_2,0}_0&d_2=0&1&2\\\hline
		d_1=0& &28&0\\
		1&28&48&4\\
		2&0&4&0\\\hline
	\end{array}\,,\qquad\widehat{X}^{\rm r}_4:\quad
    	\begin{array}{|c|ccc|}\hline
		n^{d_1,d_2,0}_0&d_2=0&1&2\\\hline
		d_1=0& &28&0\\
		1&24&48&8\\
		2&0&4&0\\\hline
	\end{array}
\end{align}
Comparing again with the discussion in Section~\ref{sec:hyperdeterminants}, we deduce that the numbers $n_{{\rm A}_i}$, $n_{{\rm B}_j}$ of nodes of Type A and B are
\begin{align}
    (n_{{\rm A}_1},n_{{\rm A}_2},n_{{\rm A}_3},n_{{\rm B}_1},n_{{\rm B}_2})=\left\{\begin{array}{cc}
        (28,\,28,\,48,\,4,\,4)&\,\,\text{for}\,\,\,\widehat{X}^{\rm r}_3\,,\\
        (28,\,24,\,48,\,4,\,8)&\,\,\text{for}\,\,\,\widehat{X}^{\rm r}_4\,.\\
    \end{array}\right.
    \label{eqn:ex2nodesAB}
\end{align}
Using a random choice of coefficients we have verified numerically that $\widehat{X}^{\rm r}_3$ and $\widehat{X}^{\rm r}_4$ indeed both have 112 isolated nodes, and that $n_{{\rm B}_1}+n_{{\rm B}_2}$ of them correspond to the points where $B(C)$ or $C(z)$ has rank 1.

\subsection{Almost generic octic threefolds $X_3$ and $X_4$}
To obtain the geometries $X_3$ and $X_4$ we start again with the corresponding octic threefolds $X^{\rm r}_a$ from~\eqref{eqn:detQuintic} and deform the equation such that we preserve only the nodes of Type A while smoothing the nodes of Type B.
Using~\eqref{eqn:Ndet}, and
\begin{align}
    \det\left(\begin{array}{cc}
        2&1\\
        1&2
    \end{array}\right)=3\,,
\end{align}
we see that the resulting torsion is $B(X_3)=B(X_4)=\mathbb{Z}_3$.
The existence of such a deformation follows again from the discussion in Section~\ref{sec:conifoldTransitions} and the fact that in both cases $n_{{\rm B}_1}$ and $n_{{\rm B}_2}$ are greater than one.

The number of nodes of the different types was given in~\eqref{eqn:ex2nodesAB}.
After the transition we find that the remaining number of nodes of Type A in $X_3$ and $X_4$ is respectively 104 and 100.
The exceptional curves resolving each of these nodes represent a non-trivial 3-torsion class in $H_2(\widehat{X}_a,\mathbb{Z})=\mathbb{Z}\times\mathbb{Z}_3$.

\subsection{Mirror symmetry and the conifold transitions $\widehat{X}^{\rm r}\rightarrow (X,[\pm1]_3)$}
As in Section~\ref{sec:conifoldTransitions}, we will now find a limit in the complex structure moduli space of the mirror of $\widehat{X}^{\rm r}_a$ that is dual to the transition $\widehat{X}^{\rm r}_a\rightarrow (X_a,[\pm1]_3)$ for $a=3,4$.
We will first discuss the transition for $X_3$ and then state the corresponding results for $X_4$.

\paragraph{The transition $\widehat{X}^{\rm r}_3\rightarrow (X_3,[\pm 1]_3)$}
Using again the techniques from~\cite{Batyrev:1993oya,Hosono:1993qy,Hosono:1994ax} we obtain the fundamental period of the mirror of $\widehat{X}^{\rm r}_3$, which takes the form
\begin{align}
	\begin{split}
	\varpi_0^{\widehat{X}^{\rm r}_3}(v,z_1,z_2)=&\sum\limits_{n,l_1,l_2\ge 0}\binom{l_1+l_2}{l_1}^2\binom{n+l_1}{l_1}^2\binom{n+l_2}{l_2}^2z_1^{l_1}z_2^{l_2}v^n\,.
	\end{split}
\end{align}
We denote the three single-logarithmic periods by
\begin{align}
	\varpi_{1,i}=\varpi_0 \log z_i+\mathcal{O}(z,v)\,,\quad i=1,2,3\,,
\end{align}
and observe that they satisfy
\begin{align}
	\varpi_{1,1}/\varpi_0=&\log z_1+2z_2\sum\limits_{l_1,l_2\ge 0}\frac{1}{l_2+1}\binom{l_1+l_2}{l_1}^2z_1^{l_1}z_2^{l_2}+\mathcal{O}(z_3)\,,\\
	\varpi_{1,2}/\varpi_0=&\log z_2+2z_1\sum\limits_{l_1,l_2\ge 0}\frac{1}{l_1+1}\binom{l_1+l_2}{l_1}^2z_1^{l_1}z_2^{l_2}+\mathcal{O}(z_3)\,.
\end{align}
We focus on the slice $z_1=z_2=z$ and in the limit $v\rightarrow 0$ one obtains
\begin{align}
	t_e(z):=&\frac{1}{2\pi i}\frac12\frac{\varpi_{1,1}+\varpi_{1,2}}{\varpi_0}\big\vert_{z_1\rightarrow z,\,z_2\rightarrow z,v\rightarrow 0}=\frac{1}{2\pi i}\log\left(\frac{4z}{(1+\sqrt{1-4z})^2}\right)\,.
\end{align}
In the limit $v\rightarrow 0$, the complexified volume of a curve of degree $(d_1,d_2,0)$ is therefore given by $(d_1+d_2)t_e(z)$.
This means that the exceptional curve $C_p$ that resolves a node $p\in X^{\rm r}$ has complexified volume
\begin{align}
    \text{Vol}_{\mathbb{C}}(C_p)=\left\{\begin{array}{cl}
        t_e&\text{ for }\,p\in S^{{\rm r},{\rm A}_1}\cup S^{{\rm r},{\rm A}_2}\\
        2t_e&\text{ for }\,p\in S^{{\rm r},{\rm A}_3}\\
        3t_e&\text{ for }\,p\in S^{{\rm r},{\rm B}_1}\cup S^{{\rm r},{\rm B}_2}
    \end{array}\right.\,,
\end{align}
where $S^{{\rm r},{\rm A}_i}$ and $S^{{\rm r},{\rm B}_j}$ respectively denote the subsets of nodes of Type A$_i$ and B$_j$ in $X^{\rm r}_3$.

In order to perform the conifold transition from $\widehat{X}^{\rm r}_3$ to $X_3$ we want to smoothen the nodes of Type B and therefore need to take a limit where $3t_e(z)\in \mathbb{Z}$.
The two relevant limits are therefore
\begin{align}
	\lim\limits_{z\rightarrow 1/4}t_e(z)=0\,,\quad \lim\limits_{z\rightarrow 1}t_e(z)=-\frac13\,.
\end{align}

In the first limit, $z\rightarrow 1/4$, we observe that $t_e(z)\rightarrow 0$.
Therefore the complexified volume of all of the exceptional curves is zero and we can deform away all of the nodes in $X^{\rm r}_3$, leading to a conifold transition from $\widehat{X}^{\rm r}_3$ to the generic smooth octic $\widetilde{X}_3\subset\mathbb{P}^4_{1,1,1,1,4}$.
At the level of the fundamental period, this corresponds to the regularized limit
\begin{align}
    \begin{split}
    &\lim_{z\rightarrow 1/4}\sqrt{1-4z}\,\varpi_0^{\widehat{X}^{\rm r}_3}\left(\frac{(1-4z)^4}{z^6}w,z,z\right)\\
    =&\sum\limits_{n\ge 0}\frac{(8n)!}{(n!)^4(4n)!}w^n
    =1+1680w+32432400w^2+\mathcal{O}(w^3)\,,
    \end{split}
\end{align}
which is indeed the fundamental period of the mirror of $\widetilde{X}_3$.

Taking on the other hand the limit $z\rightarrow 1$, we observe that only the complexified volume along the curves that resolve the nodes of Type B is trivial.
The other exceptional curves measure a non-trivial B-field holonomy of $\pm 1/3$ and the nodes of Type A are therefore protected against deformation.
The mirror fundamental period that one obtains after the transition can be obtained by taking the limit
\begin{align}
	\begin{split}
	&\lim_{z\rightarrow 1}\sqrt{1-4z}\,\varpi_0^{\widehat{X}^{\rm r}_3}\left(\frac{(1-4z)^4}{z^6}w,z,z\right)\\
		=&1+15w+567w^2+28113w^3+\mathcal{O}(w^4)\,,
	\end{split}
    \label{eqn:AESZ199fundamental}
\end{align}
and is annihilated by the Picard-Fuchs operator AESZ~199~\eqref{eqn:AESZ199}.
We can therefore interpret this operator as annihilating the periods of the mirror of the Calabi-Yau background $(X_3,[\pm 1]_3)$.

\paragraph{The transition $\widehat{X}^{\rm r}_4\rightarrow (X_4,[\pm 1]_3)$}
The mirror transition of the conifold transition $\widehat{X}^{\rm r}_4\rightarrow (X_4,[\pm 1]_3)$ can again be understood in a completely analogous fashion.
The fundamental period of the mirror of $\widehat{X}^{\rm r}_4$ takes the form
\begin{align}
    \begin{split}
    \varpi_0^{\widehat{X}^{\rm r}_4}(v,z_1,z_2,z_3)
    = \binom{2 n+l_1}{l_1} \binom{n+l_2}{l_2}^2\binom{l_1+l_2}{l_1}^2 \binom{2 n}{n}z_1^{l_1}z_2^{l_2}v^n\,.
    \end{split}
\end{align}
In the limit $z\rightarrow 1/4$ this can be regularized to yield the fundamental period of the generic mirror octic
\begin{align}
    \begin{split}
    &\lim\limits_{z\rightarrow 1/4}\,\sqrt{1-4z}\varpi_0^{\widehat{X}^{\rm r}_4}\left(\frac{(1-4z)^4}{z^6}w,z,z\right)=1+1680w+32432400w^2+\mathcal{O}(w^3)\,.
    \end{split}
\end{align}
However, in the limit $z\rightarrow 1$ we obtain the fundamental period
\begin{align}
    \begin{split}
    &\lim\limits_{z\rightarrow 1}\sqrt{1-4z}\,\varpi_0^{\widehat{X}^{\rm r}_4}\left(\frac{(1-4z)^4}{z^6}w,z,z\right)\\
    =&1+24 w+1944 w^2+232800 w^3+34133400 w^4+5649061824 w^5+\mathcal{O}(w^6)\,,
    \end{split}
    \label{eqn:AESZ350fundamental}
\end{align}
that is annihilated by the Picard-Fuchs operator AESZ~350~\eqref{eqn:AESZ350}.
We can therefore interpret this as corresponding to the mirror of the Calabi-Yau background $(X_4,[\pm 1]_3)$.

\section{A degeneration of $X_{(6,6)}$ with 50 nodes and $\mathbb{Z}_5$-torsion}
\label{sec:example3}
Our final example, the almost generic Calabi-Yau threefold $X_5$ with torsion $B(X_5)=\mathbb{Z}_5$, will not be constructed by following a conifold transition.
Instead, we will deduce the existence of this $\mathbb{Q}$-factorial degeneration of the generic complete intersection $X_{(6,6)}\subset\mathbb{P}^5_{1,1,2,2,3,3}$, as well as its properties, directly from an irrational Picard-Fuchs operator.

Using $v_\pm=-\frac12(11\pm 5\sqrt{5})$, which are the roots of the monic polynomial $\widetilde{\Delta}=v^2+11v-1$, the Picard-Fuchs operator takes the form
\begin{align}
    \begin{split}
    \mathcal{D}_-=&\mathcal{D}_-^A+z\,v_-\,\mathcal{D}_-^B\,,\\
            \mathcal{D}_-^A=&2 \theta ^4-4 z \left(5+36 \theta +101 \theta ^2+130 \theta ^3+62 \theta ^4\right)\\
        &+2 z^2 \left(660+2422 \theta +3181 \theta ^2+746 \theta^3-373 \theta ^4\right)\\
        &-250 z^3 \left(304+402 \theta -609 \theta ^2-1482 \theta ^3-247 \theta ^4\right)\\
        &-500 z^4 \left(18850+69340\theta +84589 \theta ^2+30498 \theta ^3-7626 \theta ^4\right)\\
        &-472718750 z^5 (1+\theta )^4\,,\\
        \mathcal{D}_-^B=&2 \left(5+26 \theta +59 \theta ^2+66 \theta ^3\right)\\
        &-2 z \left(5880+20446 \theta +26183 \theta ^2+5478 \theta ^3-2739 \theta^4\right)\\
        &+1000 z^2 \left(843+1119 \theta -1673 \theta ^2-4092 \theta ^3-682 \theta ^4\right)\\
        &+250 z^3 \left(418100+1537990 \theta+1876229 \theta ^2+676478 \theta ^3-169136 \theta ^4\right)\\
        &+5242531250 z^4 (1+\theta )^4\,.
    \end{split}
    \label{eqn:irrationalOpOrder4}
\end{align}
As usual, $\theta=z\partial_z$.
A corresponding operator $\mathcal{D}_+$ is obtained by exchanging $v_-$ with $v_+$.
The discriminant polynomial of the operator~\eqref{eqn:irrationalOpOrder4} is given by
\begin{align}
    \Delta(z)=\Delta_1\Delta_2\Delta_3\,,\quad \Delta_i=1-\frac{z}{z_{{\rm c},i}}\,,\quad i=1,2,3\,,
\end{align}
in terms of the points
\begin{align}
    z_{{\rm c},1}=\frac{1}{125}\,,\quad z_{{\rm c},2}=\frac{1}{50} \left(25+11 \sqrt{5}\right)\,,\quad z_{{\rm c},3}=\frac{1}{2} \left(123+55 \sqrt{5}\right)\,,
\end{align}
and the Riemann symbol is as follows:
\begin{align}
    \left\{
    \begin{array}{ccccc}
    0&z_{{\rm c},1}&z_{{\rm c},2}&z_{{\rm c},3}&\infty\\\hline
    0&0&0&0&1\\
    0&1&1&1&1\\
    0&1&1&1&1\\
    0&2&2&2&1
    \end{array}
    \right\}
\end{align}
All three points $z_{{\rm c},i}$ for $i=1,2,3$ are conifold points and we will later show that the point at $z=\infty$ is related by a $\mathbb{Z}_2$-symmetry of the moduli space to the point $z=0$.
Before that, let us first explain how we obtained this operator.

\subsection{Constructing the irrational differential operator}
Consider first the Grassmanian $\text{Gr}(2,5)$, denote the tautological bundle by $\mathcal{S}$ and write $\mathcal{O}(1)=\det \mathcal{S}^\vee$.
The vanishing locus of a generic section of $\mathcal{O}(1)^{\oplus 5}$ is a genus one curve and the restriction of $\mathcal{O}(1)$ induces a polarization of degree 5.
The curve can equivalently be obtained as the vanishing locus of the $4\times 4$ Pfaffians of a $5\times 5$ skew-symmetric matrix with entries that are linear polynomials in the homogeneous coordinates on $\mathbb{P}^4$.
For more details see for example~\cite{Knapp:2021vkm}.

Using conifold transitions, as described in~\cite{Batyrev:1998kx}, one finds that the mirror family of elliptic curves takes the form
\begin{align}
    \begin{split}
        y^2=&x^3-3 \left(1-12 v+14 v^2+12 v^3+v^4\right)x\\
        &+2 \left(1+v^2\right) \left(1-18 v+74 v^2+18 v^3+v^4\right)\,.
    \end{split}
\end{align}
Comparing with the results from~\cite{Aspinwall:1998xj} one can see that this curve has a 5-torsion point.
The fundamental period takes the form~\footnote{In~\cite{Hori:2013gga} this has also been obtained from a localization calculation of the sphere partition function of a non-Abelian gauged linear sigma model that flows to a non-linear sigma model on the complete intersection curve in $\text{Gr}(2,5)$.}
\begin{align}
    \widetilde{\varpi}_0(v)=\sum\limits_{n,k\ge 0}\binom{n}{k}^2\binom{n+k}{k}v^n\,,
\end{align}
and is annihilated by the Picard-Fuchs operator
\begin{align}
    \widetilde{\mathcal{D}}=\theta ^2-v \left(3+11 \theta +11 \theta ^2\right)-v^2 (1+\theta )^2\,,\quad \theta=v\partial_v\,.
\end{align}
The discriminant polynomial is given by $\widetilde{\Delta}=v^2+11v-1$ and there are four points of maximally unipotent monodromy~\footnote{Note that conifold points have maximally unipotent monodromy for Calabi-Yau 1-folds.}
\begin{align}
    v_0=0\,,\quad v_\pm=-\frac12(11\pm 5\sqrt{5})\,,\quad v_\infty=\infty\,.
\end{align}

We want to expand around $v_-$ and to this end introduce the coordinate
\begin{align}
    u=-\frac{1}{5\sqrt{5}}\frac{1}{v_-}(v-v_-)\,.
\end{align}
The transformed Picard-Fuchs operator then takes the form
\begin{align}
    \widetilde{\mathcal{D}}_-=251 \theta ^2-u \left(69+251 \theta +251 \theta ^2-2 v_-\right) \left(11+3v_-\right)+251 u^2 (1+\theta )^2 \left(2-11 v_-\right)\,,
\end{align}
in terms of $\theta=u\partial_u$ and the leading terms of the corresponding regular period are
\begin{align}
    \widetilde{\varpi}_-(u)=1+u \left(3+v_-\right)+u^2 \left(20-5 v_-\right)+u^3 \left(145+80 v_-\right)+\mathcal{O}(u^4)\,.
\end{align}
Let us denote the coefficients of $\widetilde{\varpi}_-$ by $a_n$, $n\in\mathbb{N}$ such that $\widetilde{\varpi}_-(u)=\sum_{n\ge 0}a_n u^n$.
We have checked to order 300 that $a_n\in \mathbb{Z}[v_-]$.

We now take the Hadamard product of $\widetilde{\varpi}_-(u)$ with itself and define
\begin{align}
    \varpi_-(z)=\sum_{n\ge 0}a_n^2 z^n\,.
\end{align}
This is the fundamental period that is annihilated by the Picard-Fuchs operator~\eqref{eqn:irrationalOpOrder4}.
Again, we observe that the coefficients take values $a_n^2\in \mathbb{Z}[v_-]$.

\subsection{Integral basis and monodromies}
Using a basis of solutions~\eqref{eqn:Qbasis} that is annihilated by the Picard-Fuchs operator~\eqref{eqn:irrationalOpOrder4}, we obtain the flat coordinate~\eqref{eqn:flatCoordinate} and in terms of $q=e^{2\pi {\rm i} t}$ we find that the inverted mirror map takes the form
\begin{align}
    z(q)=q-q^2 \left(32-6 v_-\right)+q^3 \left(345-805 v_-\right)-q^4 \left(11650-77220v_-\right)+\mathcal{O}(q^5)\,.
\end{align}
We have verified to order 300 that the coefficients take values in $\mathbb{Z}[v_-]$.

We now want to assume that there is a basis of solutions that takes the form~\eqref{eqn:integralBasis},~\eqref{eqn:prepotential} for some values $N,\kappa,\sigma,b$ and that has integral monodromy around all of the singular points.
The monodromy matrices can be calculated using numerical analytic continuation and we denote the monodromy around $z=z_{{\rm c},1}$ by $\widetilde{M}_{\text{C},1}$.
It turns out that this is only real if we set
\begin{align}
        \epsilon=-4.7490038227443091059086866245\ldots\,.
        \label{eqn:irrationalEpsilon}
\end{align}
The resulting matrix then takes the form
\begin{align}
    \widetilde{M}_{\text{C},1}=\left(
\begin{array}{cccc}
 1 & 0 & 0 & 0 \\
 \frac{5 (14 \kappa  N-5 b)}{24 \kappa  N^2} & 1 & 0 & \frac{(14 \kappa  N-5 b)^2}{576 \kappa  N^2} \\
 -\frac{25}{\kappa  N^2} & 0 & 1 & -\frac{5 (14 \kappa  N-5 b)}{24 \kappa  N^2} \\
 0 & 0 & 0 & 1 \\
\end{array}
\right)\,.
\end{align}
One immediately sees that the only possible values for $(N,\kappa)$ are $(1,1)$, $(1,5)$, $(1,25)$ and $(5,1)$.
However, if we assume that $N=1$ then we do not know how to interpret the value for $\epsilon$.
Therefore we fix $N=5$ and $\kappa=1$.
We then observe that the value~\eqref{eqn:irrationalEpsilon} can be written as~\eqref{eqn:zeta3term},
\begin{align}
    \begin{split}
    \epsilon=&\zeta(3)\left(\chi(\widetilde{X})+2(n_1+n_2)\right)\\
    &-n_1\left(\text{Li}_3(e^{\frac{2\pi i}{5}})+\text{Li}_3(e^{-\frac{2\pi i}{5}})\right)-n_2\left(\text{Li}_3(e^{\frac{4\pi i}{5}})+\text{Li}_3(e^{-\frac{4\pi i}{5}})\right)\,,
    \end{split}
\end{align}
with $\chi(\widetilde{X})=-120$, $n_1=30$ and $n_2=20$.
The Euler characteristic $\chi=-120$ and the triple intersection number $\kappa=1$ can be recognized as being those of the complete intersection
\begin{align}
    X_{(6,6)}\subset \mathbb{P}^5_{1,1,2,2,3,3}\,.
\end{align}
We therefore propose that the Picard-Fuchs operator~\eqref{eqn:irrationalOpOrder4} can be interpreted as annihilating the periods of the mirror of a $\mathbb{Q}$-factorial degeneration $X$ of $X_{(6,6)}$ with $n_1+n_2=50$ isolated nodes that carries a $\mathbb{Z}_5$ B-field.
Moreover, we claim that there is an analytic small resolution $\rho:\widehat{X}\rightarrow X$ such that $H_2(\widehat{X},\mathbb{Z})=\mathbb{Z}\times\mathbb{Z}_5$ and the exceptional curves resolving $30$ of the nodes represent the class $(0,[1]_5)$ while $20$ nodes are resolved by exceptional curves that represent the class $(0,[2]_5)$.

We also fix $\sigma=-1/2$ and $b=14$ such that the corresponding basis of periods~\eqref{eqn:integralBasis} takes the form
\begin{align}
    \vec{\Pi}=\varpi_0\left(
\begin{array}{c}
 \frac{5}{6}t^3+\frac{7}{12}t+\frac{5 {\rm i} \epsilon }{8 \pi ^3} +\mathcal{O}(q)\\
 -\frac{t^2}{2}-\frac{t}{2}+\frac{7}{60} +\mathcal{O}(q) \\
 \frac{1}{5} \\
 t \\
\end{array}
\right)\,.
\end{align}
Performing the numerical analytic continuation to the remaining singular points in the moduli space, we then obtain the monodromy matrices
\begin{align}
\begin{split}
    M_{\text{LR}}=&\left(
\begin{array}{cccc}
 1 & -5 & 10 & 0 \\
 0 & 1 & -5 & -1 \\
 0 & 0 & 1 & 0 \\
 0 & 0 & 5 & 1 \\
\end{array}
\right)\,,\quad M_{\text{C},1}=\left(
\begin{array}{cccc}
 1 & 0 & 0 & 0 \\
 0 & 1 & 0 & 0 \\
 -1 & 0 & 1 & 0 \\
 0 & 0 & 0 & 1 \\
\end{array}
\right)\,,\quad M_{\text{C},2}=\left(
\begin{array}{cccc}
 -19 & -40 & 40 & -20 \\
 10 & 21 & -20 & 10 \\
 -10 & -20 & 21 & -10 \\
 -20 & -40 & 40 & -19 \\
\end{array}
\right)\,,\\
M_{\text{C},3}=&\left(
\begin{array}{cccc}
 -99 & -400 & 400 & -280 \\
 70 & 281 & -280 & 196 \\
 -25 & -100 & 101 & -70 \\
 -100 & -400 & 400 & -279 \\
\end{array}
\right)\,,\quad M_{\text{I}}=\left(
\begin{array}{cccc}
 81 & 45 & 50 & -55 \\
 -60 & -39 & -30 & 35 \\
 16 & 0 & 21 & -20 \\
 80 & 40 & 55 & -59 \\
\end{array}
\right)\,.
\end{split}
\end{align}
They satisfy the topological relation
\begin{align}
    M_{\text{LR}}M_{\text{C},1}M_{\text{C},2}M_{\text{C},3}M_{\text{I}}=1\,.
\end{align}

It turns out that the point $z=\infty$ is a second point of maximally unipotent monodromy.
After changing coordinates
\begin{align}
    v= \frac{1}{250} \left(123+55 \sqrt{5}\right)\frac{1}{z}\,,
\end{align}
one obtains an operator that annihilates the periods $\vec{\Pi}'(v)=v\vec{\Pi}(v)$.
Numerical analytic continuation shows that
\begin{align}
    \vec{\Pi}(v)=\frac{1}{f_K}T\,\vec{\Pi}\left(\frac{1}{250} \left(123+55 \sqrt{5}\right)\frac{1}{v}\right)\,,
\end{align}
in terms of the transfer matrix $T$ and the K\"ahler transformation $f_K$ given by
\begin{align}
    T=\left(
\begin{array}{cccc}
 5 & 20 & -20 & 14 \\
 0 & -5 & 6 & -5 \\
 0 & -4 & 5 & -4 \\
 4 & 20 & -20 & 15 \\
\end{array}
\right)\,,\quad \quad f_K=\frac{1}{5 \sqrt{\frac{5}{2} \left(123-55 \sqrt{5}\right)}}\frac{1}{v}\,.
\end{align}
Since $\det T=1$, we conclude that $\vec{\Pi}'(v)$ is an integral basis around $v=0$ and therefore this point should admit the same geometric interpretation as $z=0$.
In other words, the moduli space is $\mathbb{Z}_2$ symmetric.

We observe that the matrix
\begin{align}
    \widetilde{T}=TM_{\text{LR}}=\left(
\begin{array}{cccc}
 5 & -5 & 0 & -6 \\
 0 & -5 & 6 & 0 \\
 0 & -4 & 5 & 0 \\
 4 & 0 & -5 & -5 \\
\end{array}
\right)\,,
\end{align}
satisfies $\widetilde{T}^2=1$, which is again just a consequence of the topology of the corresponding paths.

\subsection{$\mathbb{Z}_5$-refined Gopakumar-Vafa invariants}

\begin{table}[ht!]
\begin{align*}
    \begin{array}{|c|cccc|}\hline
    n^{d,0}_g&d=1&2&3&4\\\hline
    g=0&13454 & 169457024 & 5716649675230 & 286377027842279920 \\
    1&60 & 8139096 & 991240932000 & 123239826623645330 \\
    2&1 & 58028 & 50804486175 & 20596996668558428 \\\hline
    \end{array}\\[.1em]
    \begin{array}{|c|cccc|}\hline
    n^{d,\pm 1}_g&d=1&2&3&4\\\hline
    g=0 & 13400 & 169458200 & 5716649623300 & 286377027845065300 \\
    1 & 80 & 8137840 & 991241024060 & 123239826616365525 \\
    2 & 0 & 58450 & 50804421105 & 20596996676934280 \\\hline
    \end{array}\\[.1em]
    \begin{array}{|c|cccc|}\hline
    n^{d,\pm 2}_g&d=1&2&3&4\\\hline
    g=0 & 13425 & 169457400 & 5716649653725 & 286377027843293700 \\
    1 & 70 & 8138660 & 991240969240 & 123239826620972450 \\
    2 & 0 & 58200 & 50804460270 & 20596996671667570 \\\hline
    \end{array}
\end{align*}
\caption{Some $\mathbb{Z}_5$-refined Gopakumar-Vafa invariants of the almost generic Calabi-Yau 3-fold $X_5$.}
    \label{tab:gvX5}
\end{table}
Let us now calculate the torsion refined Gopakumar-Vafa invariants.
The coefficients of $-\log(\Delta_i)/12$, $i=1,2,3$ in the genus one free energy~\eqref{eqn:f1ansatz} are
\begin{align}
    c_1=c_3=1\,,\quad c_2=10\,,
\end{align}
and we choose the propagator ambiguities to be
\begin{align}
    \begin{split}
    s^z_{zz}=&-\frac{2}{z}\,,\quad h^{zz}_z=0\,,\quad h^z_z=0\,,\\
    h_{zz}=&\frac{2-140 z+610 z^2-30750 z^3+10 v_- z \left(3-523 z+34100 z^2\right)}{4z^2\Delta_1\Delta_2\Delta_3}\,,\\
    h_z=&\frac{2-200 z+52500 z^3-3812500 z^4+10 v_- z \left(2-58000 z^2+4228125 z^3\right)}{16z\Delta_1\Delta_3}\,.
    \end{split}
    \label{eqn:propagatorAmbiguitiesX5}
\end{align}
The holomorphic ambiguity at genus $g=2$ then takes the form
\begin{align}\small
    \begin{split}
    f_2=&\frac{1}{14400(\Delta_1\Delta_2\Delta_3)^2}\left[3768-898524 z+59075578 z^2-1071200350 z^3+124882663800 z^4\right.\\
    &\left.-15467556568750 z^5+1429239402781250 z^6-78400873117187500 z^7\right.\\
    &\left.+1817632218750000000 z^8+v_- \left(-24+160202 z-62279454 z^2+11151750200 z^3\right.\right.\\
    &\left.\left.-1447090150900 z^4+171537091031250 z^5-15850507254750000 z^6\right.\right.\\
    &\left.\left.+869479006597656250 z^7-20157850201171875000 z^8\right)\right]\,.
    \end{split}
    \label{eqn:genus2ambiguityX5}
\end{align}
In this way we obtain the topological string free energies on $(X_5,[\pm1]_5)$ for genus $g=0,1,2$.
The free energies on $(X_5,[\pm2]_5)$ are obtained by replacing $v_-$ with $v_+$, as has first been observed in~\cite{Schimannek:2021pau}.
The corresponding free energies for the generic smooth complete intersection $X_{(6,6)}$ are easily obtained as well and have also been calculated for example in~\cite{Huang:2006hq}.
Combining the information from all of these free energies, and using~\eqref{eqn:GVex}, we can then extract the $\mathbb{Z}_5$-refined Gopakumar-Vafa invariants.
Some of the invariants are listed in Table~\ref{tab:gvX5}.

\section{Outlook}
By using relatively elementary geometric constructions, we have obtained several new almost geometric Calabi-Yau threefolds, as well as some new smooth geometries, and deduced the existence of twisted derived equivalences between them.
Above all, we hope to have convinced the reader that such geometries -- as well as the corresponding string compactifications -- are both interesting to study and pose a vast number of open questions that want to be explored.
Let us briefly outline some of the interesting problems that will hopefully be addressed in future work.

Although we have shown the existence of the geometries $X_a$, with $a=1,\ldots,4$ using conifold transitions, we have not been able to identify explicit defining equations.
This is in contrast to the examples in~\cite{Katz:2023zan}, where both the geometries $X^{\rm r}$ and $X$ admitted a description in terms of symmetric determinantal double covers.
At the beginning of the work on this paper, we had hoped to find an analogous construction also for the hyperdeterminantal double covers discussed in Section~\ref{sec:example2}.
Unfortunately, we have so far been unable to realize this idea.
It would be very interesting to explicitly construct the deformations of our geometries $X^{\rm r}$ that remove the nodes of Type B while preserving those of Type A.

Similarly, it would be very interesting to find realizations of the worldsheet theories of strings propagating in the Calabi-Yau backgrounds $(X,\alpha)$ using gauged linear sigma models (GLSM).
In the context of symmetric determinantal double covers, such a GLSM description is also closely related to the interpretation of $(X,\alpha)$ in terms of non-commutative geometry~\cite{Cldraru2009,Katz:2022lyl,Katz:2023zan}.
The relation to non-commutative resolutions is another subject that we have not touched on in this paper and that would be interesting to explore in the future.

The integral structure of the mirror periods, described in Section~\ref{sec:BmodelGenus0}, depends on a choice of values $b$ and $\sigma$.
In the smooth case, that is $N=1$, it is possible to use the Gamma class~\cite{libgober1998chernclassesperiodsmirrors,iritani2009realintegralstructuresquantum,Iritani_2009,Katzarkov:2008hs,Halverson:2013qca} to determine an integral basis of periods on the mirror in terms of the topological invariants of $X$.
This fixes $b$ and the ambiguity in $\sigma$ is only related to the change to a symplectic basis.
For Calabi-Yau backgrounds $(X,\alpha)$ it should be possible to find a generalization of this.

The example $X_5$ discussed in Section~\ref{sec:example3} relied on taking the regular period of the mirror of a degree $5$-polarized genus one curve, at the conifold point that is closest to the large volume limit.
In~\cite{Schimannek:2021pau} it was observed that after fibering the curve over a surface, the conifold point induces a MUM-point in the stringy K\"ahler moduli space of the genus one fibered threefold that is associated with the relative Jacobian fibration together with a non-trivial B-field topology.
The fibers of the Jacobian are sextic curves in $\mathbb{P}^2_{1,2,3}$.
Note that the Hadamard product of the fundamental period of the mirror of this sextic curve with itself is precisely the fundamental period of the mirror of $X_{(6,6)}$.
We have verified that applying the construction from Section~\ref{sec:example3} to families of degree $N$-polarized curves, for $N\le 5$, one obtains a Picard-Fuchs operator that is associated to an almost degeneration of $X_{(6,6)}$ with torsion $B(X)=\mathbb{Z}_N$.
This suggests that every torsion that can be realized by taking the relative Jacobian of a genus one fibration with an $N$-section is also realized for a degeneration of $X_{(6,6)}$.
It would be very interesting to understand this phenomenon and in particular to find the explicit geometries.
One hint is perhaps given by the observation that the Hadamard products can be interpreted as periods of fiber products of elliptic surfaces, see for example~\cite{Schoen1988,kapustka2008fiberproductsellipticsurfaces,Golyshev2023,Elmi:2023hof}.

Whether or not a given smooth Calabi-Yau threefold exhibits a conifold transition to an almost generic geometry can be deduced from the genus zero Gopakumar-Vafa invariants, as is also illustrated in our examples in Section~\ref{sec:example2}.
In order to find new almost generic Calabi-Yau threefolds, a brute force approach that one can take is therefore to just scan over a set of geometries and use the Gopakumar-Vafa invariants to identify interesting examples.
For CICY Calabi-Yau threefolds with $h^{1,1}\le 9$, the genus zero invariants have for example been calculated in~\cite{Carta:2021sms}.
More generally, and this is perhaps the most ambitious question to answer, it would be interesting to know what are the possible torsion groups that can appear in the set of all almost generic Calabi-Yau threefolds.

\appendix
\section{The BCOV ring}
\label{sec:BCOVring}

We will now review the construction of the propagators, the BCOV ring and the polynomial structure of the topological string free energies observed in~\cite{Yamaguchi:2004bt,Alim:2007qj} that allows us to integrate the holomorphic anomaly equations~\eqref{eqn:holAnomaly2}.

Let us first collect some useful relations.
The Christoffel symbols that are associated to the Weil-Petersson metric on $\mathcal{M}_{\text{cs}}$ are given by
\begin{align}
	\Gamma^{z}_{zz}=G^{z\bar{z}}\partial_{\bar{z}}\partial_z\partial_z K\,,\quad \Gamma^{\bar{z}}_{\bar{z}\bar{z}}=G^{z\bar{z}}\partial_{z}\partial_{\bar{z}}\partial_{\bar{z}} K\,.
\end{align}
To illustrate, the action of the covariant derivatives $D_z,D_{\bar{z}}$ on a section
\begin{align}
	V_{z\bar{z}}\in \left(T^*\mathcal{M}_{\text{cs}}\right)^{1,0}\otimes \left(T^*\mathcal{M}_{\text{cs}}\right)^{0,1}\otimes \mathcal{L}^m\otimes\overline{\mathcal{L}}^n\,,
\end{align}
takes the following form
\begin{align}
	D_zV_{z\bar{z}}=\partial_zV_{z\bar{z}}-\Gamma^z_{zz}V_{z\bar{z}}+m(\partial_z K)V_{z\bar{z}}\,,\quad D_{\bar{z}}V_{z\bar{z}}=\partial_{\bar{z}}V_{z\bar{z}}-\Gamma^{\bar{z}}_{\bar{z}\bar{z}}V_{z\bar{z}}+n(\partial_{\bar{z}} K)V_{z\bar{z}}\,.
\end{align}
Using the fact that the connection is metric compatible, i.e. $\left(\partial_z+\Gamma^z_{zz}\right)G^{z\bar{z}}=0$, one can check that
\begin{align}
	D_zC^{zz}_{\bar{z}}=\left(\partial_z+2\Gamma^z_{zz}-2K_z\right) C^{zz}_{\bar{z}}=0\,.
\end{align}

In order to integrate the holomorphic anomaly equations~\eqref{eqn:holAnomaly2}, one first introduces the so-called propagators $S^{zz},S^z$ and $S$ that satisfy the relations
\begin{align}
	\partial_{\bar{z}}S^{zz}=C^{zz}_{\bar{z}}\,,\quad \partial_{\bar{z}}S^z=G_{z\bar{z}}S^{zz}\,,\quad \partial_{\bar{z}}S=G_{z\bar{z}}S^z\,.
\end{align}
They are sections of $\mathcal{L}^{-2}\otimes \text{Sym}^\bullet\left(\left(T\mathcal{M}_{\text{cs}}\right)^{1,0}\right)$.
The Christoffel symbols satisfy~\cite{Bershadsky:1993cx}
\begin{align}
	\partial_{\bar{z}}\Gamma^z_{zz}=2G_{z\bar{z}}-C^{zz}_{\bar{z}}C_{zzz}\,.
\end{align}
This can be integrated with respect to $\bar{z}$ to yield
\begin{align}
	\Gamma^z_{zz}=2K_z-C_{zzz}S^{zz}+s^{z}_{zz}\,,
\end{align}
in terms of $K_z=\partial_zK$ and an ambiguity $s^z_{zz}$ that, at least in all known examples, can be chosen to be a rational function in $z$.
Calculating now
\begin{align}
	\begin{split}
	\partial_{\bar{z}}D_zS^{zz}=&\partial_{\bar{z}}\left(\partial_zS^{zz}+2\Gamma^z_{zz}S^{zz}-2K_zS^{zz}\right)\\
		=&\left(\partial_z+2\Gamma^z_{zz}-2K_z\right) C^{zz}_{\bar{z}}+2(\partial_{\bar{z}}\Gamma^z_{zz})S^{zz}-2G_{z\bar{z}}S^{zz}\\
		=&2\partial_{\bar{z}}S^z-\partial_{\bar{z}}\left(C_{zzz}S^{zz}S^{zz}\right)\,,
	\end{split}
\end{align}
and integrating with respect to $\partial_{\bar{z}}$ gives the relation
\begin{align}
	D_zS^{zz}=2S^z-C_{zzz}S^{zz}S^{zz}+h^{zz}_{z}\,.
\end{align}
Here $h^{zz}_{z}$ is another ambiguity that, at least for suitable choices of $s^{z}_{zz}$, can also be chosen to be a rational function in $z$.
Analogous calculations for $D_z S^z$, $D_z S$ and $D_zK_z$ complete the BCOV ring with the relations
\begin{align}
	\begin{split}
		D_zS^{z}=&2S-C_{zzz}S^zS^{zz}+h^{zz}_zK_z+h^z_z\,,\\
		D_zS=&-\frac12C_{zzz}S^zS^z+\frac12h^{zz}_zK_zK_z+h^z_zK_z+h_z\,,\\
		D_zK_z=&-K_zK_z-C_{zzz}S^z+C_{zzz}S^{zz}K_z+h_{zz}\,,
	\end{split}
\end{align}
in terms of further ambiguities $h^z_z$, $h_z$ and $h_{zz}$.

As was observed in~\cite{Alim:2007qj}, it is useful to introduce the shifted propagators
\begin{align}
    \widetilde{S}^{zz}=S^{zz}\,,\quad \widetilde{S}^z=S^z-S^{zz}K_z\,,\quad \widetilde{S}=S-S^zK_z+\frac12S^{zz}K_zK_z\,.
\end{align}
In terms of the shifted propagators, the BCOV ring takes the form
\begin{align}
    \begin{split}
        \partial_z \widetilde{S}^{zz}=&C_{zzz}\widetilde{S}^{zz}\widetilde{S}^{zz}+2\widetilde{S}^z-2s^z_{zz}\widetilde{S}^{zz}+h^{zz}_z\,,\\
        \partial_z \widetilde{S}^z=&C_{zzz}\widetilde{S}^{zz}\widetilde{S}^z+2\widetilde{S}-s^z_{zz}\widetilde{S}^z-h_{zz}\widetilde{S}^{zz}+h^z_z\,,\\
        \partial_z \widetilde{S}=&\frac12 C_{zzz}\widetilde{S}^z\widetilde{S}^z-h_{zz}\widetilde{S}^z+h_z\,,\\
        \partial_z K_z=&K_zK_z-C_{zzz}\widetilde{S}^{zz}K_z+s^z_{zz}K_z-C_{zzz}\widetilde{S}^z+h_{zz}\,.
    \end{split}
\end{align}

The equation~\eqref{eqn:holAnomaly1} can be rewritten and integrated to obtain
\begin{align}
    \partial_z\mathcal{F}_1(z,\bar{z})=\frac12 C_{zzz}\widetilde{S}^{zz}-\left(\frac{\chi(\widehat{X})}{24}-1\right)K_z+\partial_zf_1(z)\,.
\end{align}
The holomorphic anomaly equations for genus $g\ge 2$~\eqref{eqn:holAnomaly2} can be expressed as
\begin{align}
    \begin{split}
    &\frac{\partial\mathcal{F}_g}{\partial\widetilde{S}^{zz}}-K_z\frac{\partial\mathcal{F}_g}{\partial\widetilde{S}^z}+\frac12K_zK_z\frac{\partial\mathcal{F}_g}{\partial\widetilde{S}}
    =\frac12 \left(D_zD_z\mathcal{F}_{g-1}+\sum\limits_{h=1}^{g-1}D_z\mathcal{F}_hD_z\mathcal{F}_{g-h}\right)\,,
    \end{split}
    \label{eqn:holomorphicAnomalyPropagators}
\end{align}
together with $\partial\mathcal{F}_g/\partial K_z=0$.
The equation can be integrated after calculating the right-hand side of~\eqref{eqn:holomorphicAnomalyPropagators} and collecting the terms with the same order in $K_z$.

The entire anti-holomorphic dependence of the free energies is then absorbed in the propagators $\widetilde{S}^{zz}$, $\widetilde{S}^z$ and $\widetilde{S}$.
Moreover, if one associates to these propagators respective weights $1,2,3$, one can show that the genus $g$ free energy $\mathcal{F}_g$ for $g\ge 2$ is a polynomial in the propagators of weight $3g-3$ with coefficients that are rational functions in $z$~\cite{Yamaguchi:2004bt,Alim:2007qj}.
The term of weight zero in $\mathcal{F}_g$ is called the holomorphic ambiguity $f_g(z)$ and has to be determined, for example, using the behavior of $\mathcal{F}_g$ at boundaries of the moduli space and known Gopakumar-Vafa invariants, as described in Section~\ref{sec:boundaryConditions}.

Finally, let us summarize the behavior of the ambiguities under K\"ahler transformations
\begin{align}
    K_z\rightarrow K_z-\partial_z \log f_K(z)\,,\quad \varpi_0(z)\rightarrow f_K(z)\varpi_0(z)\,,
\end{align}
that has been worked out in~\cite[Appendix B]{Katz:2022lyl}.
The propagator ambiguities transform as
\begin{align}
\begin{split}
    s^z_{zz}\rightarrow& s^z_{zz}+2\partial_z\log f_K\,,\quad h^{zz}_z\rightarrow f_K^{-2}h^{zz}_z\,,\quad h^z_z\rightarrow f_K^{-2}\left(h^z_z+h^{zz}_z\partial_z\log f_K\right)\,,\\
    h_z\rightarrow&f_K^{-2}\left(h_z+\frac12h^{zz}_z\left(\partial_z\log f_K\right)^2+h^z_z\partial_z\log f_K\right)\,,\quad h_{zz}\rightarrow h_{zz}+s^z_{zz}\partial_z\log f_K\,,
\end{split}
\end{align}
while the transformation of the shifted propagators themselves is given by
\begin{align}
\begin{split}
    &\tilde{S}^{zz}\rightarrow f_K^{-2}\tilde{S}^{zz}\,,\quad \tilde{S}^z\rightarrow  f_K^{-2}\left(\tilde{S}^z-\tilde{S}^{zz}\partial_z\log f_K\right)\,,\\
    &\tilde{S}\rightarrow f_K^{-2}\left(\tilde{S}-\tilde{S}^z\partial_z\log f_K(z)+\frac12(\partial_z\log f_K)^2\tilde{S}^{zz}\right)\,.
    \end{split}
\end{align}
The holomorphic ambiguities $f_g(z)$ for $g\ge 1$ transform as
\begin{align}
    f_1\rightarrow& f_1+\left(1-\frac{\chi(\widehat{X})}{24}\right)\log f_K\,,\quad f_{g\ge 2}\rightarrow f_K^{2-2g}f_{g\ge 2}\,.
\end{align}

\section{B-model data, integral period bases and monodromies}
\label{sec:dataXs}
\subsection{Almost generic quintic $X_1$ with $54$ nodes and torsion $B(X_1)=\mathbb{Z}_2$}
\label{sec:dataX1}
The geometry $X_1\subset\mathbb{P}^4$ has been constructed in Section~\ref{sec:example1} and is an almost generic quintic with 54 isolated nodes and torsion $B(X_1)=\mathbb{Z}_2$.
The relevant topological invariants are listed in Table~\ref{tab:geometries}.
The periods of the mirror of $(X_1,[1]_2)$ are annihilated by the Picard-Fuchs operator AESZ~203,
\begin{align}
\begin{split}
    \mathcal{D}_{X_1}=&25 \theta ^4-5 w \theta  \left(10+53 \theta +86 \theta ^2+499 \theta ^3\right)\\
    &+16 w^2\left(-2200-11020 \theta -19776 \theta ^2-13183 \theta ^3+1649 \theta ^4\right)\\
    &+64 w^3\left(6540+51540 \theta +142095 \theta ^2+162000 \theta ^3+39521 \theta ^4\right)\\
    &-38912 w^4\left(174+1019 \theta +2449 \theta ^2+2860 \theta ^3+1370 \theta ^4\right)\\
    &+23658496 w^5 (1+\theta)^4\,,\quad \theta=w\partial_w\,.
    \end{split}
    \label{eqn:AESZ203}
\end{align}
The fundamental period has been obtained in~\eqref{eqn:AESZ203fundamental} and the leading terms are
\begin{align}
    \varpi_0^{X_1}(w)=1+88 w^2+1728 w^3+99576 w^4+4104000 w^5+\mathcal{O}(w^6)\,,
\end{align}
The discriminant polynomial takes the form $\Delta=\Delta_1\Delta_2$ in terms of
\begin{align}
    \Delta_1=1-71w+32w^2\,,\quad \Delta_2=1+32w\,,
\end{align}
and we denote the singular points by
\begin{align}
    w_{{\rm c},1}=\frac{1}{64} \left(71-17 \sqrt{17}\right)\,,\quad w_{{\rm c},2}=\frac{1}{64} \left(71+17 \sqrt{17}\right)\,,\quad w_{{\rm c},3}=-\frac{1}{32}\,,
\end{align}
The Riemann symbol that is associated to the operator~\eqref{eqn:AESZ203} is as follows:
    \begin{align}
    \left\{
    \begin{array}{ccccc}
    0&w_{{\rm c},1}&w_{{\rm c},2}&w_{{\rm c},3}&\infty\\\hline
    0&0&0&0&1\\
    0&1&1&1&1\\
    0&1&1&1&1\\
    0&2&2&2&1
    \end{array}
    \right\}
\end{align}
One can see that the points $w_{{\rm r},i}$, $i=1,2,3$ all correspond to conifold singularities while $w=0$ and $w=\infty$ are both MUM-points.

The leading terms of the inverted mirror map are
\begin{align}
    w(q)=q-2 q^2-275 q^3-4288 q^4-95386 q^5+\mathcal{O}(q^6)\,,
\end{align}
in terms of $q=e^{2\pi{\rm i}t}$. Since the triple intersection number of the smooth quintic is $\kappa=5$, we find that the correctly normalized Yukawa coupling takes the form
\begin{align}
    C_{www}=\frac{5-152 w}{w^3 (1+32 w) \left(1-71 w+32 w^2\right)}\,.
    \label{eqn:yukawaX1}
\end{align}

Using~\eqref{eqn:integralBasis},~\eqref{eqn:prepotential},~\eqref{eqn:zeta3term}, together with the invariants
\begin{align}
    N=2\,,\quad \kappa=5\,,\quad \chi(\widetilde{X}_1)=-200\,,\quad n_1=54\,,
\end{align}
and fixing $b=22$ and $\sigma=0$, we obtain the integral basis of periods
\begin{align}
    \vec{\Pi}_{X_1}=\varpi_0^{X_1}(w)\left(
\begin{array}{c}
 \frac{5}{3}t^3+\frac{11}{12}t-\frac{11 {\rm i} \zeta (3)}{4 \pi ^3}+\mathcal{O}(q) \\
 -\frac{5}{2}t^2+\frac{11}{24}+\mathcal{O}(q) \\
 \frac{1}{2} \\
 t \\
\end{array}
\right)\,.
\label{eqn:X1periods}
\end{align}

In order to verify the choice for $b$ and $\sigma$ we use numerical analytic continuation and calculate the monodromies around the singular points as well as the continuation to the second MUM-point at $w=\infty$.
We denote the monodromies around the points $w=0$, $w=w_{{\rm c},i}$ and $w=\infty$ respectively by $M_{\text{LR}}$, $M_{\text{C},i}$ and $M_{\text{C},i}$ for $i=1,2,3$.
All of the paths are based at the point $w={\rm i}$ and follow a lasso around the singularity in counterclockwise direction with the corresponding action on the period vector being $\vec{\Pi}_{X_1}\rightarrow M\vec{\Pi}_{X_1}$.

The resulting matrices are
\begin{align}
\begin{split}
    M_{\text{LR}}=&\left(
\begin{array}{cccc}
 1 & -2 & 7 & 5 \\
 0 & 1 & -5 & -5 \\
 0 & 0 & 1 & 0 \\
 0 & 0 & 2 & 1 \\
\end{array}
\right)\,,\quad M_{\text{C},1}=\left(
\begin{array}{cccc}
 1 & 0 & 0 & 0 \\
 0 & 1 & 0 & 0 \\
 -1 & 0 & 1 & 0 \\
 0 & 0 & 0 & 1 \\
\end{array}
\right)\,,\quad M_{\text{C},2}=\left(
\begin{array}{cccc}
 1 & 0 & 0 & 0 \\
 -20 & 1 & 0 & 16 \\
 -25 & 0 & 1 & 20 \\
 0 & 0 & 0 & 1 \\
\end{array}
\right)\,,\\
M_{\text{C},3}=&\left(
\begin{array}{cccc}
 -5 & -6 & 12 & -6 \\
 3 & 4 & -6 & 3 \\
 -3 & -3 & 7 & -3 \\
 -3 & -3 & 6 & -2 \\
\end{array}
\right)\,,\quad M_{I}=\left(
\begin{array}{cccc}
 -5 & -4 & 5 & -1 \\
 -55 & -34 & 37 & 14 \\
 -67 & -41 & 45 & 17 \\
 -3 & -3 & 4 & -2 \\
\end{array}
\right)\,,
\end{split}
\end{align}
and satisfy the topological relation
\begin{align}
    M_{\text{C},3}M_{\text{LR}}M_{\text{C},1}M_{\text{C},2}M_{\text{I}}=\text{Id}\,.
\end{align}
The transfer matrix, that connects the basis~\eqref{eqn:X1periods} to the basis of periods~\eqref{eqn:Y1periods} that will be introduced in Appendix~\ref{sec:dataY1}, takes the form
\begin{align}
    \vec{\Pi}_{Y_1}=\frac{1}{2v}T\,\vec{\Pi}_{X_1}\,,\quad T=\left(
\begin{array}{cccc}
 -5 & 0 & 0 & 4 \\
 2 & 5 & -4 & 0 \\
 3 & 1 & -1 & -2 \\
 -1 & 0 & 0 & 1 \\
\end{array}
\right)\,.
    \label{eqn:XY1transfer}
\end{align}

The coefficients of $-\log(\Delta_i)/12$, $i=1,2$ in the genus one free energy~\eqref{eqn:f1ansatz} are
\begin{align}
    c_1=1\,,\quad c_2=3\,,
\end{align}
and we choose the propagator ambiguities to be
\begin{align}
    \begin{split}
    s^w_{ww}=&-\frac{32}{17w}\,,\quad h^{ww}_w=\frac{3}{68}w\,,\quad h^w_w=0\,,\quad h_{ww}=\frac{645-41528 w+37504 w^2}{2312 w^2\Delta_1}\,,\\
    h_w=&\frac{83205-5591864 w-20949504 w^2+832751616 w^3-449511424 w^4}{10690688 w \Delta_1}\,.
    \end{split}
    \label{eqn:propagatorAmbiguitiesX1}
\end{align}
The holomorphic ambiguity at genus $g=2$ is
\begin{align}\small
    \begin{split}
    f_2=&\frac{1}{14149440(\Delta_1\Delta_2)^2}\left(3990346-316218805 w-10021700027 w^2+657164169760 w^3\right.\\
    &\left.+16936021857280 w^4-1748458176512w^5+1705251440689152 w^6\right.\\
    &\left.-1574813915152384 w^7+363341348339712 w^8\right)\,.
    \end{split}
    \label{eqn:genus2ambiguityX1}
\end{align}
Some of the resulting $\mathbb{Z}_2$-refined Gopakumar-Vafa invariants up to genus $g=8$ are listed in Table~\ref{tab:gvX1}.

\subsection{Smooth Calabi-Yau threefold $Y_1$ dual to $(X_1,[1]_2)$}
\label{sec:dataY1}
The Calabi-Yau threefold $Y_1$ is smooth and we conjecture it to be twisted derived equivalent to $(X_1,[1]_2)$.
The relevant topological invariants are listed in Table~\ref{tab:geometries}.

The mirror periods~\eqref{eqn:AESZ202fundamental} are annihilated by the Picard-Fuchs operator AESZ~202,
\begin{align}
\begin{split}
    \mathcal{D}_{Y_1}=&361 \theta ^4-19 v \left(114+779 \theta +2089 \theta ^2+2620 \theta ^3+1370 \theta ^4\right)\\
    &-v^2 \left(25384+95266 \theta +106779 \theta ^2+3916 \theta ^3-39521 \theta ^4\right)\\
    &+8 v^3 \left(3876+17613 \theta +29667 \theta ^2+19779 \theta ^3+1649 \theta ^4\right)\\
    &-80 v^4 (1+\theta) \left(456+1378 \theta +1411 \theta ^2+499 \theta ^3\right)+12800 v^5 (1+\theta )^4\,,\quad \theta=v\partial_v\,.
   \end{split}
   \label{eqn:AESZ202}
\end{align}
This is obtained from the operator AESZ~203~\eqref{eqn:AESZ203} by performing a coordinate and K\"ahler transformation
\begin{align}
    v=\frac{1}{32w}\,,\quad f_K=\frac{1}{2v}\,.
\end{align}
The corresponding fundamental period can be written as
\begin{align}
	\begin{split}
	\varpi_0^{Y_1}(v)=&\sum\limits_{l_1,l_2,n\ge 0}\binom{l_1+l_2}{l_1}\binom{l_1+l_2}{n}\binom{n}{l_1}^2\binom{n}{l_2}^2v^n\\
		=&1+6 v+142 v^2+4920 v^3+205326 v^4+\mathcal{O}(v^5)\,,
        \label{eqn:AESZ202fundamental}
	\end{split}
\end{align}
and the discriminant polynomial takes the form $\Delta=\Delta_1\Delta_2$ in terms of
\begin{align}
    \Delta_1=1-71v+32v^2\,,\quad \Delta_2=1+v\,.
\end{align}
The leading terms of the inverted mirror map are
\begin{align}
    v(q)=q-17 q^2+52 q^3-447 q^4-11351 q^5+\mathcal{O}(q^6)\,,
\end{align}
in terms of $q=e^{2\pi{\rm i}t}$.
The Yukawa coupling takes the form
\begin{align}
    C_{vvv}=\frac{2 (19-20 v)}{v^3 (1+v) \left(1-71 v+32 v^2\right)}\,.
\end{align}
One can check that this is related to~\eqref{eqn:yukawaX1} via
\begin{align}
    C_{vvv}=f_K^2\left(\frac{\partial w}{\partial v}\right)^3C_{www}\bigg\vert_{w\rightarrow 1/(32 v)}\,.
\end{align}

Using~\eqref{eqn:integralBasis},~\eqref{eqn:prepotential},~\eqref{eqn:zeta3term}, together with the invariants
\begin{align}
    N=1\,,\quad \kappa=38\,,\quad b=80\,,\quad \chi(Y_1)=-92\,,
\end{align}
and fixing $\sigma=0$ gives an integral basis of periods
\begin{align}
    \vec{\Pi}_{Y_1}=\varpi_0^{Y_1}(v)\left(\begin{array}{c}
    \frac{19}{3}t^3+\frac{10}{3}t-\frac{23 {\rm i} \zeta (3)}{2 \pi ^3}+\mathcal{O}(q)\\
    -19 t^2+\frac{10}{3}+\mathcal{O}(q)\\
    1\\
    t
    \end{array}
    \right)\,.
\label{eqn:Y1periods}
\end{align}
This is related to the basis~\eqref{eqn:X1periods} via the transfer matrix~\eqref{eqn:XY1transfer}.

The coefficients of $-\log(\Delta_i)/12$, $i=1,2$ in the genus one free energy~\eqref{eqn:f1ansatz} are
\begin{align}
    c_1=1\,,\quad c_2=3\,.
\end{align}
and we choose the propagator ambiguities to be
\begin{align}
    \begin{split}
    s^v_{vv}=&-\frac{36}{17 v}\,,\quad h^{vv}_v=-\frac{3}{17}v^3\,,\quad h^v_v=\frac{3 v^2}{17}\,,\quad h_{vv}=\frac{361-15210 v+7336 v^2}{578 v^2 \Delta_1}\,,\\
    h_v=&\frac{6859-406617 v+209424 v^2+11167684 v^3-5104464 v^4}{1336336 v \Delta _1}\,.
    \end{split}
    \label{eqn:propagatorAmbiguitiesY1}
\end{align}
The holomorphic ambiguity at genus $g=2$ is then given by
\begin{align}
    \begin{split}
    f_2=&\frac{1}{3537360(\Delta_1\Delta_2)^2}\left(338388-46933112 v+1626254502 v^2-53358709 v^3\right.\\
    &\left.+16539083845 v^4+20536380305 v^5-10021700027v^6\right.\\
    &\left.-10119001760 v^7+4086114304 v^8\right)\,.
    \end{split}
    \label{eqn:genus2ambiguityY1}
\end{align}
One can check that~\eqref{eqn:propagatorAmbiguitiesY1} and~\eqref{eqn:genus2ambiguityY1} are respectively obtained from~\eqref{eqn:propagatorAmbiguitiesX1} and~\eqref{eqn:genus2ambiguityX1} by applying the coordinate and K\"ahler transformation as described in Appendix~\ref{sec:BCOVring}.
Some of the resulting Gopakumar-Vafa invariants up to genus $g=8$ are listed in Table~\ref{tab:gvY1}.

\subsection{Almost generic quintic $X_2$ with $48$ nodes and torsion $B(X_2)=\mathbb{Z}_2$}
\label{sec:dataX2}
The geometry $X_2\subset\mathbb{P}^4$ has been constructed in Section~\ref{sec:example1} and is an almost generic quintic with 48 isolated nodes and torsion $B(X_2)=\mathbb{Z}_2$.
As usual, the relevant topological invariants are listed in Table~\ref{tab:geometries}.
The periods of the mirror of $(X_2,[1]_2)$ are annihilated by the Picard-Fuchs operator AESZ~222,
\begin{align}
\begin{split}
    \mathcal{D}=&125 \theta ^4-25 w \left(40+310 \theta +909 \theta ^2+1198 \theta ^3+2578 \theta ^4\right)\\
    &+5 w^2\left(-200+154610 \theta +903615 \theta ^2+1704986 \theta ^3+614413 \theta ^4\right)\\
    &+128 w^3\left(1460280+8638950 \theta +17652258 \theta ^2+14580429 \theta ^3+4343693 \theta^4\right)\\
    &+4096 w^4 \left(214540+2063680 \theta +6627411 \theta ^2+8254720 \theta ^3+2970217\theta ^4\right)\\
    &+4552916992 w^5 (1+2 \theta )^4\,,\quad \theta=w\partial_w\,.
\end{split}
\label{eqn:AESZ222}
\end{align}
The leading terms of the fundamental period are
\begin{align}
    \varpi_0^{X_2}(w)=1+8 w+504 w^2+36800 w^3+3518200 w^4+365275008 w^5+\mathcal{O}(w^6)\,.
\end{align}
The discriminant polynomial takes the form
\begin{align}
    \Delta=1-107 w-8192 w^2\,,
\end{align}
and we denote the singular points by
\begin{align}
    w_{{\rm c},1}=-\frac{107-51 \sqrt{17}}{16384}\,,\quad w_{{\rm c},2}=-\frac{107+51 \sqrt{17}}{16384}\,.
\end{align}
The Riemann symbol that is associated to the operator~\eqref{eqn:AESZ203} is as follows:
    \begin{align}
    \left\{
    \begin{array}{cccc}
    0&w_{{\rm c},1}&w_{{\rm c},2}&\infty\\\hline
    0&0&0&1/2\\
    0&1&1&1/2\\
    0&1&1&1/2\\
    0&2&2&1/2
    \end{array}
    \right\}
\end{align}
One can see that the points $w_{{\rm r},i}$, $i=1,2$ both correspond to conifold singularities while $w=0$ and $w=\infty$ are both MUM-points.

The leading terms of the inverted mirror map are
\begin{align}
    w(q)=q-30 q^2-507 q^3+15896 q^4-864690 q^5+\mathcal{O}(q^6)\,,
\end{align}
in terms of $q=e^{2\pi{\rm i}t}$, and the correctly normalized Yukawa coupling takes the form
\begin{align}
    C_{www}=\frac{5+64 w}{w^3 \left(1-107 w-8192 w^2\right)}\,.
    \label{eqn:yukawaX2}
\end{align}

Using~\eqref{eqn:integralBasis},~\eqref{eqn:prepotential},~\eqref{eqn:zeta3term}, together with the invariants
\begin{align}
    N=2\,,\quad \kappa=5\,,\quad \chi(\widetilde{X}_2)=-200\,,\quad n_1=48\,,
\end{align}
and fixing $b=34$ and $\sigma=0$, we obtain the integral basis of periods
\begin{align}
    \vec{\Pi}_{X_2}=\varpi_0^{X_2}(w)\left(
\begin{array}{c}
 \frac{5}{3}t^3+\frac{17}{12}t-\frac{8 {\rm i} \zeta (3)}{\pi ^3}+\mathcal{O}(q) \\
 -\frac{5}{2}t^2+\frac{17}{24}+\mathcal{O}(q) \\
 \frac{1}{2} \\
 t \\
\end{array}
\right)\,.
\label{eqn:X2periods}
\end{align}

In order to verify the choice for $b$ and $\sigma$ we use again numerical analytic continuation.
We choose the paths as in Appendix~\ref{sec:dataX1} and denoting the monodromy matrices around the points $0$, $\infty$ and $w_{{\rm c},i}$ respectively by $M_{\text{LR}}$, $M_{\text{I}}$ and $M_{\text{C},i}$ for $i=1,2$.
The resulting matrices are
\begin{align}
\begin{split}
    M_{\text{LR}}=&\left(
\begin{array}{cccc}
 1 & -2 & 9 & 5 \\
 0 & 1 & -5 & -5 \\
 0 & 0 & 1 & 0 \\
 0 & 0 & 2 & 1 \\
\end{array}
\right)\,,\quad M_{\text{C},1}=\left(
\begin{array}{cccc}
 1 & 0 & 0 & 0 \\
 0 & 1 & 0 & 0 \\
 -1 & 0 & 1 & 0 \\
 0 & 0 & 0 & 1 \\
\end{array}
\right)\,,\\
M_{\text{C},2}=&\left(
\begin{array}{cccc}
 -2 & -3 & 9 & -3 \\
 1 & 2 & -3 & 1 \\
 -1 & -1 & 4 & -1 \\
 -1 & -1 & 3 & 0 \\
\end{array}
\right)\,,\quad M_{I}=\left(
\begin{array}{cccc}
 2 & 1 & 0 & -2 \\
 1 & 0 & 2 & -4 \\
 1 & 0 & 2 & -3 \\
 1 & 1 & -1 & 0 \\
\end{array}
\right)\,.
\end{split}
\end{align}
They satisfy the relation
\begin{align}
    M_{\text{C},2}M_{\text{LR}}M_{\text{C},1}M_{\text{I}}=-\text{Id}\,,
    \label{eqn:topRelMonX2}
\end{align}
with the sign apparently being a consequence of the orbifold singularity of the moduli space.

The transfer matrix, that connects the basis~\eqref{eqn:X2periods} to the basis of periods~\eqref{eqn:Y2periods} that will be introduced in Appendix~\ref{sec:dataY1}, takes the form
\begin{align}
    \vec{\Pi}_{Y_2}=\frac{1}{2^8v}T\,\vec{\Pi}_{X_2}\,,\quad T=\left(
\begin{array}{cccc}
 -1 & 0 & 0 & 1 \\
 0 & 1 & -1 & 1 \\
 1 & 1 & -2 & 0 \\
 -1 & 0 & 0 & 2 \\
\end{array}
\right)\,.
    \label{eqn:XY2transfer}
\end{align}

The coefficients of $-\log(\Delta)/12$ in the genus one free energy~\eqref{eqn:f1ansatz} is $c_1=1$ and we choose the propagator ambiguities to be
\begin{align}
    \begin{split}
    s^w_{ww}=&-\frac{32}{17}\frac{1}{w}\,,\quad h^{ww}_w=\frac{45}{289}w\,,\quad h^w_w=0\,,\\
    h_{ww}=&-\frac{2 (3581+234496 w)}{289 w \Delta }\,,\quad h_w=\frac{2 \left(21233+1361850 w+1048576 w^2\right)}{83521 \Delta }\,.
    \end{split}
    \label{eqn:propagatorAmbiguitiesX2}
\end{align}
The holomorphic ambiguity at genus $g=2$ is
\begin{align}
    \begin{split}
    f_2=&\frac{1}{7074720\Delta^2}\left(1282901-286296173 w-5645077728 w^2\right.\\
    &\left.+2444748352000 w^3+93682510856192 w^4-146308060938240 w^5\right)\,.
    \end{split}
    \label{eqn:genus2ambiguityX2}
\end{align}
Some of the resulting $\mathbb{Z}_2$-refined Gopakumar-Vafa invariants up to genus $g=2$ are listed in Table~\ref{tab:gvX2}.

\subsection{Almost generic octic $Y_2$ with torsion $B(Y_2)=\mathbb{Z}_2$, dual to $(X_2,[1]_2)$}
\label{sec:dataY2}
The Calabi-Yau threefold $Y_2\subset\mathbb{P}^4_{1,1,1,1,4}$ is an almost generic octic with $96$ isolated nodes and torsion $B(Y_2)=\mathbb{Z}_2$.
We conjecture $(Y_2,[1]_2)$ to be twisted derived equivalent to $(X_2,[1]_2)$.
The relevant topological invariants are listed in Table~\ref{tab:geometries}.

The mirror periods are annihilated by the Picard-Fuchs operator AESZ~225,~\footnote{Strictly speaking, this operator is obtained from AESZ~225 by pullback along a two-to-one covering map.}
\begin{align}\small
\begin{split}
    \mathcal{D}_{Y_2}=&49 \theta ^4-v^2 \left(614656+6573056 \theta +30030336 \theta ^2+26743808 \theta ^3-18512689\theta ^4\right)\\
    &-256 v^4 \left(2401-2248514484 \theta -10357585350 \theta ^2+20313511956 \theta^3\right.\\
    &\left.-8298629499 \theta ^4\right)-1374389534720 v^8 \left(728835313+2910884608 \theta\right.\\
    &\left.+4319532150\theta ^2+2910884608 \theta ^3+60338713 \theta ^4\right)\\
    &-16777216 v^6\left(755776703+2100123438 \theta -63408216 \theta ^2\right.\\
    &\left.+9306516786 \theta ^3-3976811143 \theta^4\right)-16056752206178365420339200 v^{10} (1+\theta )^4\,,
\end{split}
\label{eqn:AESZ225}
\end{align}
where $\theta=v\partial_v$. This is obtained from the operator AESZ~222~\eqref{eqn:AESZ222} by performing a coordinate and K\"ahler transformation
\begin{align}
    w=\frac{1}{2^{21}v^2}\,,\quad f_K=\frac{1}{2^8v}\,.
\end{align}
The leading terms of the corresponding fundamental period are
\begin{align}
	\begin{split}
	\varpi_0^{Y_2}(v)=&1+784 v^2+3226896 v^4+20413907200 v^6+\mathcal{O}(v^8)\,,
        \label{eqn:AESZ225fundamental}
	\end{split}
\end{align}
and the discriminant polynomial takes the form
\begin{align}
    \Delta=1+27392 v^2-536870912 v^4\,.
\end{align}
The leading terms of the inverted mirror map are
\begin{align}
    v(q)=q-6816 q^3+151598640 q^5-5204580617728 q^7+\mathcal{O}(q^9)\,.
\end{align}
The Yukawa coupling takes the form
\begin{align}
    C_{vvv}=\frac{2 \left(1+163840 v^2\right)}{v^3 \left(1+27392 v^2-536870912 v^4\right)}\,.
\end{align}
One can check that this is related to~\eqref{eqn:yukawaX2} via
\begin{align}
    C_{vvv}=f_K^2\left(\frac{\partial w}{\partial v}\right)^3C_{www}\bigg\vert_{w\rightarrow 1/(2^{21}v^2)}\,.
\end{align}

Using~\eqref{eqn:integralBasis},~\eqref{eqn:prepotential},~\eqref{eqn:zeta3term}, together with the invariants
\begin{align}
    N=2\,,\quad \kappa=2\,,\quad \chi(\widehat{Y}_2)=-92\,,
\end{align}
and fixing $b=10$ as well as $\sigma=1/2$ gives an integral basis of periods
\begin{align}
    \vec{\Pi}_{Y_2}=\varpi_0^{Y_2}(v)\left(\begin{array}{c}
    \frac{2}{3}t^3+\frac{5}{12}t+\frac{10 {\rm i} \zeta (3)}{\pi ^3}+\mathcal{O}(q)\\
    -t^2+\frac12 t+\frac{5}{24}+\mathcal{O}(q)\\
    1/2\\
    t
    \end{array}
    \right)\,.
\label{eqn:Y2periods}
\end{align}
This is related to the basis~\eqref{eqn:X2periods} via the transfer matrix~\eqref{eqn:XY2transfer}.

 The coefficients of $-\log(\Delta)/12$ in the genus one free energy~\eqref{eqn:f1ansatz} is $c_1=1$ and we choose the propagator ambiguities to be
\begin{align}\small
    \begin{split}
    s^v_{vv}=&-\frac{21}{17 v}\,,\quad h^{vv}_v=-\frac{1474560}{289}v^3\,,\quad h_{vv}=\frac{8 \left(1+160032 v^2+14831058944 v^4\right)}{289 v^2 \Delta }\,,\\
    h^v_v=&\frac{1474560 v^2}{289}\,,\quad h_v=\frac{16 \left(1-10593420 v^2-275724894208 v^4+7149574359613440 v^6\right)}{83521 v \Delta }\,.
    \end{split}
    \label{eqn:propagatorAmbiguitiesY2}
\end{align}
The holomorphic ambiguity at genus $g=2$ is then given by
\begin{align}\small
    \begin{split}
    f_2=&\frac{1}{3537360\Delta^2}\left(-34065+45743413504 v^2+2503422312448000 v^4\right.\\
    &\left.-12122712112568991744 v^6-1289363338040399057911808v^8\right.\\
    &\left.+12116657366479901428591624192 v^{10}\right)\,.
    \end{split}
    \label{eqn:genus2ambiguityY2}
\end{align}
One can check that~\eqref{eqn:propagatorAmbiguitiesY2} and~\eqref{eqn:genus2ambiguityY2} are respectively obtained from~\eqref{eqn:propagatorAmbiguitiesX2} and~\eqref{eqn:genus2ambiguityX2} by applying the coordinate and K\"ahler transformation as described in Appendix~\ref{sec:BCOVring}.
Some of the resulting Gopakumar-Vafa invariants up to genus $g=2$ are listed in Table~\ref{tab:gvY2}.

\subsection{Almost generic octic $X_3$ with $104$ nodes and torsion $B(X_3)=\mathbb{Z}_3$}
\label{sec:dataX3}
The geometry $X_3\subset\mathbb{P}^4_{1,1,1,1,4}$ has been constructed in Section~\ref{sec:example2} and is an almost generic octic with 104 isolated nodes and torsion $B(X_3)=\mathbb{Z}_3$.
The relevant topological invariants are again listed in Table~\ref{tab:geometries}.

The mirror periods of $(X_3,[\pm 1]_3)$ are annihilated by the Picard-Fuchs operator AESZ~199,
\begin{align}
	\begin{split}
		\mathcal{D}_{X_3}=&\theta ^4-w \left(15+88 \theta +200 \theta ^2+224 \theta ^3+265 \theta ^4\right)\\
		&+6 w^2 \left(468+2718 \theta +6011 \theta ^2+6386 \theta ^3+4325 \theta ^4\right)\\
		&-18 w^3 \left(4824+37422 \theta +102361 \theta ^2+116478 \theta ^3+62015 \theta ^4\right)\\
		&+12393 w^4 \left(200+1140 \theta +2686 \theta ^2+3092 \theta ^3+1465 \theta ^4\right)\\
  &-17065161 w^5 (1+\theta )^4\,,\quad \theta=w\partial_w\,.
	\end{split}
    \label{eqn:AESZ199}
\end{align}
The fundamental period of the mirror of $(X_3,[\pm 1]_3)$ has been obtained in~\eqref{eqn:AESZ199fundamental} and the leading terms are
\begin{align}
    \varpi_0^{X_3}(w)=1+15 w+567 w^2+28113 w^3+1584279 w^4+96217065 w^5+\mathcal{O}(w^6)\,.
\end{align}
The discriminant polynomial takes the form $\Delta=\Delta_1\Delta_2$ in terms of
\begin{align}
    \Delta_1=1-81w\,,\quad \Delta_2=1-w\,,
\end{align}
and we denote the singular points by
\begin{align}
    w_{{\rm c},1}=\frac{1}{81}\,,\quad w_{{\rm c},2}=1\,.
\end{align}
The Riemann symbol that is associated to the operator~\eqref{eqn:AESZ199} is as follows:
    \begin{align}
    \left\{
    \begin{array}{cccc}
    0&w_{{\rm c},1}&w_{{\rm c},2}&\infty\\\hline
    0&0&0&1\\
    0&1/2&1&1\\
    0&1/2&1&1\\
    0&1&2&1
    \end{array}
    \right\}
\end{align}
We therefore see that $w_{{\rm c},2}$ is a conifold point, while $w_{{\rm c},1}$ has the characteristic properties of a hyperconifold point where a cycle $S^3/\mathbb{Z}_2$ shrinks in the mirror.
The points $w=0$ and $w=\infty$ are both MUM-points.

The leading terms of the inverted mirror map are
\begin{align}
    w(q)=q-28 q^2+282 q^3-100 q^4-909 q^5+\mathcal{O}(q^6)\,,
\end{align}
in terms of $q=e^{2\pi{\rm i}t}$, and the Yukawa coupling takes the form
\begin{align}
    C_{www}=\frac{2 (1-51 w)}{w^3 (1-w) (1-81 w)^2}\,.
    \label{eqn:yukawaX3}
\end{align}

Using~\eqref{eqn:integralBasis},~\eqref{eqn:prepotential},~\eqref{eqn:zeta3term}, and the invariants
\begin{align}
    N=3\,,\quad \kappa=2\,,\quad \sigma=0\,,\quad b=4\,,\quad \chi(\widetilde{X})=-296\,,\quad n_1=104\,,
\end{align}
we obtain the basis of periods
\begin{align}
    \vec{\Pi}_{X_3}=\varpi_0^{X_3}(w)\left(
\begin{array}{c}
 t^3+\frac{t}{6}+\frac{5 {\rm i} \zeta (3)}{3 \pi ^3}+\mathcal{O}(q) \\
 -t^2+\frac{1}{18}+\mathcal{O}(q) \\
 \frac{1}{3} \\
 t \\
\end{array}
\right)\,.
\label{eqn:X3periods}
\end{align}

Again we perform numerical analytic continuations to verify that this basis is indeed integral.
We choose the paths as in Appendix~\ref{sec:dataX1} and denoting the monodromy matrices around the points $0$, $\infty$ and $w_{{\rm c},i}$ respectively by $M_{\text{LR}}$, $M_{\text{I}}$ and $M_{\text{C},i}$ for $i=1,2$.
We find that the matrices are given by
\begin{align}
\begin{split}
    M_{\text{LR}}=&\left(
\begin{array}{cccc}
 1 & -3 & 4 & 3 \\
 0 & 1 & -3 & -2 \\
 0 & 0 & 1 & 0 \\
 0 & 0 & 3 & 1 \\
\end{array}
\right)\,,\quad M_{\text{C},1}=\left(
\begin{array}{cccc}
 -1 & 0 & 0 & 0 \\
 0 & 1 & 0 & 0 \\
 1 & 2 & -1 & 0 \\
 -2 & 0 & 0 & 1 \\
\end{array}
\right)\,,\\
M_{\text{C},2}=&\left(
\begin{array}{cccc}
 1 & 0 & 0 & 0 \\
 0 & 1 & 0 & 0 \\
 -25 & -20 & 1 & 0 \\
 -20 & -16 & 0 & 1 \\
\end{array}
\right)\,,\quad
M_{I}=\left(
\begin{array}{cccc}
 -1 & -3 & 4 & -3 \\
 0 & 1 & -3 & 2 \\
 -26 & -56 & 37 & -34 \\
 -22 & -50 & 37 & -33 \\
\end{array}
\right)\,.
\end{split}
\end{align}
They satisfy again the topological relation
\begin{align}
    M_{\text{LR}}M_{\text{C},1}M_{\text{C},2}M_{\text{I}}=\text{Id}\,.
\end{align}

The transfer matrix, that connects the basis~\eqref{eqn:X3periods} to the basis of periods~\eqref{eqn:Y3periods} that will be introduced in Appendix~\ref{sec:dataY3}, takes the form
\begin{align}
    \vec{\Pi}_{Y_3}=\frac{1}{{\rm i}\sqrt{3}v}T\,\vec{\Pi}_{X_3}\,,\quad T=\left(
\begin{array}{cccc}
 -5 & -4 & 0 & 0 \\
 -3 & 4 & -4 & 5 \\
 2 & 3 & -1 & 1 \\
 -1 & -1 & 0 & 0 \\
\end{array}
\right)\,.
\label{eqn:XY3transfer}
\end{align}

The coefficients of $-\log(\Delta_i)/12$, $i=1,2$ in the genus one free energy~\eqref{eqn:f1ansatz} are
\begin{align}
    c_1=7\,,\quad c_2=1\,.
\end{align}
We choose the propagator ambiguities to be
\begin{align}
    \begin{split}
    s^w_{ww}=&-\frac{3}{w}\,,\quad h^{ww}_w=\frac{w}{3}\,,\quad h^w_w=0\,,\\
    h_{ww}=&\frac{5-318 w+243 w^2}{3 w^2\Delta_1\Delta_2}\,,\quad h_w=\frac{25-366 w+243 w^2}{36 w\Delta_2}\,.
    \end{split}
    \label{eqn:propagatorAmbiguitiesX3}
\end{align}
The holomorphic ambiguity at genus $g=2$ is
\begin{align}
    \begin{split}
    f_2=&\frac{1}{2160\Delta_1\Delta_2^2}\left(26425-3125191 w+84786042 w^2\right.\\
    &\left.-164847798 w^3+89052453 w^4-5845851 w^5\right)\,.
    \end{split}
    \label{eqn:genus2ambiguityX3}
\end{align}
Some of the resulting $\mathbb{Z}_3$-refined Gopakumar-Vafa invariants up to genus $g=7$ are listed in Table~\ref{tab:gvX3}.

\subsection{Smooth Calabi-Yau threefold $Y_3$ dual to $(X_3,[\pm 1]_3)$}
\label{sec:dataY3}
The Calabi-Yau threefold $Y_3$, constructed in Section~\ref{sec:example2}, is smooth and we conjecture it to be twisted derived equivalent to $(X_3,[\pm 1]_3)$.
The relevant topological invariants are listed in Table~\ref{tab:geometries}.

The periods of the mirror of $Y_3$ are annihilated by the Picard-Fuchs operator AESZ~194,
\begin{align}
	\begin{split}
		\mathcal{D}_{Y_3}=&289 \theta ^4-17 v \left(119+816 \theta +2200 \theta ^2+2768 \theta ^3+1465 \theta ^4\right)\\
		&+2 v^2 \left(15300+65926 \theta +125017 \theta ^2+131582 \theta ^3+62015 \theta ^4\right)\\
		&-54 v^3 \left(1700+7446 \theta +12803 \theta ^2+10914 \theta ^3+4325 \theta ^4\right)\\
		&+729 v^4 \left(168+700 \theta +1118 \theta ^2+836 \theta ^3+265 \theta ^4\right)-59049 v^5 (1+\theta )^4\,,
	\end{split}
    \label{eqn:AESZ194}
\end{align}
with $\theta=v\partial_v$. This is obtained from the operator AESZ~199~\eqref{eqn:AESZ199} by performing a coordinate and K\"ahler transformation
\begin{align}
    v=\frac{1}{81 w}\,,\quad f_K=\frac{1}{{\rm i}\sqrt{3}v}\,.
\end{align}
The fundamental period takes the form
\begin{align}
    \varpi_0^{Y_3}(v)=&\sum\limits_{l_1,l_2,n\ge 0}\binom{l_1+l_2}{l_1}^2\binom{n}{l_1}^2\binom{n}{l_2}^2v^n\\
    =&1+7v+183v^2+7225v^3+\mathcal{O}(v^4)\,.
\end{align}
The discriminant polynomial takes the form $\Delta=\Delta_1\Delta_2$ in terms of
\begin{align}
    \Delta_1=1-81v\,,\quad \Delta_2=1-v\,.
\end{align}
The leading terms of the inverted mirror map are
\begin{align}
    v(q)=q-20 q^2+94 q^3-1036 q^4-16091 q^5+\mathcal{O}(q^6)\,,
\end{align}
in terms of $q=e^{2\pi{\rm i}t}$.
The Yukawa coupling takes the form
\begin{align}
    C_{vvv}=\frac{2 (17-27 v)}{v^3 (1-81 v) (1-v)^2}\,.
\end{align}
One can check that this is related to~\eqref{eqn:yukawaX3} via
\begin{align}
    C_{vvv}=f_K^2\left(\frac{\partial w}{\partial v}\right)^3C_{www}\bigg\vert_{w\rightarrow 1/(81 v)}\,.
\end{align}

Using~\eqref{eqn:integralBasis},~\eqref{eqn:prepotential},~\eqref{eqn:zeta3term}, together with the invariants
\begin{align}
    N=1\,,\quad \kappa=34\,,\quad b=76\,,\quad \chi(Y_3)=-88\,,
\end{align}
and fixing $\sigma=0$ gives an integral basis of periods
\begin{align}
    \vec{\Pi}_{Y_3}=\varpi_0^{Y_3}(v)\left(\begin{array}{c}
    \frac{17}{3}t^3+\frac{19}{6}t-\frac{11 i \zeta (3)}{\pi ^3}+\mathcal{O}(q)\\
    -17 t^2+\frac{19}{6}+\mathcal{O}(q)\\
    1\\
    t
    \end{array}
    \right)\,.
\label{eqn:Y3periods}
\end{align}
This is related to the basis~\eqref{eqn:X3periods} via the transfer matrix~\eqref{eqn:XY3transfer}.

The coefficients of $-\log(\Delta_i)/12$, $i=1,2$ in the genus one free energy~\eqref{eqn:f1ansatz} are
\begin{align}
    c_1=1\,,\quad c_2=7\,.
\end{align}
and we choose the propagator ambiguities to be
\begin{align}
    \begin{split}
    s^v_{vv}=&-\frac{1}{v}\,,\quad h^{vv}_v=v^3\,,\quad h^v_v=-v^2\,,\\
    h_{vv}=&\frac{6 (-4+9 v)}{v\Delta_1\Delta_2}\,,\quad h_v=\frac{-1+124 v-837 v^2}{4 \Delta_1}\,.
    \end{split}
    \label{eqn:propagatorAmbiguitiesY3}
\end{align}
The holomorphic ambiguity at genus $g=2$ is then given by
\begin{align}
    \begin{split}
    f_2=&\frac{1}{720\Delta_1^2\Delta_2}\left(-11+13573 v-2035158 v^2+84786042 v^3\right.\\
    &\left.-253140471 v^4+173374425 v^5\right)\,.
    \end{split}
    \label{eqn:genus2ambiguityY3}
\end{align}
One can check that~\eqref{eqn:propagatorAmbiguitiesY3} and~\eqref{eqn:genus2ambiguityY3} are respectively obtained from~\eqref{eqn:propagatorAmbiguitiesX3} and~\eqref{eqn:genus2ambiguityX3} by applying the coordinate and K\"ahler transformation as described in Appendix~\ref{sec:BCOVring}.
Some of the resulting Gopakumar-Vafa invariants up to genus $g=7$ are listed in Table~\ref{tab:gvY3}.

\subsection{Almost generic octic $X_4$ with $100$ nodes and torsion $B(X_4)=\mathbb{Z}_3$}
\label{sec:dataX4}
The geometry $X_4\subset\mathbb{P}^4_{1,1,1,1,4}$ has been constructed in Section~\ref{sec:example2} and is an almost generic octic with 100 isolated nodes and torsion $B(X_4)=\mathbb{Z}_3$.
The relevant topological invariants are again listed in Table~\ref{tab:geometries}.

The mirror periods of $(X_4,[\pm 1]_3)$ are annihilated by the Picard-Fuchs operator AESZ~350,
\begin{align}
    \begin{split}
    \mathcal{D}_{X_4}=&\theta ^4-w \left(24+184 \theta +545 \theta ^2+722 \theta ^3+289 \theta ^4\right)\\
    &+24 w^2\left(468+2640 \theta +4861 \theta ^2+2734 \theta ^3+214 \theta ^4\right)\\
    &+576 w^3\left(126+1296 \theta +4252 \theta ^2+5184 \theta ^3+1391 \theta ^4\right)\\
    &+746496 w^4 (1+2\theta )^4\,,\quad \theta=w\partial_w\,.
   \end{split}
   \label{eqn:AESZ350}
\end{align}
The fundamental period of the mirror of $(X_3,[\pm 1]_3)$ has been obtained in~\eqref{eqn:AESZ350fundamental} and the leading terms are
\begin{align}
    \varpi_0^{X_4}(w)=1+24 w+1944 w^2+232800 w^3+34133400 w^4+5649061824 w^5+\mathcal{O}(w^6)\,.
\end{align}
The discriminant polynomial takes the form $\Delta=\Delta_1\Delta_2$ in terms of
\begin{align}
    \Delta_1=1-256w\,,\quad \Delta_2=1-81w\,,
\end{align}
and we denote the singular points by
\begin{align}
    w_{{\rm c},1}=\frac{1}{256}\,,\quad w_{{\rm c},2}=\frac{1}{81}\,.
\end{align}
The Riemann symbol that is associated to the operator~\eqref{eqn:AESZ350} is as follows:
    \begin{align}
    \left\{
    \begin{array}{cccc}
    0&w_{{\rm c},1}&w_{{\rm c},2}&\infty\\\hline
    0&0&0&1/2\\
    0&1&1&1/2\\
    0&1&1&1/2\\
    0&2&2&1/2
    \end{array}
    \right\}
\end{align}
We see that $w_{{\rm c},i}$, $i=1,2$ are both conifold points while $w=0$ and $w=\infty$ are both MUM-points.

The leading terms of the inverted mirror map are
\begin{align}
    w(q)=q-88 q^2+4980 q^3-282064 q^4+12865650 q^5+\mathcal{O}(q^6)\,,
\end{align}
in terms of $q=e^{2\pi{\rm i}t}$, and the Yukawa coupling takes the form
\begin{align}
    C_{www}=\frac{2 (1+24 w)}{w^3 (1-81 w) (1-256 w)}\,.
    \label{eqn:yukawaX4}
\end{align}

Using~\eqref{eqn:integralBasis},~\eqref{eqn:prepotential},~\eqref{eqn:zeta3term}, and the invariants
\begin{align}
    N=3\,,\quad \kappa=2\,,\quad \sigma=0\,,\quad b=20\,,\quad \chi(\widetilde{X})=-296\,,\quad n_1=100\,,
\end{align}
we obtain the basis of periods
\begin{align}
    \vec{\Pi}_{X_4}=\varpi_0^{X_4}(w)\left(
\begin{array}{c}
 t^3+\frac{5}{6}t-\frac{8 {\rm i} \zeta (3)}{3 \pi ^3}+\mathcal{O}(q) \\
 -t^2+\frac{5}{18}+\mathcal{O}(q) \\
 \frac{1}{3} \\
 t \\
\end{array}
\right)\,.
\label{eqn:X4periods}
\end{align}

Again we perform numerical analytic continuations to verify that this basis is indeed integral.
We choose the paths as in Appendix~\ref{sec:dataX1} and denoting the monodromy matrices around the points $0$, $\infty$ and $w_{{\rm c},i}$ respectively by $M_{\text{LR}}$, $M_{\text{I}}$ and $M_{\text{C},i}$ for $i=1,2$.
We find that the matrices are given by
\begin{align}
\begin{split}
    M_{\text{LR}}=&\left(
\begin{array}{cccc}
 1 & -3 & 8 & 3 \\
 0 & 1 & -3 & -2 \\
 0 & 0 & 1 & 0 \\
 0 & 0 & 3 & 1 \\
\end{array}
\right)\,,\quad M_{\text{C},1}=\left(
\begin{array}{cccc}
 1 & 0 & 0 & 0 \\
 0 & 1 & 0 & 0 \\
 -1 & 0 & 1 & 0 \\
 0 & 0 & 0 & 1 \\
\end{array}
\right)\,,\\
M_{\text{C},2}=&\left(
\begin{array}{cccc}
 -3 & -4 & 8 & 0 \\
 0 & 1 & 0 & 0 \\
 -2 & -2 & 5 & 0 \\
 -2 & -2 & 4 & 1 \\
\end{array}
\right)\,,\quad
M_{I}=\left(
\begin{array}{cccc}
 3 & 5 & -4 & 1 \\
 0 & -1 & 3 & -2 \\
 1 & 1 & 1 & -1 \\
 2 & 4 & -3 & 1 \\
\end{array}
\right)\,.
\end{split}
\end{align}
They satisfy the relation
\begin{align}
    M_{\text{LR}}M_{\text{C},1}M_{\text{C},2}M_{\text{I}}=-\text{Id}\,.
\end{align}
We have observed the sign already for $(X_2,[\pm 1]_2)$ in~\eqref{eqn:topRelMonX2} and it is again related to a $\mathbb{Z}_2$ orbifold singularity in the moduli space.

The transfer matrix, that connects the basis~\eqref{eqn:X4periods} to the basis of periods~\eqref{eqn:Y4periods} that will be introduced in Appendix~\ref{sec:dataY4}, takes the form
\begin{align}
    \vec{\Pi}_{Y_4}=\frac{1}{128\sqrt{3}{\rm i}v}T\,\vec{\Pi}_{X_4}\,,\quad T=\left(
\begin{array}{cccc}
 -1 & -1 & 2 & 0 \\
 -1 & -1 & 1 & 1 \\
 0 & 1 & -2 & 1 \\
 -1 & -2 & 2 & 0 \\
\end{array}
\right)\,.
\label{eqn:XY4transfer}
\end{align}

The coefficients of $-\log(\Delta_i)/12$, $i=1,2$ in the genus one free energy~\eqref{eqn:f1ansatz} are
\begin{align}
    c_1=1\,,\quad c_2=2\,.
\end{align}
We choose the propagator ambiguities to be
\begin{align}
    \begin{split}
    s^w_{ww}=&-\frac{9}{5}\frac{1}{w}\,,\quad h^{ww}_w=0\,,\quad h^w_w=0\,,\\
    h_{ww}=&\frac{2 \left(4-693 w+44064 w^2\right)}{25 w^2 \Delta _1 \Delta _2}\,,\quad h_w=\frac{32-8301 w+482031 w^2+2239488 w^3}{1250 w \Delta _1 \Delta _2}\,.
    \end{split}
    \label{eqn:propagatorAmbiguitiesX4}
\end{align}
The holomorphic ambiguity at genus $g=2$ is
\begin{align}
    \begin{split}
    f_2=&\frac{1}{9000(\Delta_1\Delta_2)^2}\left(5754-3810575 w+868299560 w^2-79347971520 w^3\right.\\
    &\left.+2726503856640 w^4-19759378857984 w^5\right)\,.
    \end{split}
    \label{eqn:genus2ambiguityX4}
\end{align}
Some of the resulting $\mathbb{Z}_3$-refined Gopakumar-Vafa invariants up to genus $g=2$ are listed in Table~\ref{tab:gvX4}.

\subsection{Almost generic octic $Y_4$ with $100$ nodes and torsion $B(Y_4)=\mathbb{Z}_2$}
\label{sec:dataY4}
The Calabi-Yau threefold $Y_4\subset\mathbb{P}^4_{1,1,1,1,4}$ is an almost generic octic with $100$ isolated nodes and torsion $B(Y_4)=\mathbb{Z}_2$.
We conjecture $(Y_4,[1]_2)$ to be twisted derived equivalent to $(X_4,[\pm 1]_3)$.
The relevant topological invariants are listed in Table~\ref{tab:geometries}.

The Picard-Fuchs operator that annihilates the periods is related by a two-to-one covering map to operator 4.70 in the database~\cite{cyopdatabase}.
It takes the form
\begin{align}
\begin{split}
    \mathcal{D}_{Y_4}=&\theta ^4-256 v^2 \left(321+1892 \theta +5750 \theta ^2+4804 \theta ^3-1391 \theta^4\right)\\
    &+113246208 v^4 \left(279+1110 \theta +2162 \theta ^2-2306 \theta ^3+107 \theta^4\right)\\
    &+12524124635136 v^6 \left(63+288 \theta +418 \theta ^2+288 \theta ^3-289 \theta^4\right)\\
    &+66483263599150104576 v^8 (1+\theta )^4\,,\quad \theta=v\partial_v\,.
\end{split}
\label{eqn:op470}
\end{align}
This is obtained from the operator AESZ~350~\eqref{eqn:AESZ350} by performing a coordinate and K\"ahler transformation
\begin{align}
    w=\frac{1}{2^{16}3^4v^2}\,,\quad f_K=\frac{{\rm i}}{128\sqrt{3}v}\,.
\end{align}
The leading terms of the corresponding fundamental period are
\begin{align}
	\begin{split}
	\varpi_0^{Y_4}(v)=&1+5136 v^2+98870544 v^4+2900370796800 v^6+\mathcal{O}(v^8)\,,
        \label{eqn:op470fundamental}
	\end{split}
\end{align}
and the discriminant polynomial takes the form $\Delta=\Delta_1\Delta_2\Delta_3\Delta_4$ with
\begin{align}
    \Delta_1=1-256v\,,\quad \Delta_2=1+256v\,,\quad \Delta_3=1-144v\,,\quad \Delta_4=1+144v\,.
\end{align}
The leading terms of the inverted mirror map are
\begin{align}
    v(q)=q-20000 q^3+631687216 q^5-24706844960256 q^7+\mathcal{O}(q^9)\,.
\end{align}
The Yukawa coupling takes the form
\begin{align}
    C_{vvv}=\frac{2 \left(1+221184 v^2\right)}{v^3 (1-144 v) (1+144 v) (1-256 v) (1+256 v)}\,.
\end{align}
One can check that this is related to~\eqref{eqn:yukawaX4} via
\begin{align}
    C_{vvv}=f_K^2\left(\frac{\partial w}{\partial v}\right)^3C_{www}\bigg\vert_{w\rightarrow 1/(2^{16}2^4v^2)}\,.
\end{align}

Using~\eqref{eqn:integralBasis},~\eqref{eqn:prepotential},~\eqref{eqn:zeta3term}, together with the invariants
\begin{align}
    N=2\,,\quad \kappa=2\,,\quad \chi(\widehat{Y}_4)=-96\,,
\end{align}
and fixing $b=-2$ as well as $\sigma=1/2$ gives an integral basis of periods
\begin{align}
    \vec{\Pi}_{Y_4}=\varpi_0^{Y_4}(v)\left(\begin{array}{c}
    \frac{2}{3}t^3-\frac{1}{12}t+\frac{27 {\rm i} \zeta (3)}{2\pi ^3}+\mathcal{O}(q)\\
    -t^2+\frac12 t-\frac{1}{24}+\mathcal{O}(q)\\
    1/2\\
    t
    \end{array}
    \right)\,.
\label{eqn:Y4periods}
\end{align}
This is related to the basis~\eqref{eqn:X4periods} via the transfer matrix~\eqref{eqn:XY4transfer}.

The coefficients of $-\log(\Delta_i)/12$ in the genus one free energy~\eqref{eqn:f1ansatz} are
\begin{align}
    c_1=c_2=2\,,\quad c_3=c_4=1\,,
\end{align}
and we choose the propagator ambiguities to be
\begin{align}\small
    \begin{split}
    s^v_{vv}=&-\frac{7}{5}\frac{1}{v}\,,\quad h^{vv}_v=0\,,\quad h_{vv}=\frac{2 \left(1-62592 v^2+11551113216 v^4\right)}{25 v^2 \Delta _1 \Delta _2 \Delta_3\Delta_4}\,,\\
    h^v_v=&0\,,\quad h_v=\frac{1+1142592 v^2-104450752512 v^4+2137450604396544 v^6}{625 v \Delta _1 \Delta _2 \Delta _3
   \Delta _4}\,.
    \end{split}
    \label{eqn:propagatorAmbiguitiesY4}
\end{align}
The holomorphic ambiguity at genus $g=2$ is then given by
\begin{align}\small
    \begin{split}
    f_2=&\frac{1}{18000(\Delta_1\Delta_2\Delta_3\Delta_4)^2}\left(851-623343360 v^2+96299049287680 v^4\right.\\
    &\left.-5593977319996784640 v^6+130318653389573770444800v^8\right.\\
    &\left.-1044602057385327825524883456 v^{10}\right)\,.
    \end{split}
    \label{eqn:genus2ambiguityY4}
\end{align}
One can check that~\eqref{eqn:propagatorAmbiguitiesY4} and~\eqref{eqn:genus2ambiguityY4} are respectively obtained from~\eqref{eqn:propagatorAmbiguitiesX4} and~\eqref{eqn:genus2ambiguityX4} by applying the coordinate and K\"ahler transformation as described in Appendix~\ref{sec:BCOVring}.
Some of the resulting $\mathbb{Z}_2$-refined Gopakumar-Vafa invariants up to genus $g=2$ are listed in Table~\ref{tab:gvY4}.

\newpage
\section{Gopakumar-Vafa invariants}
\label{sec:gvinvariants}

\begin{table}[ht!]
\centering
\begin{align*}\small
    \begin{array}{|c|c|ccc|}\hline
     \multicolumn{5}{|c|}{\text{$\mathbb{Z}_2$-charge 0:}} \\\hline
\multicolumn{2}{|c|}{n^{d_{\text{m}+\delta},0}_g} & \delta=0 & 1 & 2 \\\hline
 g=0 & d_{\text{m}}=1 & 1444 & 304658 & 158606552 \\
 1 & 3 & 304592 & 1860732685 & 6064955360224 \\
 2 & 4 & 267570 & 37739424984 & 435854072807610 \\
 3 & 4 & 4293 & -7830072 & 1578222789291 \\
 4 & 5 & 23976 & -3764690858 & 122738714371292 \\
 5 & 5 & 648 & -1541025 & -958991871380 \\
 6 & 5 & 0 & -17172 & 650496672 \\
 7 & 7 & 2436736 & 1438165906657 & -67598754138919396 \\
 8 & 8 & -835174680 & 968826151718392 & -12650516382161084226 \\\hline
 \multicolumn{5}{c}{}\\\hline
 \multicolumn{5}{|c|}{\text{$\mathbb{Z}_2$-charge 1:}} \\\hline
\multicolumn{2}{|c|}{n^{d_{\text{m}+\delta},1}_g} & \delta=0 & 1 & 2 \\\hline
 g=0 & d_{\text{m}}=1 & 1431 & 304592 & 158599823 \\
 1 & 3 & 304658 & 1860698940 & 6064954339976 \\
 2 & 4 & 267180 & 37739562916 & 435854066830640 \\
 3 & 4 & 4332 & -7833678 & 1578223373584 \\
 4 & 5 & 25274 & -3764640892 & 122738716243958 \\
 5 & 5 & 452 & -1538100 & -958992660120 \\
 6 & 5 & 10 & -17328 & 650458578 \\
 7 & 7 & 2437264 & 1438164754468 & -67598754038521354 \\
 8 & 8 & -835222320 & 968826139252733 & -12650516383922218924 \\\hline
\end{array}
\end{align*}
\caption{Some $\mathbb{Z}_2$-refined Gopakumar-Vafa invariants of the nodal quintic Calabi-Yau 3-fold $X_1$.}
\label{tab:gvX1}
\end{table}

\begin{table}[ht!]
\centering
\begin{align*}\small
\begin{array}{|c|c|cccc|}\hline
\multicolumn{2}{|c|}{n^{d_{\text{m}+\delta}}_g} & \delta=0 & 1 & 2 & 3 \\\hline
 g=0 & d_{\text{m}}=1 & 226 & 1404 & 17030 & 293738 \\
 1 & 4 & 1 & 1444 & 258932 & 24277736 \\
 2 & 7 & 324 & 793866 & 248716386 & 39631543454 \\
 3 & 9 & 4332 & 17924450 & 10102180988 & 2609846475293 \\
 4 & 10 & -6 & 356360 & 1706381078 & 1219805719268 \\
 5 & 12 & -6993 & 188368868 & 498019451760 & 358127114449608 \\
 6 & 13 & 2034 & 7890152 & 179503623998 & 304240706452288 \\
 7 & 14 & -852 & -355248 & 57096033755 & 263447849158200 \\
 8 & 14 & -18 & 3390 & -39744 & 15250273652 \\\hline
\end{array}
\end{align*}
\caption{Some Gopakumar-Vafa invariants of the smooth Calabi-Yau 3-fold $Y_1$.}
\label{tab:gvY1}
\end{table}

\begin{table}[ht!]
\begin{align*}
    \begin{array}{|c|cccccc|}\hline
    n^{d,0}_g&d=1&2&3&4&5&6\\\hline
    g=0& 1472 & 305314 & 158617728 & 121234592208 & 114652980455616 & 124124873404229616 \\
    1& 0 & 0 & 303936 & 1860664521 & 6064948752704 & 15573649525019620 \\
    2& 0 & 0 & 0 & 267582 & 37739440512 & 435854054403178 \\\hline
    \end{array}\\[.1em]
    \begin{array}{|c|cccccc|}\hline
    n^{d,1}_g&d=1&2&3&4&5&6\\\hline
    g=0& 1403 & 303936 & 158588647 & 121232937792 & 114652908432009 & 124124868713792384 \\
    1 & 0 & 0 & 305314 & 1860767104 & 6064960947496 & 15573650208266880 \\
    2 & 0 & 0 & 0 & 267168 & 37739547388 & 435854085235072 \\\hline
    \end{array}
\end{align*}
\caption{Some $\mathbb{Z}_2$-refined Gopakumar-Vafa invariants of the nodal quintic Calabi-Yau 3-fold $X_2$.}
\label{tab:gvX2}
\end{table}

\begin{table}[ht!]
\begin{align*}
    \begin{array}{|c|ccccc|}\hline
    n^{d,0}_g&d=1&2&3&4&5\\\hline
    g=0&  14752 & 64427360 & 711860273440 & 11596527868780512 & 233938237312624658400 \\
    1 & 0 & 18432 & 10732175296 & 902646121253184 & 50712027457008177856 \\
    2 & 0 & 576 & -8275872 & 6249814150960 & 2700746768622436448 \\\hline
    \end{array}\\[.1em]
    \begin{array}{|c|ccccc|}\hline
    n^{d,1}_g&d=1&2&3&4&5\\\hline
    g=0&   14752 & 64407552 & 711860273440 & 11596528156012800 & 233938237312624658400 \\
    1 & 0 & 22880 & 10732175296 & 902645971452672 & 50712027457008177856 \\
    2 & 0 & 288 & -8275872 & 6249853126784 & 2700746768622436448 \\\hline
    \end{array}
\end{align*}
\caption{Some $\mathbb{Z}_2$-refined Gopakumar-Vafa invariants of the nodal octic Calabi-Yau 3-fold $Y_2$.}
\label{tab:gvY2}
\end{table}

\begin{table}[ht!]
\centering
\begin{align*}\small
    \begin{array}{|c|c|ccc|}\hline
    \multicolumn{5}{|c|}{\text{$\mathbb{Z}_3$-charge 0:}} \\\hline
\multicolumn{2}{|c|}{n^{d_{\text{m}+\delta},0}_g} & \delta=0 & 1 & 2 \\\hline
 g=0 & d_{\text{m}}=1 & 9832 & 42945008 & 474573515464 \\
 1 & 2 & 13960 & 7154786592 & 601764030967148 \\
 2 & 2 & 240 & -5517416 & 4166555676032 \\
 3 & 2 & 6 & -58992 & -58286454232 \\
 4 & 4 & 132054824 & -224854076074112 & 85146750281537603384 \\
 5 & 4 & 103850 & 4021309638688 & -1173847999477576812 \\
 6 & 4 & 1032 & -28769288272 & 48358822676551208 \\
 7 & 4 & 24 & 12510464 & -1342965026795834 \\\hline
 \multicolumn{5}{c}{}\\\hline
 \multicolumn{5}{|c|}{\text{$\mathbb{Z}_3$-charge $\pm 1$:}} \\\hline
\multicolumn{2}{|c|}{n^{d_{\text{m}+\delta},\pm 1}_g} & \delta=0 & 1 & 2 \\\hline
 g=0 & d_{\text{m}}=1 & 9836 & 42944952 & 474573515708 \\
 1 & 2 & 13676 & 7154782000 & 601764030869354 \\
 2 & 2 & 312 & -5517164 & 4166555800856 \\
 3 & 2 & 0 & -59016 & -58286524796  \\
 4 & 4 & 132080488 & -224854074322368 & 85146750281820261952 \\
 5 & 4 & 98800 & 4021309315184 & -1173847999642876344 \\
 6 & 4 & 1560 & -28769261080 & 48358822746595696 \\
 7 & 4 & 0 & 12509312 & -1342965050359712 \\\hline
\end{array}
\end{align*}
\caption{Some $\mathbb{Z}_3$-refined Gopakumar-Vafa invariants of the nodal octic Calabi-Yau 3-fold $X_3$.}
\label{tab:gvX3}
\end{table}

\begin{table}[ht!]
\centering
\begin{align*}\small
\begin{array}{|c|c|cccc|}\hline
 \multicolumn{2}{|c|}{n^{d_{\text{m}+\delta}}_g} & \delta=0 & 1 & 2 & 3 \\\hline
 0 & 1 & 252 & 1696 & 23400 & 459616 \\
 1 & 4 & 10 & 5480 & 817664 & 77409128 \\
 2 & 7 & 9836 & 6333608 & 1585457592 & 245248052408 \\
 3 & 9 & 238784 & 287043880 & 118456675816 & 28192349847392 \\
 4 & 10 & 1040 & 33097216 & 46952799328 & 24514941284396 \\
 5 & 11 & 16 & 1905093 & 16547353944 & 20657473268036 \\
 6 & 13 & -15448 & 5046139144 & 17624475017732 & 18377310005237912 \\
 7 & 14 & 12112 & 1291163144 & 15800875137056 & 31706036574234992 \\\hline
\end{array}
\end{align*}
\caption{Some Gopakumar-Vafa invariants of the smooth Calabi-Yau 3-fold $Y_3$.}
\label{tab:gvY3}
\end{table}

\begin{table}[ht!]
\begin{align*}
    \begin{array}{|c|ccccc|}\hline
    n^{d,0}_g&d=1&2&3&4&5\\\hline
    g=0& 9900 & 42945152 & 474573540660 & 7731018675972640 & 155958824875159222600 \\
    1 & 0 & 13812 & 7154772160 & 601764030615196 & 33808018304612861912 \\
    2 & 0 & 264 & -5514700 & 4166555815600 & 1800497845771437772 \\\hline
    \end{array}\\[.1em]
    \begin{array}{|c|ccccc|}\hline
    n^{d,\pm1}_g&d=1&2&3&4&5\\\hline
    g=0& 9802 & 42944880 & 474573503110 & 7731018674410336 & 155958824875045047100 \\
    1 & 0 & 13750 & 7154789216 & 601764031045330 & 33808018304701746900 \\
    2 & 0 & 300 & -5518522 & 4166555731072 & 1800497845736717562 \\\hline
    \end{array}
\end{align*}
\caption{Some $\mathbb{Z}_3$-refined Gopakumar-Vafa invariants of the nodal octic Calabi-Yau 3-fold $X_4$.}
\label{tab:gvX4}
\end{table}

\begin{table}[ht!]
\begin{align*}
    \begin{array}{|c|ccccc|}\hline
    n^{d,0}_g&d=1&2&3&4&5\\\hline
    g=0&  14752 & 64437760 & 711860273440 & 11596527759342720 & 233938237312624658400 \\
    1 & 0 & 17600 & 10732175296 & 902646194897696 & 50712027457008177856 \\
    2 & 0 & 600 & -8275872 & 6249795070720 & 2700746768622436448 \\\hline
    \end{array}\\[.1em]
    \begin{array}{|c|ccccc|}\hline
    n^{d,1}_g&d=1&2&3&4&5\\\hline
    g=0&  14752 & 64397152 & 711860273440 & 11596528265450592 & 233938237312624658400 \\
    1 & 0 & 23712 & 10732175296 & 902645897808160 & 50712027457008177856 \\
    2 & 0 & 264 & -8275872 & 6249872207024 & 2700746768622436448 \\\hline
    \end{array}
\end{align*}
\caption{Some $\mathbb{Z}_2$-refined Gopakumar-Vafa invariants of the nodal octic Calabi-Yau 3-fold $Y_4$.}
\label{tab:gvY4}
\end{table}

\newpage
\addcontentsline{toc}{section}{References}
\bibliographystyle{utphys}
\bibliography{names}
\end{document}